\documentclass{JHEP3}
\usepackage{psfrag,graphicx}

\usepackage{amsmath,amssymb,amsfonts}
\usepackage{epsfig}
\usepackage{lscape}
\usepackage{multirow}
\usepackage{cite}

\makeatletter

\@addtoreset{equation}{section}

\newcommand\Ref[1]     {Ref.\,\cite{#1}}
\newcommand\Refs[1]    {Refs.\,\cite{#1}}
\newcommand\eqn[1]     {Eq.\,(\ref{#1})}
\newcommand\eqns[2]    {Eqs.\,(\ref{#1}) and~(\ref{#2})}
\newcommand\eqnss[2]   {Eqs.\,(\ref{#1})--(\ref{#2})}
\newcommand\fig[1]     {Fig.\,{\ref{#1}}}
\newcommand\figs[2]     {Figs.\,{\ref{#1}} and~{\ref{#2}}}
\newcommand\figss[2]     {Figs.\,{\ref{#1}}--{\ref{#2}}}
\newcommand\sect[1]    {Sect.\,{\ref{#1}}}
\newcommand\sects[2]   {Sects.\,\ref{#1} and~\ref{#2}}
\newcommand\appx[1]    {Appendix~\ref{#1}}
\newcommand\tab[1]     {Table~\ref{#1}}
\newcommand\nn         {\nonumber}
\newcommand\nt         {\notag}

\def\beq{\begin{equation}}
\def\eeq{\end{equation}}
\def\bsp#1\esp{\begin{split}#1\end{split}}
\def\bal#1\eal{\begin{align}#1\end{align}}
\def\beeq{\begin{eqnarray}}
\def\eeeq{\end{eqnarray}}
\newcommand\bom[1]     {{\mbox{\boldmath $#1$}}}

\newcommand\as         {\ensuremath{\alpha_{\mathrm{s}}}}

\newcommand\aeps{\frac{\as}{2\pi}\,S_\ep\left(\frac{\mu^2}{Q^2}\right)^{\ep}}

\newcommand{\CF}       {C_{\mathrm{F}}}
\newcommand{\CA}       {C_{\mathrm{A}}}
\newcommand{\TR}       {T_{\mathrm{R}}}
\newcommand{\Nc}       {N_{\mathrm{c}}}
\newcommand{\Nf}       {n_{\mathrm{f}}}
\newcommand{\Ns}       {n_{\mathrm{s}}}
\newcommand{\bT}       {\bom{T}}

\newcommand\qb         {{\bar q}}

\newcommand\Oe[1]      {\ensuremath{\mathrm O(\ep^{#1})}}
\newcommand{\ep}       {\epsilon}
\newcommand{\eps}      {\varepsilon}

\newcommand\Real       {\mathop{\mathrm{Re}}\nolimits}

\newcommand\ldot       {\!\cdot\!}

\newcommand{\rd}       {{\mathrm{d}}}
\newcommand{\PS}[1]    {\rd\phi_{#1}}
\newcommand\tsig[1]    {\sigma^{\mathrm{#1}}}
\newcommand\dsig[1]    {\rd\sigma^{{\rm #1}}}
\newcommand\dsiga[2]   {\rd\sigma^{{\rm #1,A}_{\scriptscriptstyle #2}}}

\newcommand\la         {\langle}
\newcommand\ra         {\rangle}

\newcommand{\cM}       {{\cal M}}

\newcommand{\cII}[1] {{\cal I}\kern-4pt *\kern-4pt{\cal I}_{#1}}
\newcommand{\cIJ}     {{\cal I}\kern-4pt *\kern-4pt{\cal J}}
\newcommand{\cJJ}[1] {{\cal J}\kern-4pt *\kern-4pt{\cal J}_{#1}}
\newcommand{\cJI}     {{\cal J}\kern-4pt *\kern-4pt{\cal I}}
\newcommand{\cJK}     {{\cal J}\kern-4pt *\kern-4pt{\cal K}}
\newcommand{\cKJ}[1] {{\cal K}\kern-4pt *\kern-4pt{\cal J}_{#1}}
\newcommand{\cKI}     {{\cal K}\kern-4pt *\kern-4pt{\cal I}}
\newcommand{\cKK}     {{\cal K}\kern-4pt *\kern-4pt{\cal K}}

\newcommand\SME[3]     {|{\cal M}_{#1}^{(#2)}{(#3)}|^2}
\newcommand\M[2]       {\ensuremath{|{\cal{M}}_{#1}^{#2}|^2}}
\newcommand\bra[3]     {\la {\cal M}_{#1}^{#2}#3|}
\newcommand\ket[3]     {|{\cal M}_{#1}^{#2}#3\ra}

\newcommand{\mom}[1]   {\{p\}^{#1}}
\newcommand{\momt}[1]   {\{\ti{p}\}}%^{#1}}
\newcommand{\momti}[1]  {\{\ti{p}\}^{#1}}

\newcommand{\cmap}[1]   {\stackrel{{\mathsf C}_{#1}}{\longrightarrow}}
\newcommand{\smap}[1]   {\stackrel{{\mathsf S}_{#1}}{\longrightarrow}}

\newcommand{\bC}[1]    {\bom{\mathrm C}_{#1}}

\newcommand{\bS}[1]    {\bom{\mathrm S}_{#1}}

\newcommand{\bSCS}[1]  {\bom{\mathrm C}\kern-2pt\bom{\mathrm S}_{#1}}

\def\hP{\hat{P}}

\newcommand{\calS}     {{\cal S}}
\newcommand{\bcA}[1]   {{\bom{\cal A}}_{#1}}%^{#2}}
\newcommand{\cC}[2]    {{\cal C}_{#1}^{#2}}
\newcommand{\cS}[2]    {{\cal S}_{#1}^{#2}}
\newcommand{\cCS}[3]   {{\cal C}_{#1}^{~}{\cal S}_{#2}^{#3}}

\newcommand{\cSCS}[2]  {{\cal C}\kern-2pt{\cal S}_{#1}^{#2}}

\newcommand{\IcC}[2]   {{\mathrm C}_{#1}^{#2}}
\newcommand{\IcS}[2]   {{\mathrm S}_{#1}^{#2}}
\newcommand{\IcCS}[1]  {\mathrm{CS}^{#1}}
\newcommand{\R}[1]     {{\mathrm R}_{#1}}
\newcommand{\tiR}[1]   {\widetilde{{\mathrm R}}_{#1}}
\newcommand{\bI}       {\bom{I}}

\newcommand{\TcS}[2]   {\widetilde{{\mathrm S}}_{#1}^{#2}}

\newcommand{\Y}[2]     {Y_{\ti{#1}\ti{#2},Q}}

\newcommand{\ti}[1]    {\tilde{#1}}
\newcommand{\wti}[1]   {\widetilde{\,#1\,}}

\newcommand\tzz[2]     {z_{#1,#2}}

\newcommand\kT[1]      {k_{\perp,#1}}
\newcommand\kTt[1]     {k_{\perp,#1}}

\renewcommand\S        {{\scriptscriptstyle\rm S\!.}}
\newcommand\NS         {{\scriptscriptstyle\rm N\!.S\!.}}

\title{A subtraction scheme for computing QCD jet cross sections
at NNLO: integrating the subtraction terms I.}

\author{G\'abor Somogyi \\
Institute for Theoretical Physics, University of
Z\"urich\\ Winterthurerstrasse 190, CH-8057 Z\"urich, Switzerland\\
E-mail: \email{sgabi@physik.unizh.ch}}

\author{Zolt\'an Tr\'ocs\'anyi\\
University of Debrecen and Institute of Nuclear Research of the 
Hungarian Academy of Sciences, H-4001 Debrecen P.O.Box 51, Hungary\\
E-mail: \email{Zoltan.Trocsanyi@cern.ch}}

\abstract{In previous articles we outlined a subtraction scheme for 
regularizing doubly-real emission and real-virtual emission in
next-to-next-to-leading order (NNLO) calculations of jet cross sections
in electron-positron annihilation. In order to find the NNLO correction
these subtraction terms have to be integrated over the factorized
unresolved phase space and combined with the two-loop corrections. In
this paper we perform the integration of all one-parton unresolved 
subtraction terms.}

\preprint{arXiv:yymm.nnnn [hep-ph]\\ ZU-TH 12/08}

\begin{document}

\renewcommand{\thefootnote}{\fnsymbol{footnote}}

%%%
%%% Introduction
%%%

\section{Introduction}
\label{sec:intro}

In recent years a lot of effort has been devoted to extending
the subtraction method of computing QCD corrections at the
next-to-leading order (NLO) accuracy to the computation of the radiative
corrections at the next-to-next-to-leading order (NNLO) 
\cite{Gehrmann-DeRidder:2004tv,Gehrmann-DeRidder:2004xe,Weinzierl:2003fx,%
Weinzierl:2003ra,Frixione:2004is,Gehrmann-DeRidder:2005hi,Ridder:2005aw,%
Somogyi:2005xz,Weinzierl:2006ij,Somogyi:2006da,GehrmannDeRidder:2007hr}.
In particular, in \Ref{Somogyi:2006da}, a subtraction scheme was defined
for computing NNLO corrections to QCD jet cross sections to processes
without coloured partons in the initial state and arbitrary number of
massless particles (coloured or colourless) in the final state. That scheme 
can be summarized as follows.

The NNLO correction to any $m$-jet cross section of processes without
coloured partons in the initial state is a sum of three
contributions, the doubly-real, the one-loop singly-unresolved
real-virtual and the two-loop doubly-virtual terms,
\beq
\tsig{NNLO} =
\int_{m+2}\!\dsig{RR}_{m+2} J_{m+2}
+ \int_{m+1}\!\dsig{RV}_{m+1} J_{m+1}
+ \int_m\!\dsig{VV}_m J_m\:.
\label{eq:sigmaNNLO}
\eeq
Here the notation for the integrals indicates that the doubly-real
corrections involve the fully-differential cross section
$\dsig{RR}_{m+2}$ of $m+2$ final-state partons, the real-virtual
contribution involves the fully-differential cross section for the
production of $m+1$ final-state partons at one-loop and the doubly-virtual 
term is an integral of the fully-differential cross section for
the production of $m$ final-state partons at two-loops.  The phase
spaces are restricted by the corresponding jet functions $J_n$ that
define the physical quantity.

In $d=4$ dimensions the three contributions in \eqn{eq:sigmaNNLO} are
separately divergent, but their sum is finite for infrared-safe
observables. (The requirement of infrared safety implies certain analytic
properties of the jet functions $J_n$ that are spelled out in
\Ref{Somogyi:2005xz}.) As explained in \Ref{Somogyi:2006da} we first
continue analytically all integrals to $d=4-2\ep$ dimensions and then
rewrite \eqn{eq:sigmaNNLO} as 
\beq
\tsig{NNLO} =
\int_{m+2}\!\dsig{NNLO}_{m+2}
+ \int_{m+1}\!\dsig{NNLO}_{m+1}
+ \int_m\!\dsig{NNLO}_m\,,
\label{eq:sigmaNNLOfin}
\eeq
that is a sum of integrals,
\beeq
\dsig{NNLO}_{m+2} &=&
\Big\{\dsig{RR}_{m+2} J_{m+2} - \dsiga{RR}{2}_{m+2} J_{m}
     -\Big[\dsiga{RR}{1}_{m+2} J_{m+1} - \dsiga{RR}{12}_{m+2} J_{m}\Big]
\Big\}_{\ep=0}\,,
\label{eq:sigmaNNLOm+2}
\\
\dsig{NNLO}_{m+1} &=&
\Big\{\Big[\dsig{RV}_{m+1} + \int_1\dsiga{RR}{1}_{m+2}\Big] J_{m+1} 
     -\Big[\dsiga{RV}{1}_{m+1} + \Big(\int_1\dsiga{RR}{1}_{m+2}\Big)
\strut^{{\rm A}_{\scriptscriptstyle 1}}
\Big] J_{m} \Big\}_{\ep=0}\,, \;\;\;\;\;\;
\label{eq:sigmaNNLOm+1}
\eeeq
and
\beq
\dsig{NNLO}_{m} =
\Big\{\dsig{VV}_m + \int_2\Big[\dsiga{RR}{2}_{m+2} - \dsiga{RR}{12}_{m+2}\Big]
     +\int_1\Big[\dsiga{RV}{1}_{m+1} + \Big(\int_1\dsiga{RR}{1}_{m+2}\Big)
\strut^{{\rm A}_{\scriptscriptstyle 1}}
\Big]\Big\}_{\ep=0} J_{m}\,,
\label{eq:sigmaNNLOm}
\eeq
each integrable in four dimensions by construction. The forms of the
subtraction terms in \eqns{eq:sigmaNNLOm+2}{eq:sigmaNNLOm+1} are symbolic
in the sense that each approximate cross section is actually a sum of
many terms. The jet function depends on different momenta in each of
those terms; the exact set of momenta for each term can be found in
\Refs{Somogyi:2006da,Somogyi:2006db}.  In \eqn{eq:sigmaNNLOm+2}
$\dsiga{RR}{1}_{m+2}$ and $\dsiga{RR}{2}_{m+2}$ are approximate cross
sections that regularize the doubly-real emission cross section in the
one- and two-parton infrared regions of the phase space, respectively.
The double subtraction due to the overlap of these two terms is
compensated by $\dsiga{RR}{12}_{m+2}$. These terms are defined in
\Ref{Somogyi:2006da} explicitly, where the finiteness of
$\dsig{NNLO}_{m+2}$ is demonstrated also numerically for the case of
$e^+e^- \to 3$ jets ($m=3$). In \Ref{Somogyi:2006cz}, we computed the
integral $\int_1\dsiga{RR}{1}_{m+2}$ and showed that the terms in the
first bracket in \eqn{eq:sigmaNNLOm+1} do not contain $\ep$ poles.
Nevertheless, those terms still lead to divergent integrals due to
kinematical singularities in the one-parton unresolved parts of the
phase space. In \Ref{Somogyi:2006db} we defined explicitly
$\dsiga{RV}{1}_{m+1}$ and
$\Big(\int_1\dsiga{RR}{1}_{m+2}\Big) \strut^{{\rm A}_{\scriptscriptstyle 1}}$,
that regularize the singly-unresolved limits of the real-virtual cross
section and $\int_1\dsiga{RR}{1}_{m+2}$ in turn and demonstrated the
finiteness of the regularized cross section $\dsig{NNLO}_{m+1}$ for the
example of $e^+e^- \to 3$ jets. Thus all formulae relevant for
constructing $\dsig{NNLO}_{m+2}$ and $\dsig{NNLO}_{m+1}$ explicitly are
available. 

In order to finish the definition of the subtraction scheme, one has to
compute the integrals over the factorized one- and two-parton phase
spaces, indicated in \eqn{eq:sigmaNNLOm}. Those integrals have to be
computed in $d$ dimensions and the results have to be presented in the
form of a Laurent expansion in $\ep$. According to the KLN theorem, the 
$\ep$ poles in the expansions have to cancel those in the two-loop
contribution $\dsig{VV}_m$, leading to a finite cross section
$\dsig{NNLO}_m$, that can be integrated in four dimensions. In this paper
we compute the $\ep$-expansion of the one-particle integrals, denoted
formally by $\int_1$, that appear in \eqns{eq:sigmaNNLOm+1}{eq:sigmaNNLOm}.
Once these are known, the only missing ingredient for a complete scheme
for computing NNLO corrections is the $\ep$-expansion of the
two-particle integrals, denoted formally by $\int_2$ in
\eqn{eq:sigmaNNLOm}.

There are several ways to compute the one-particle integrals. If the
singular  integrals in the chosen integration variables are
non-overlapping and  occur in a single point in the integration region
(which can always be mapped to the origin), then one can isolate the
poles using standard residuum subtraction, leading to integrals of
smooth functions that can easily be evaluated numerically. In most
integrals we encounter, there are overlapping singularities in some
variables and/or some variables lead to singular integrals in two
points of the integration region. In the latter case, the integral can
be written as a sum of two  integrals with singularity in a single
point, while the overlaps can be disentangled using sector
decomposition \cite{Heinrich:2008si}, so that residuum subtraction can
be applied. We have written a {\em Mathematica} program for the
extraction of the poles employing these techniques.  The program
produces {\sc Fortran} codes that can be immediately used in a Monte
Carlo integration program such as those in the {\sc Cuba} library
\cite{Hahn:2004fe}.  We used the program {\sc SectorDecompisition} of
\Ref{Bogner:2007cr} to check our integrations.  The integrated
subtraction terms are smooth functions of a single kinematical variable
(in one case three variables) and some parameters.

One can also extend the method of differential equations
\cite{Kotikov:1991pm}, developed for computing multi-loop Feynman
integrals \cite{Remiddi:1997ny}, to the relevant phase-space
integrations. This leads to $\ep$-expansions with analytic, though
rather cumbersome coefficients.  We have used this method for computing
some of the singly-unresolved integrals of the approximate real-virtual
cross section and found numerical agreement with the results of
residuum subtraction \cite{Aglietti:2008}.

A third way to obtain the one-particle integrals is to use the
Mellin-Barnes (MB) technique \cite{Smirnov:1999gc,Tausk:1999vh,Smirnov}
to compute them, leading to $\ep$-expansions with analytic
coefficients. The singly-unresolved integrals defined in this paper
have also been computed via deriving MB representations and using
the MB \cite{Czakon:2005rk} {\em Mathematica} program for obtaining the
analytic continuations and then the coefficients of $\ep$-expansions
\cite{Bolzoni:2008}.

All integrals presented in this paper have been obtained by at least two
independent computations. 

In the next section we set some basic notation used in the rest of the
paper. In general we adopt the colour- and spin-state notation of
\Ref{Catani:1996vz} and the notation for defining the various cross
sections introduced in \Refs{Somogyi:2005xz,Somogyi:2006da,Somogyi:2006db}.
In \sect{sec:RRA1} we compute the integral of the doubly-real 
singly-unresolved counterterms up to $\Oe{2}$ accuracy. 
\sects{sec:RVA1}{sec:IRRA1_A1} repeat the same computations to $\Oe{0}$
for the real-virtual counterterms and for the counterterms of the
integrated approximate cross section, respectively. 
These integrations are very similar to that in \sect{sec:RRA1}, therefore, 
the structure of these sections is the same as that of \sect{sec:RRA1}, 
only the actual expressions differ. \sect{sec:conclude} contains our
conclusions.

%
% Notation
%

\section{Notation}
\label{sec:notation}

\subsection{Matrix elements}
\label{sec:ME}

We consider processes with coloured particles (partons) in the final
states, while the initial-state particles are colourless (typically
electron-positron annihilation into hadrons).  Any number of additional
non-coloured final-state particles is allowed, too, but they will be
suppressed in the notation.  Resolved partons in the final state are
labeled by $i,k,l,\dots$, the unresolved one is denoted by $r$.

We adopt the colour- and spin-state notation of \Ref{Catani:1996vz}. In
this notation the amplitude for a scattering process involving the
final-state momenta $\mom{}$, $\ket{m}{}{(\mom{})}$, is an abstract
vector in colour and spin space, and its normalization is fixed such
that the squared amplitude summed over colours and spins is
\beq
\label{eq:M2}
|\cM_m|^2 = \bra{m}{}{}\ket{m}{}{}\:.
\eeq
This matrix element has the following formal loop expansion:
\beq
\ket{}{}{} = \ket{}{(0)}{} + \ket{}{(1)}{} + \dots\,,
\label{FormalLoopExpansion}
\eeq
where $\ket{}{(0)}{}$ denotes the tree-level contribution,
$\ket{}{(1)}{}$ is the one-loop contribution and the dots stand for
higher-loop contributions, which are not used in this paper. 

Colour interactions at QCD vertices are represented by associating
colour charges $\bT_i$ with the emission of a gluon from each
parton $i$.  In the colour-state notation, each vector $\ket{}{}{}$ is
a colour-singlet state, so colour conservation is simply
\beq
\biggl(\sum_j \bT_j \biggr) \,\ket{}{}{} = 0\,,
\label{eq:colourcons}
\eeq
where the sum over $j$ extends over all the external partons of the
state vector $\ket{}{}{}$, and the equation is valid order by order in
the loop expansion of \eqn{FormalLoopExpansion}. 

Using the colour-state notation, we define the two-parton colour-correlated
squared tree amplitudes as
\beq
|\cM^{(0)}_{(i,k)}(\{p\})|^2 \equiv
\bra{}{(0)}{(\{p\})} \,\bT_i \ldot \bT_k \, \ket{}{(0)}{(\{p\})}
\label{eq:colam2}
\eeq
and similarly the three-parton colour-correlated squared
tree amplitudes, $|\cM^{(0)}_{(i,k,l)}|^2$ for $i$, $k$ and $l$ being
different,
\beq
|\cM^{(0)}_{(i,k,l)}|^2 \equiv
\sum_{a,b,c} f_{abc} \bra{}{(0)}{} T^a_i T^b_k T^c_l \ket{}{(0)}{}
\,.
\label{eq:colam3}
\eeq

The colour-charge algebra for the product 
$(\bT_i)^n (\bT_k)^n \equiv \bT_i \ldot \bT_k$ is:
\beq
\bT_i \ldot \bT_k =\bT_k \ldot \bT_i \quad  {\rm if}
\quad i \neq k; \qquad \bT_i^2= C_i\:.
\label{eq:colalg}
\eeq
Here $C_i$ is the quadratic Casimir operator in the representation of
particle $i$ and we have $\CF= \TR(\Nc^2-1)/\Nc= (\Nc^2-1)/(2\Nc)$ in
the fundamental and $\CA=2\,\TR \Nc=\Nc$ in the adjoint representation,
i.e.~we are using the customary normalization $\TR=1/2$.

\subsection{Cross sections}
\label{sec:xsecs}

In writing the final results in our notation we shall use the following
cross sections:
\begin{itemize}
\item
the Born cross section of producing $m$ final-state partons,
\beq
\dsig{B}_{m} = {\cal N}\;\sum_{\{m\}}\PS{m}{(\mom{})}\frac{1}{S_{\{m\}}}
\SME{m}{0}{\mom{}}
\,,
\label{eq:dsigBm}
\eeq
\item
the Born cross section of producing $m+1$ final-state partons,
called real correction,
\beq
\dsig{R}_{m+1} = {\cal N}\sum_{\{m+1\}}\PS{m+1}{(\mom{})}\frac{1}{S_{\{m+1\}}}
\SME{m+1}{0}{\mom{}}
\,,
\label{eq:dsigRm+1}
\eeq
\item
and the one-loop correction $\dsig{V}_{m}$ to the $m$-parton Born cross
section, called virtual correction,
\beq
\dsig{V}_{m} = {\cal N}\;\sum_{\{m\}}\PS{m}{(\mom{})}\frac{1}{S_{\{m\}}}
\:2\Real \la {\cal M}_{m}^{(0)}(\mom{})|{\cal M}_{m}^{(1)}(\mom{}) \ra\,,
\label{eq:dsigRVm+1}
\eeq
\end{itemize}
where ${\cal N}$ includes all QCD-independent factors and
$\PS{n}{(\mom{})}$ is the $d$-dimensional phase space for
$n$ outgoing particles with momenta
$\mom{} \equiv \{p_1,\dots,p_{n}\}$ and total momentum $Q$,
\beq
\PS{n}(p_1,\ldots,p_n;Q) = 
\prod_{i=1}^{n}\frac{\rd^d p_i}{(2\pi)^{d-1}}\,\delta_+(p_i^2)\,
(2\pi)^d \delta^{(d)}\left(Q-\sum_{i=1}^{n}p_i\right)\,.
\label{eq:PSn}
\eeq
The symbol $\sum_{\{n\}}$ denotes summation over the different 
subprocesses and $S_{\{n\}}$ is the Bose symmetry factor for identical
particles in the final state.

%%%
%%% Integrals of the doubly-real singly-unresolved counterterms
%%%

\section{Integrals of the doubly-real singly-unresolved counterterms}
\label{sec:RRA1}

The doubly-real singly-unresolved approximate cross section times
the jet function reads
\beeq
&&
\dsiga{RR}{1}_{m+2}J_{m+1} = {\cal N}\sum_{\{m+2\}}
\PS{m+2}(\mom{})\frac{1}{S_{\{m+2\}}}
\nn\\&&\qquad\times
\sum_{r} \Bigg[\sum_{i\ne r} 
\frac{1}{2} \cC{ir}{(0,0)}(\mom{})J_{m+1}(\momti{(ir)})
\nn\\&&\qquad\qquad\qquad
+ \left(\cS{r}{(0,0)}(\mom{})
- \sum_{i\ne r} \cCS{ir}{r}{(0,0)}(\mom{})\right)J_{m+1}(\momti{(r)})
\Bigg]\,.
\label{eq:dsigRRA1}
\eeeq
The precise meaning of the counterterms $\cC{ir}{(0,0)}$,
$\cS{r}{(0,0)}$ and $\cCS{ir}{r}{(0,0)}$ and the definition of
the tilded sets of momenta $\momti{(ir)}$ and $\momti{(r)}$ (arguments of
the jet functions) were spelled out explicitly in \Ref{Somogyi:2006da}.
The integral of this approximate cross section over the one-particle
factorized phase space was presented in \Ref{Somogyi:2006cz}.
Nevertheless, we recompute it here for two reasons. On the one hand, we
use slightly generalized subtraction terms, while on the other hand in
a NNLO computation we need the expansion of the integrals up to
$\Oe{2}$ accuracy. In the following, we recall the definitions of the
subtraction terms only to the extent needed to compute their integrals
over the factorized phase spaces.

%
% Integral of the collinear counterterm
%

\subsection{Integral of the collinear counterterm}
\label{ssec:intC00}

The collinear counterterm reads
\beq
\bsp
\cC{ir}{(0,0)}(\mom{}) &=
8\pi\as\mu^{2\ep}\frac{1}{s_{ir}}
\bra{m+1}{(0)}{(\momti{(ir)})}
\hP_{f_i f_r}^{(0)}(\tzz{i}{r},\tzz{r}{i},\kTt{i,r};\ep)
\ket{m+1}{(0)}{(\momti{(ir)})}\\
&\times
(1-\alpha_{ir})^{2d_{0}-2m(1-\ep)}\Theta(\alpha_{0}-\alpha_{ir})\,,
\esp
\label{eq:Cir00}
\eeq
where $s_{ir} = 2p_i\ldot p_r$, $\hP_{f_i f_r}^{(0)}$ is the
Altarelli--Parisi splitting kernel in $d = 4-2\ep$ dimensions for the
splitting process $f \rightarrow f_i + f_r$ and $\alpha_0\in (0,1]$. 
The momentum fractions are defined as
\beq
\tzz{i}{r}=\frac{y_{iQ}}{y_{(ir)Q}}
\qquad\mbox{and}\qquad
\tzz{r}{i}=\frac{y_{rQ}}{y_{(ir)Q}}\,,
\label{eq:defzri}
\eeq
where $y_{jQ} = s_{jQ}/Q^2 = 2p_j\ldot Q/Q^2$, ($j = i,\,r$ etc.)
and $y_{(ir)Q} = y_{iQ} + y_{rQ}$.  With this definition, we clearly
have $\tzz{i}{r}+\tzz{r}{i}=1$.  In order to render the integrated
counterterm $m$-independent (see \eqn{eq:I1Cir00} below)
and to reduce the CPU time necessary for the numerical integration,
we have included two harmless factors in the second line of
\eqn{eq:Cir00} as compared to the original definitions in 
\Ref{Somogyi:2006da}.  We give a detailed discussion
of these modifications in \appx{app:modification}. 

The matrix elements on the right hand side of \eqn{eq:Cir00} are
evaluated with the $m+1$ momenta 
$\momti{(ir)} = \{\ti{p}_1,\ldots,\ti{p}_{ir},\ldots,\ti{p}_{m+2}\}$, 
that are defined by a specific mapping, 
\beq
\mom{}\cmap{ir}\momti{(ir)}\,,
\label{eq:cmap}
\eeq
of the original $m+2$ momenta $\mom{}$.  This mapping leads to an exact
factorization of the $(m+2)$-particle phase space such that we have
\beq
\PS{m+2}(\mom{};Q)=\PS{m+1}(\momti{(ir)};Q)
\: [\rd p_{1;m+1}^{(ir)}(p_r,\ti{p}_{ir};Q)]\,.
\label{eq:PSfact_Cir}
\eeq

We write the result of integrating the collinear subtraction term as
given in \eqn{eq:Cir00} over the factorized phase space
$[\rd p_{1;m+1}^{(ir)}]$ in the following
form:
\beq
\int_1\cC{ir}{(0,0)}(\mom{}) =
\aeps
\IcC{ir}{(0)}(y_{\wti{ir}Q};\ep)\, \bT_{ir}^2 \,
\SME{m+1}{0}{\momti{(ir)}}\,,
\label{eq:I1Cir00}
\eeq
where
\beq
S_\ep = \int\!\frac{\rd^{(d-3)}\Omega}{(2\pi)^{d-3}} =
\frac{(4\pi)^\ep}{\Gamma(1-\ep)}\,,
\eeq
$y_{\wti{ir}Q} \equiv s_{\wti{ir}Q}/Q^2 = 2 \ti{p}_{ir} \ldot Q/Q^2$
and the arguments of the function
$\IcC{ir}{(0)}(y_{\wti{ir}Q};\ep)$ indicate that this
function is independent of $m$. In order to lighten the notation
throughout the paper we do not show the explicit dependence of
$\IcC{ir}{(0)}$ on $\alpha_0$ and $d_0$.

The factorized phase space measure in \eqn{eq:PSfact_Cir} may be
written in several equivalent ways. In this paper choose the form of 
a convolution:
\beq
[\rd p_{1;m+1}^{(ir)}(p_{r},\ti{p}_{ir};Q)] =
\int_{0}^{1}\!\rd\alpha_{ir}\,(1-\alpha_{ir})^{2m(1-\ep)-1}
\,\frac{s_{\wti{ir}Q}}{2\pi}
\,\PS{2}(p_i,p_r; p_{(ir)})
\,,
\label{eq:dp1Cir}
\eeq
where
$p_{(ir)}^\mu = (1-\alpha_{ir}) \ti{p}_{ir}^\mu + \alpha_{ir} Q^\mu$.
The collinear momentum mapping and the implied factorization of the 
phase space measure are represented graphically in \fig{fig:PSCir}.  
\FIGURE{
\label{fig:PSCir}
\includegraphics{figs/PSCir.epsi}
\caption{Graphical representation of the collinear momentum mapping 
and the implied phase space factorization.}
}
The picture on the left shows the $(m+2)$-particle phase space 
$\PS{m+2}(\mom{};Q)$, where in the circle we have indicated the 
number of momenta. The picture on the right corresponds to 
\eqns{eq:PSfact_Cir}{eq:dp1Cir}: 
the two circles represent the $(m+1)$-particle phase space 
$\PS{m+1}(\momti{(ir)};Q)$ and the two-particle phase 
space $\PS{2}(p_i,p_r;p_{(ir)})$ respectively, while the symbol $\otimes$
stands for the convolution over $\alpha_{ir}$, as precisely defined in 
\eqn{eq:dp1Cir}.

With the chosen form of the factorized phase-space measure in
\eqn{eq:dp1Cir}, we can express the functions $\IcC{ir}{(0)}$ as
\beq
\bsp
&
\IcC{ir}{(0)}(y_{\wti{ir}Q};\ep) =
\frac{(4\pi)^2}{S_\ep}(Q^2)^{\ep}
\label{eq:ICir0}\\ &\qquad \times
\int_{0}^{\alpha_{0}}\!\rd\alpha\, (1-\alpha)^{2d_{0}-1}
\frac{s_{\wti{ir}Q}}{2\pi}\PS{2}(p_{i},p_{r};p_{(ir)})
\frac{1}{s_{ir}}P_{f_i f_r}^{(0)}(\tzz{i}{r},\tzz{r}{i};\ep)
\frac{1}{\bT_{ir}^2}\,.
\esp
\eeq
In writing the right hand side of \eqn{eq:ICir0} we have used that
because $\kT{i,r}$ as defined in \Ref{Somogyi:2006db} is orthogonal to
$\ti{p}_{ir}$, the spin correlations generally present in
\eqn{eq:Cir00} vanish after azimuthal integration.  Therefore, we may
replace the splitting kernels $\hP_{f_i f_r}^{(0)}$ by their azimuthally
averaged counterparts $P_{f_i f_r}^{(0)}$ when integrating the subtraction
term over the factorized phase space.  These spin-averaged splitting
kernels  depend, in general, on $\tzz{i}{r}$ and $\tzz{r}{i}$, with the
constraint
\beq
\tzz{i}{r} + \tzz{r}{i} = 1\,,
\label{eq:sumzizr}
\eeq
and are listed in \appx{app:APfcns}.
The phase-space measure is symmetric under the $i\leftrightarrow r$
interchange, thus integrals containing integrands that are linear
combinations of ratios of the momentum fractions, as in \eqn{eq:ICir0},
can be expressed as linear combinations of integrals with integrands
depending only on $\tzz{r}{i}^k$. Therefore, we need to evaluate the 
following integrals:
\beq
\,\frac{(4\pi)^2}{S_\ep}(Q^2)^{\ep}\,
\int_0^{\alpha_0}\!\rd \alpha_{ir}\,(1-\alpha_{ir})^{2d_0-1}
\,\frac{s_{\wti{ir}Q}}{2\pi}
\int\!\PS{2}(p_i,p_r; p_{(ir)})
\, \frac{ \tzz{r}{i}^k}{s_{ir}}
\,,
\label{eq:I0k}
\eeq
for the values $k=-1,0,1,2$.
As explained in the introduction, in this paper we use sector decomposition 
and residuum subtraction to compute the $\eps$-expansion of \eqn{eq:I0k}.
It is also possible to evaluate these integrals analytically. The
details are given in \Ref{Aglietti:2008}.

%
% Integral of the soft-type counterterms
%

\subsection{Integrals of the soft-type counterterms}
\label{ssec:intS00}

We refer to the soft and soft-collinear counterterms together
as soft-type because they both use the momentum mapping appropriate
to the soft subtraction term. We define these counterterms 
as follows (note again the harmless factors as compared to the
definitions in \Ref{Somogyi:2006da})
\bal
\cS{r}{(0,0)}(\mom{}) &=
-8\pi\as\mu^{2\ep}\sum_{i}\sum_{k\ne i} \frac12 \calS_{ik}(r)
\SME{m+1;(i,k)}{0}{\momti{(r)}}
\nt\\[2mm]
&\times
(1-y_{rQ})^{d'_{0}-m(1-\ep)}\Theta(y_{0}-y_{rQ})\,,
\label{eq:Sr00}
\\[3mm]
\cCS{ir}{r}{(0,0)}(\mom{}) &=
8\pi\as\mu^{2\ep} \frac{1}{s_{ir}}
\frac{2\tzz{i}{r}}{\tzz{r}{i}}\,\bT_i^2\,
\SME{m+1}{0}{\momti{(r)}}
\nt\\[2mm]
&\times
(1-y_{rQ})^{d'_{0}-m(1-\ep)}\Theta(y_{0}-y_{rQ})\,,
\label{eq:CirSr00}
\eal
where
\beq
\calS_{ik}(r) = \frac{2s_{ik}}{s_{ir}s_{kr}}\,,
\label{eq:defSikr}
\eeq
and $y_0 \in (0,1]$.  The two-parton colour-correlated squared  matrix
element $\M{m+1;(i,k)}{(0)}$ in \eqn{eq:Sr00} is defined in
\eqn{eq:colam2}. The momentum fractions $\tzz{i}{r}$ and $\tzz{r}{i}$ were 
defined in \eqn{eq:defzri}.
The matrix elements on the right hand sides of
\eqns{eq:Sr00}{eq:CirSr00} are evaluated with the $m+1$ momenta 
$\momti{(r)} = \{\ti{p}_1,\ldots,\ti{p}_{m+2}\}$
($p_r$ is missing from the set), obtained from the original $m+2$
momenta, $\mom{}$, by a mapping,
\beq
\mom{}\smap{r}\momti{(r)}\,.
\label{eq:smap}
\eeq
This mapping leads to an exact factorization of the $(m+2)$-particle 
phase space,
\beq
\PS{m+2}(\mom{};Q)=\PS{m+1}(\momti{(r)};Q)
\: [\rd p_{1;m+1}^{(r)}(p_r;Q)]\,.
\label{eq:PSfact_Sr}
\eeq

We write the result of integrating the soft-type subtraction terms over
the factorized phase space $[\rd p_{1;m+1}^{(r)}]$ as follows:
\bal
\int_1\cS{r}{(0,0)}(\mom{}) &=
\aeps
\sum_i\sum_{k\ne i}\IcS{ik}{(0)}(\Y{i}{k};\ep)
\SME{m+1;(i,k)}{0}{\momti{(r)}}\,,
\label{eq:I1Sr00}
\\
\int_1\cC{ir}{}\cS{r}{(0,0)}(\mom{}) &=
\aeps
\IcCS{(0)}(\ep)\, \bT_i^2\,
\SME{m+1}{0}{\momti{(r)}}\,.
\label{eq:I1CirSr00}
\eal
The functions $\IcS{ik}{(0)}(\Y{i}{k};\ep)$ and $\IcCS{(0)}(\ep)$ 
are again independent of $m$ as the arguments indicate and as before, 
we lighten the notation throughout by not showing the explicit 
dependence of the integrated subtraction terms on $y_0$ and $d'_0$.
Also, $\IcS{ik}{(0)}(\Y{i}{k};\ep)$ only depends on the
combination of variables
\beq
\Y{i}{k} = 
\frac{y_{\ti{i}\ti{k}}}{y_{\ti{i}Q}y_{\ti{k}Q}}\,.
\label{eq:defYikQ}
\eeq

As in the collinear case, the one-particle factorized phase space may
also be written in the form of a convolution
\beq
[\rd p_{1;m+1}^{(r)}(p_r;Q)] =
\int_{0}^{1}\rd y_{rQ}(1-y_{rQ})^{m(1-\ep)-1}
\frac{Q^{2}}{2\pi}\PS{2}(p_{r},K;Q)
\label{eq:dp1Sr}
\eeq
where 
the timelike momentum $K$ is massive with $K^{2}=(1-y_{rQ})Q^{2}$.
We show the soft momentum mapping and the implied phase space 
factorization in \fig{fig:PSSr}.  
\FIGURE{
\label{fig:PSSr}
\includegraphics{figs/PSSr.epsi}
\caption{Graphical representation of the soft momentum mapping 
and the implied phase space factorization.}
}
The picture on the left shows again the $(m+2)$-particle phase space 
$\PS{m+2}(\mom{};Q)$, while the picture on the right corresponds to 
\eqns{eq:PSfact_Sr}{eq:dp1Sr}: 
the two circles represent the two-particle phase 
space $\PS{2}(p_r,K;Q)$ and the $(m+1)$-particle phase space 
$\PS{m+1}(\momti{(r)};Q)$ respectively. The symbol $\otimes$
stands for the convolution over $y_{rQ}$ as defined in \eqn{eq:dp1Sr}.

Using the definition of the subtraction terms in \eqns{eq:Sr00}{eq:CirSr00}
and the chosen form of the factorized phase space, we find that the
functions $\IcS{ik}{(0)}(\Y{i}{k};\ep)$ and
$\IcCS{(0)}(\ep)$ can be written as
\bal
\IcS{ik}{(0)}(\Y{i}{k};\ep) &=
-\frac{(4\pi)^2}{S_\ep}(Q^2)^{\ep}
\int_{0}^{y_{0}}\!\rd y\,(1-y)^{d'_{0}-1}
\frac{Q^{2}}{2\pi}\PS{2}(p_{r},K;Q)
\frac{1}{2}\calS_{ik}(r)
\label{eq:ISik0}
\\[2mm]
\IcCS{(0)}(\ep) &=
\frac{(4\pi)^2}{S_\ep}(Q^2)^{\ep}
\int_{0}^{y_{0}}\!\rd y\,(1-y)^{d'_{0}-1}
\frac{Q^{2}}{2\pi}\PS{2}(p_{r},K;Q)
\frac{1}{s_{ir}}\frac{2\tzz{i}{r}}{\tzz{r}{i}}\,.
\label{eq:ICS0}
\eal
The computation of these integrals is fairly straightforward using energy
and angle variables.  The details can be found in \Ref{Aglietti:2008}.

%
% The integrated approximate cross section
%

\subsection{The integrated approximate cross section}
\label{ssec:intRRA1}

We are now in a position to compute the integral of $\dsiga{RR}{1}_{m+2}$
as given in \eqn{eq:dsigRRA1} over the one-particle factorized phase space.
In order to evaluate $\int_1\dsiga{RR}{1}_{m+2}$ we first use the
phase-space factorization properties of \eqns{eq:PSfact_Cir}{eq:PSfact_Sr},
then perform the integration to obtain
\beeq
&&
\int_1 \dsiga{RR}{1}_{m+2}J_{m+1} = {\cal N}\sum_{\{m+2\}}
\PS{m+1}(\momt{})\frac{1}{S_{\{m+2\}}}
\frac{\as}{2\pi}S_\eps\left(\frac{\mu^2}{Q^2}\right)^\eps
\nn\\&&\qquad\times
\sum_r\sum_{i\ne r}\Bigg[\frac{1}{2}\IcC{ir}{(0)}(y_{\wti{i}Q};\ep)
\bT_{ir}^2\SME{m+1}{0}{\momt{}} J_{m+1}(\momt{})
\nn\\&&\qquad\qquad\qquad\quad
+\,\sum_{k\ne i,r}\IcS{r}{(0)}(\Y{i}{k};\ep)
\SME{m+1;(i,k)}{0}{\momt{}} J_{m+1}(\momt{})
\nn\\&&\qquad\qquad\qquad\quad
\,-\IcCS{(0)}(\ep)\bT_i^2\SME{m+1}{0}{\momt{}} J_{m+1}(\momt{})
\Bigg]
\,.
\label{eq:IntdsigRRA1}
\eeeq 
This result is not in the form of an $(m+1)$-parton configuration
times a factor. In order to rewrite \eqn{eq:IntdsigRRA1} in such a 
form, we need to perform the counting of symmetry factors for
going from $m+1$ to $m+2$ partons. This was done in Appendix B of
\Ref{Somogyi:2006cz} for going from $m$ to $m+1$ partons and the 
results can readily be taken over by setting $m\to m+1$.

The final result for the integral of the doubly-real singly-unresolved
approximate cross section can be written as \cite{Somogyi:2006cz}
\beq
\int_{1}\dsiga{RR}{1}_{m+2} = \dsig{R}_{m+1}\otimes
\bI^{(0)}(\mom{}_{m+1};\ep)\,,
\label{eq:I1dsigRRA1}
\eeq
where the insertion operator reads
\beq
\bI^{(0)}(\mom{}_{m+1};\ep) =
\aeps\\
\sum_{i}\bigg[
\IcC{i}{(0)}(y_{iQ};\ep)\,\bT_{i}^{2}
+\sum_{k\ne i}
\TcS{ik}{(0)}(Y_{ik,Q};\ep)\,\bT_{i}\ldot\bT_{k}
\bigg]\,.
\label{eq:I0}
\eeq
In \eqn{eq:I0} we introduced the functions
\beq
\IcC{q}{(0)} = 
\IcC{qg}{(0)}\,,\qquad
\IcC{g}{(0)} = 
\frac{1}{2}\IcC{gg}{(0)} + \Nf \IcC{q\qb}{(0)}\,,\qquad
\TcS{ik}{(0)} = \IcS{ik}{(0)} + \IcCS{(0)}\,,
\label{eq:ICi0}
\eeq
with $\IcC{ir}{(0)}$, $\IcS{ik}{(0)}$ and $\IcCS{(0)}$ defined in
Eqns.\ (\ref{eq:ICir0}), (\ref{eq:ISik0}) and (\ref{eq:ICS0}),
respectively.%
\footnote{
Note a slight rearrangement of terms with respect to 
\Ref{Somogyi:2006cz}: here we find it more convenient to use 
colour-conservation (\eqn{eq:colourcons}) to combine $\IcCS{(0)}$
with $\IcS{ik}{(0)}$ into $\TcS{ik}{(0)}$. Then of course $\IcC{i}{(0)}$
is defined without including $\IcCS{(0)}$.
}
In this paper $\Nf$ denotes the number of light flavours, which we set
to $\Nf=5$ in all numerical results. 
Upon expanding these functions in $\ep$ we find that
the coefficients of the poles in the expansions are independent of the cut 
parameters $\alpha_0$ and $y_0$ as well as the exponents $d_0$ and $d'_0$ 
and they read
\beeq
&&
\IcC{q}{(0)}(x;\ep) 
= \frac1{\ep^2} + \frac1\ep \left(\frac32 - 2\ln x\right) + \Oe{0}\,,
\label{eq:IcCq0}
\\
&&
\IcC{g}{(0)}(x;\ep) 
= \frac1{\ep^2}
+ \frac1\ep \left(\frac{11}{6} - \frac23\Nf\frac{\TR}{\CA} - 2\ln
x\right) + \Oe{0}\,,
\label{eq:IcCg0}
\\
&&
\TcS{ik}{(0)}(Y;\ep)
= \frac1\ep \ln Y + \Oe{0}\,.
\label{eq:TcSik0}
\eeeq
For the coefficients with non-negative powers of $\ep$ we obtain
integral representations that can be evaluated numerically. The
results for four values of $\alpha_0$ or $y_0=1$, $0.3$, $0.1$ and
$0.03$ with fixed values of $d_0=d'_0=3-3\eps$ are presented in
\figs{fig:IcCg0}{fig:TcSik0}.
The dependence on $\alpha_0$ in the collinear functions is hardly
visible. Note also that the small-$x$ behaviour of the collinear
functions is dominated by the logarithmic terms that are the same in the
quark and gluon functions, therefore, the expansion coefficients of 
$\IcC{q}{(0)}(x;\eps)$ and 
$\IcC{g}{(0)}(x;\eps)$ are very similar.

Using colour-conservation (\eqn{eq:colourcons}), the definition of $\Y{i}{k}$ 
(\ref{eq:defYikQ}) and \eqnss{eq:IcCq0}{eq:TcSik0} above, it is straightforward 
to verify that our insertion operator differs from that defined in Eq.\ (7.26) of 
\Ref{Catani:1996vz} only in finite terms, i.e.
\beq
\bI^{(0)}(\mom{}_n;\eps) = \frac{\as}{2\pi}S_\eps
\left(\frac{\mu^2}{Q^2}\right)^\eps
\sum_i\left(\bT_i^2\frac{1}{\eps^2}+\gamma_i \frac{1}{\eps}
+\sum_{k\ne i}\bT_i \bT_k \frac{1}{\eps}\ln y_{ik}\right) + \Oe{0}\,,
\eeq
with the usual flavour constants
\beq
\gamma_q = \frac{3}{2}\CF\,\qquad\mbox{and}\qquad
\gamma_g = \frac{\beta_0}{2}\,.
\eeq
It follows that $\int_1\dsiga{RR}{1}_{m+2}$ correctly cancels all $\eps$-poles
of the real-virtual cross section $\dsig{RV}_{m+1}$.
%
% Collinear figures
%
\FIGURE{
\label{fig:IcCg0}
\makebox{
\hspace{-1.5em}
\hspace{-1.5em}
\psfrag{X}[ct]{$\log_{10} x$}
\makebox{
\psfrag{Y}[cb]{\raisebox{0.5em}{$\IcC{q}{(0)}(x;\eps)$}}
\psfrag{T}[b]{\raisebox{0.5em}{Order: $\eps^0$}}
\includegraphics[scale=0.42]{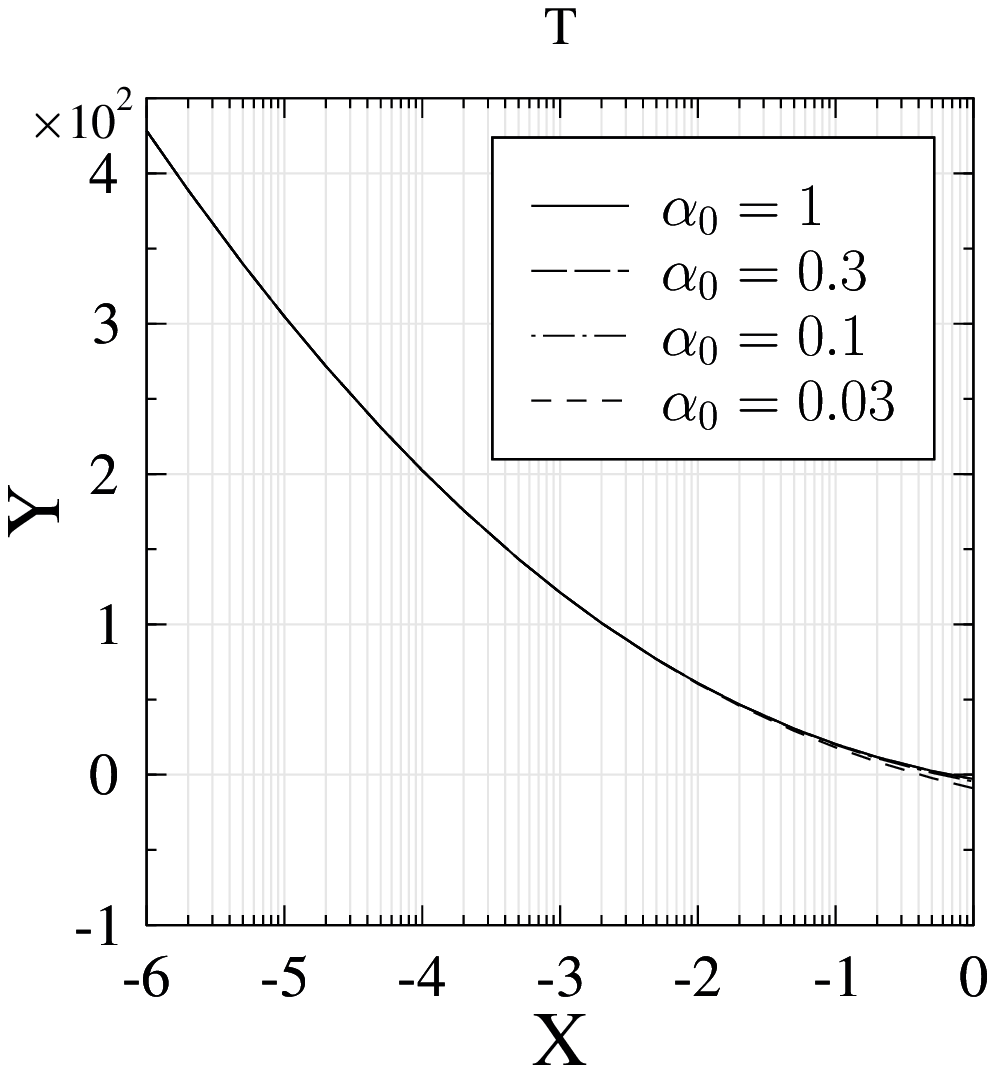}
}
\hspace{-1.5em}
\psfrag{Y}{}
\makebox{
\psfrag{T}[b]{\raisebox{0.5em}{Order: $\eps^1$}}
\includegraphics[scale=0.42]{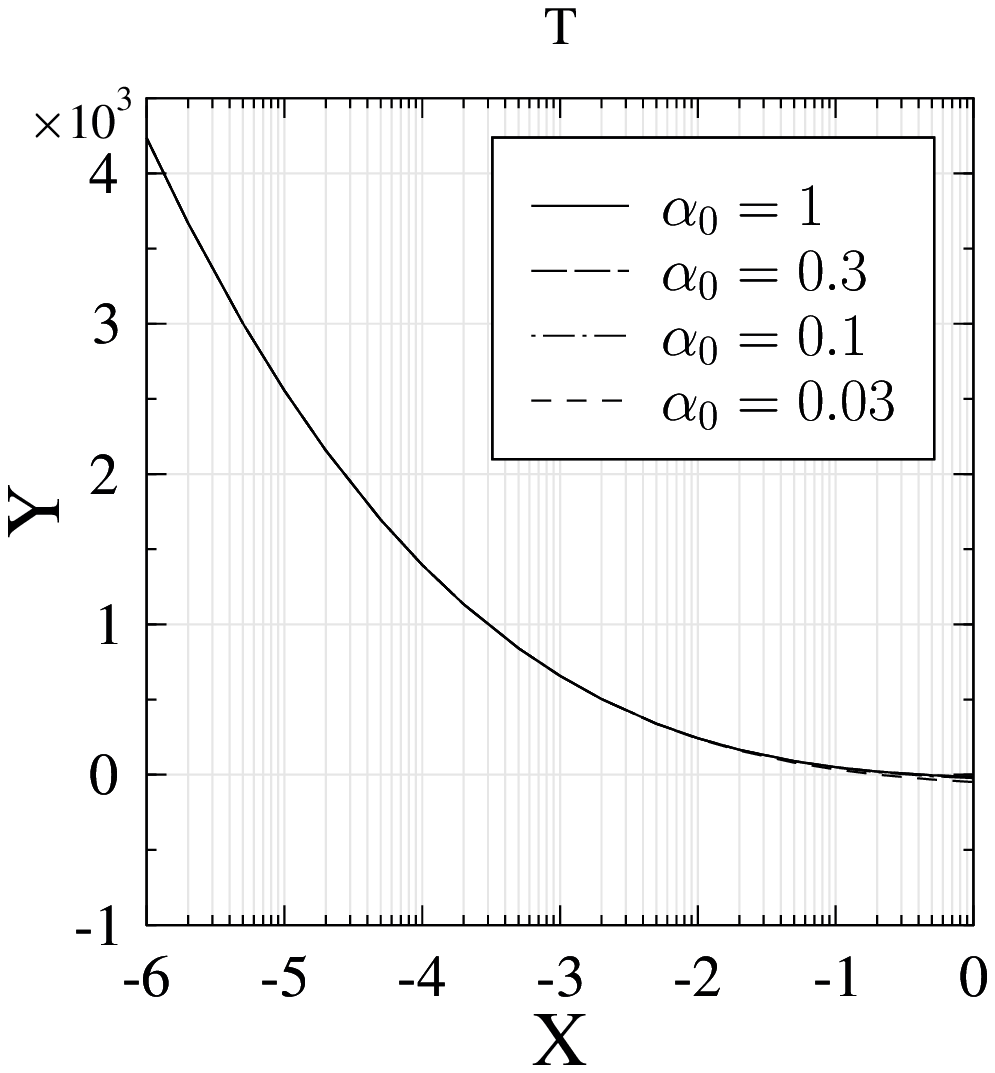}
}
\hspace{-1.5em}
\makebox{
\psfrag{T}[b]{\raisebox{0.5em}{Order: $\eps^2$}}
\includegraphics[scale=0.42]{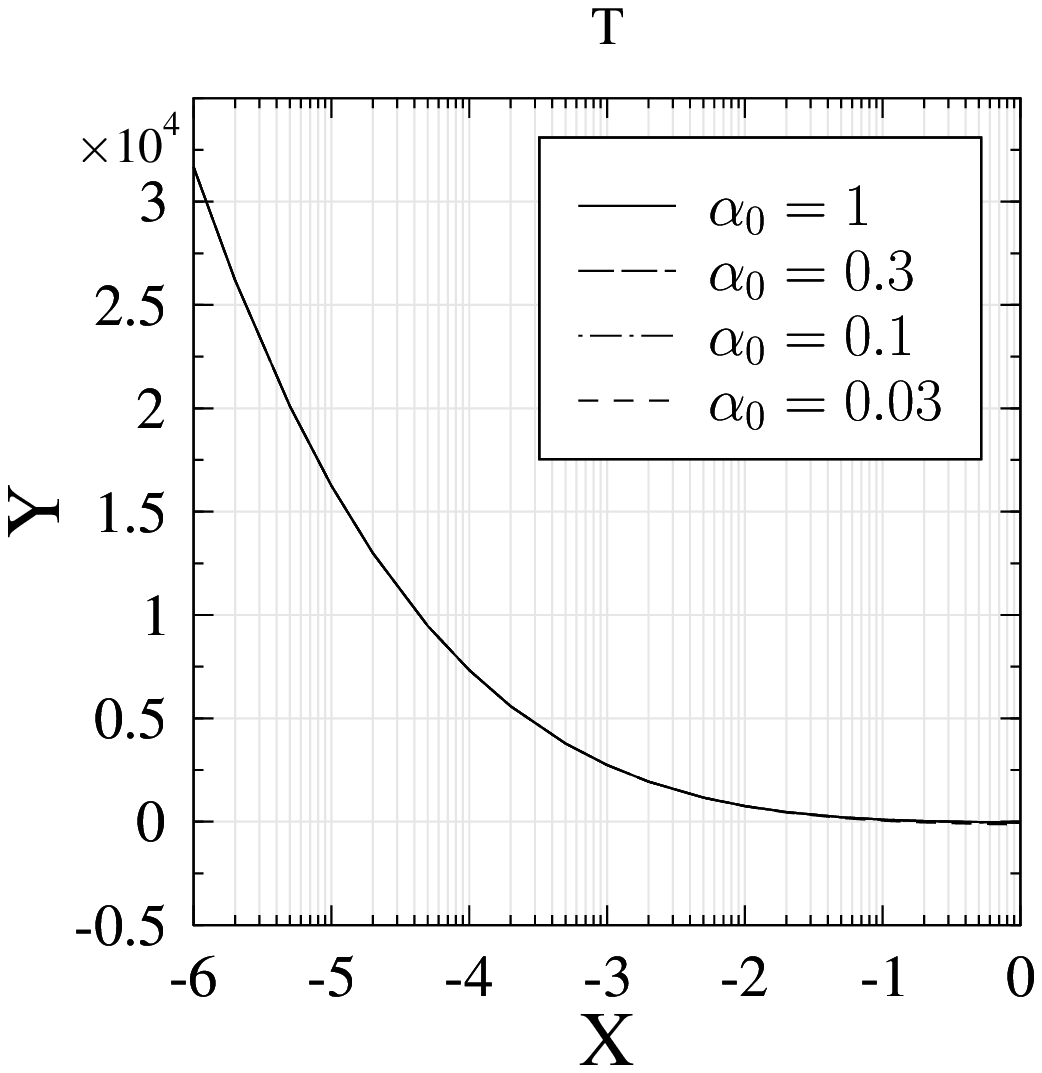}}
}
\makebox{
\hspace{-1.5em}
\hspace{-1.5em}
\psfrag{X}[ct]{$\log_{10} x$}
\psfrag{T}{}
\makebox{
\psfrag{Y}[cb]{\raisebox{0.5em}{$\IcC{g}{(0)}(x;\eps)$}}
\includegraphics[scale=0.42]{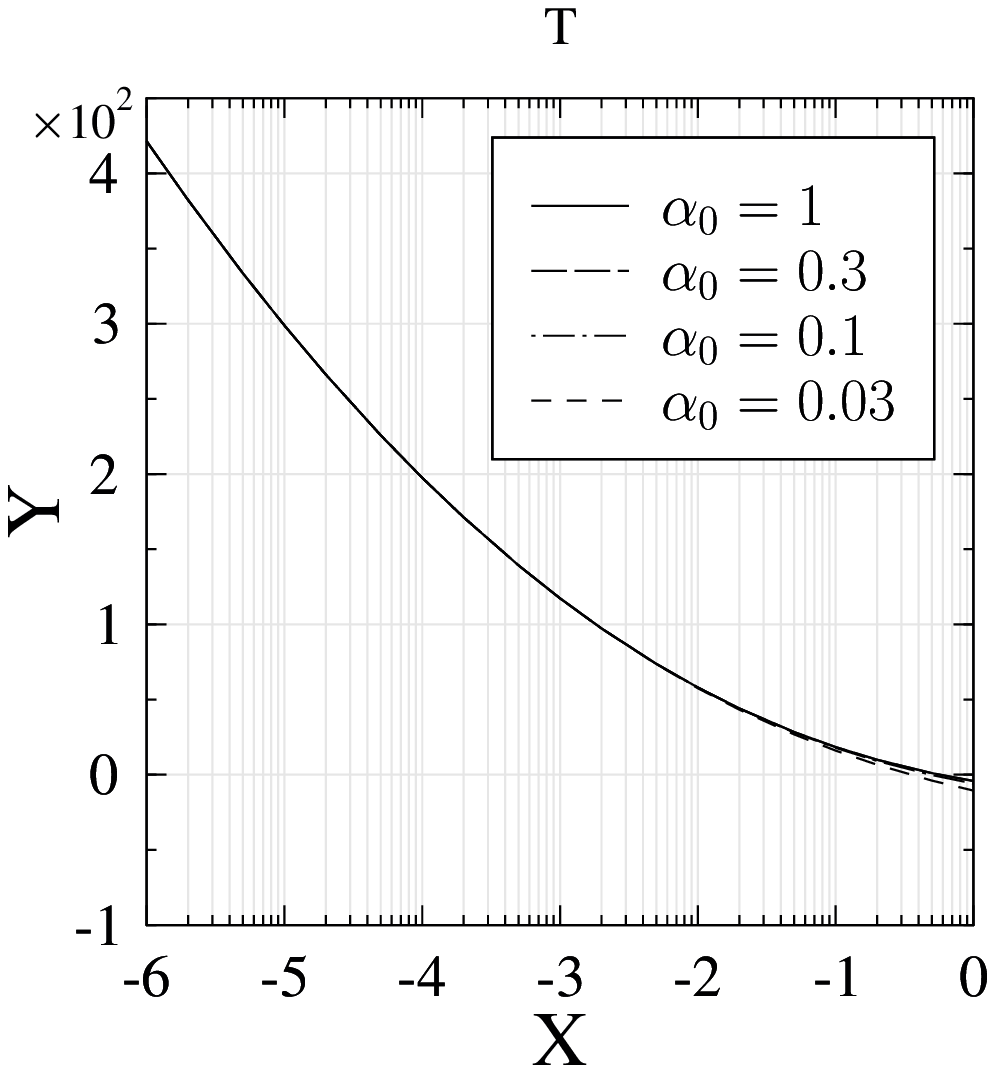}
}
\hspace{-1.5em}
\psfrag{Y}{}
\makebox{
\includegraphics[scale=0.42]{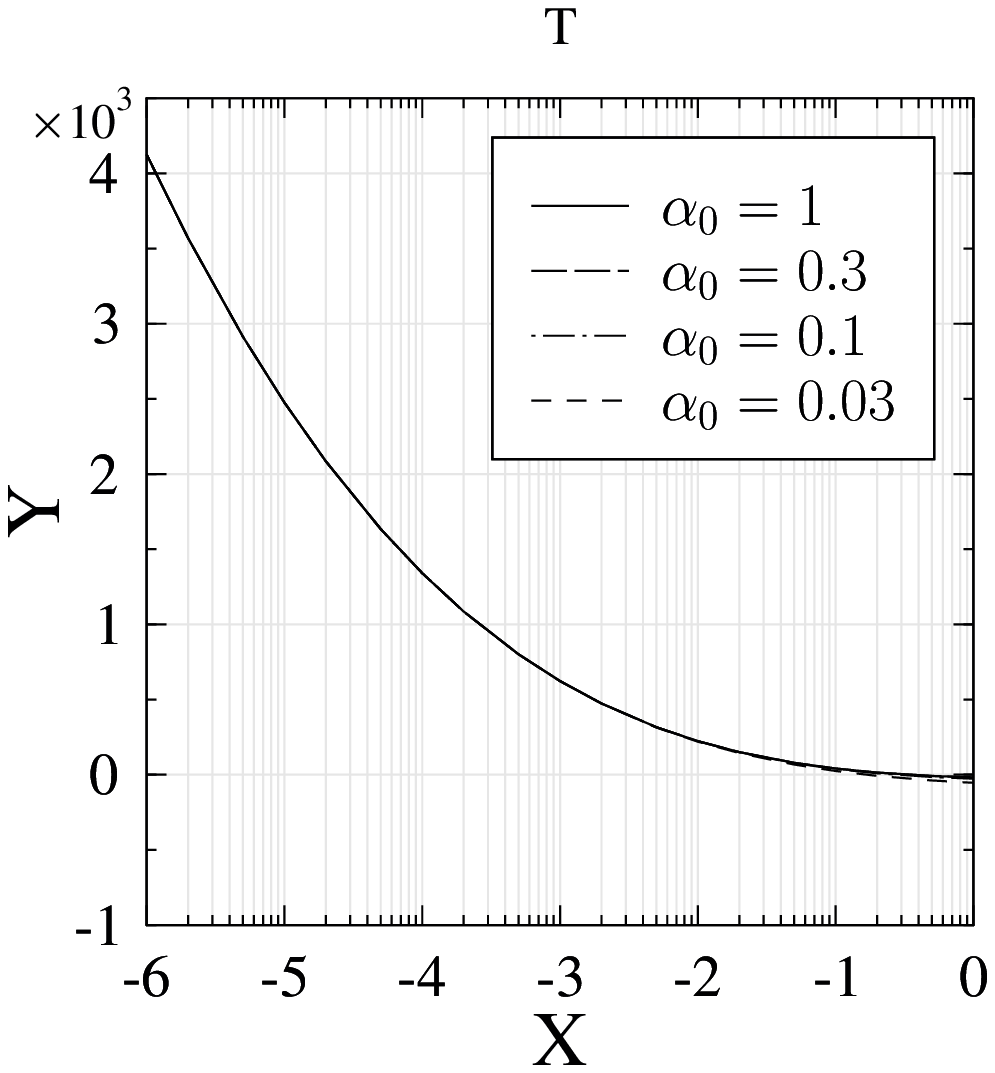}
}
\hspace{-1.5em}
\makebox{
\includegraphics[scale=0.42]{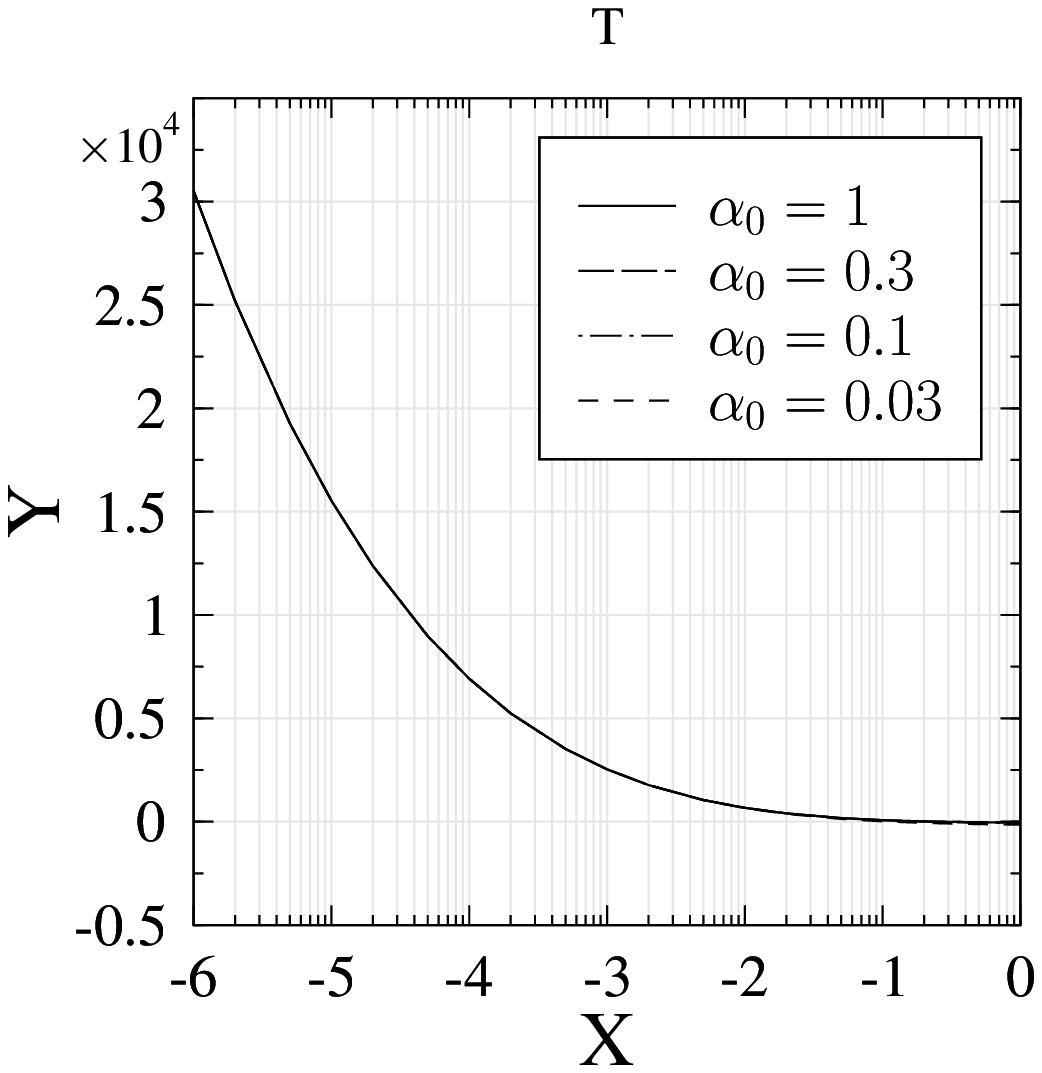}}
}
\caption{Expansion coeffiecients of the functions
$\IcC{i}{(0)}(x;\eps)$ with $d_0=3-3\eps$ and $\Nf = 5$.
Upper row: $i = q$, lower row: $i = g$.}
}
%
% Soft figures
%
\FIGURE{
\label{fig:TcSik0}
\makebox{
\hspace{-1.5em}
\hspace{-1.5em}
\psfrag{X}[ct]{$\log_{10} Y$}
\makebox{
\psfrag{Y}[cb]{\raisebox{0.5em}{$\TcS{ik}{(0)}(Y;\eps)$}}
\psfrag{T}[b]{\raisebox{0.5em}{Order: $\eps^0$}}
\includegraphics[scale=0.42]{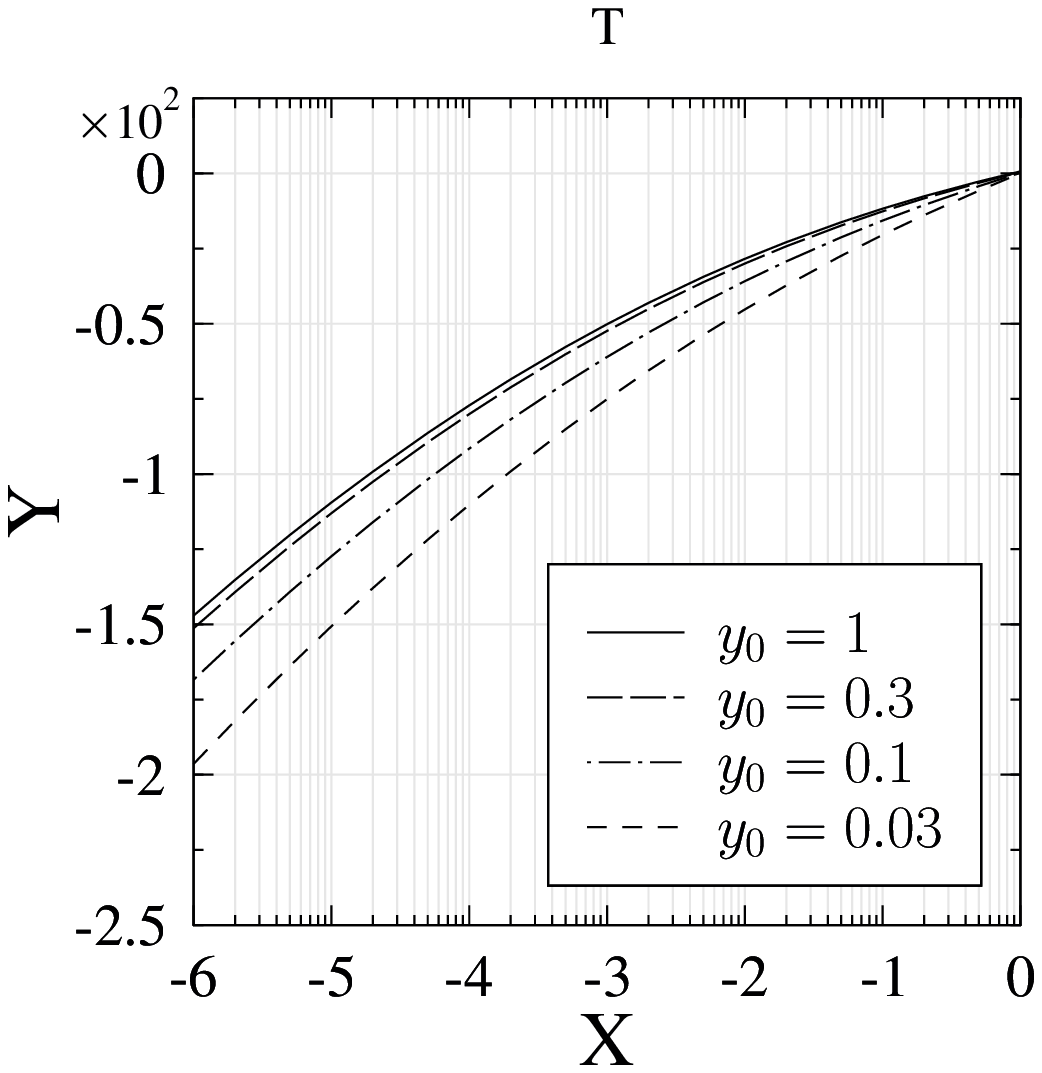}
}
\hspace{-1.5em}
\psfrag{Y}{}
\makebox{
\psfrag{T}[b]{\raisebox{0.5em}{Order: $\eps^1$}}
\includegraphics[scale=0.42]{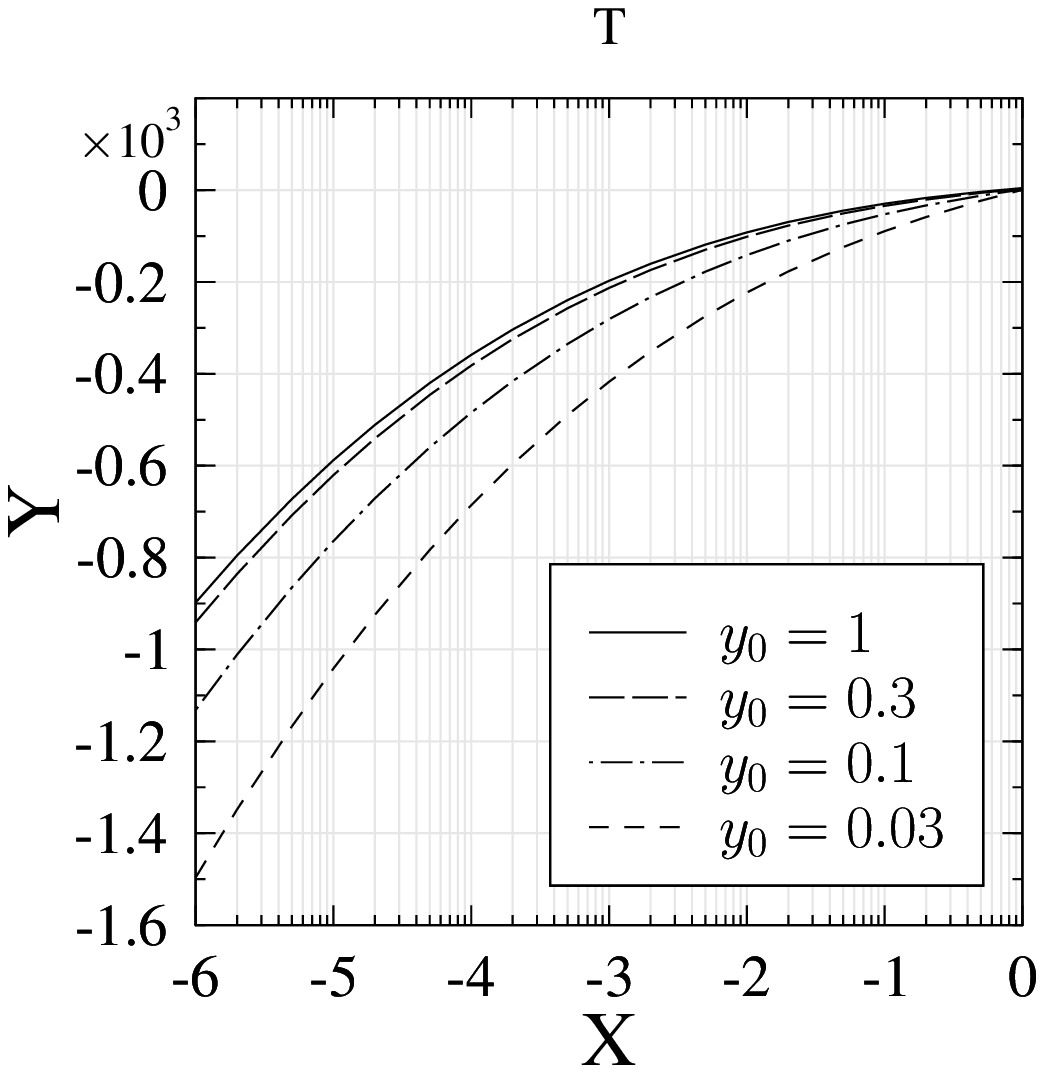}
}
\hspace{-1.5em}
\makebox{
\psfrag{T}[b]{\raisebox{0.5em}{Order: $\eps^2$}}
\includegraphics[scale=0.42]{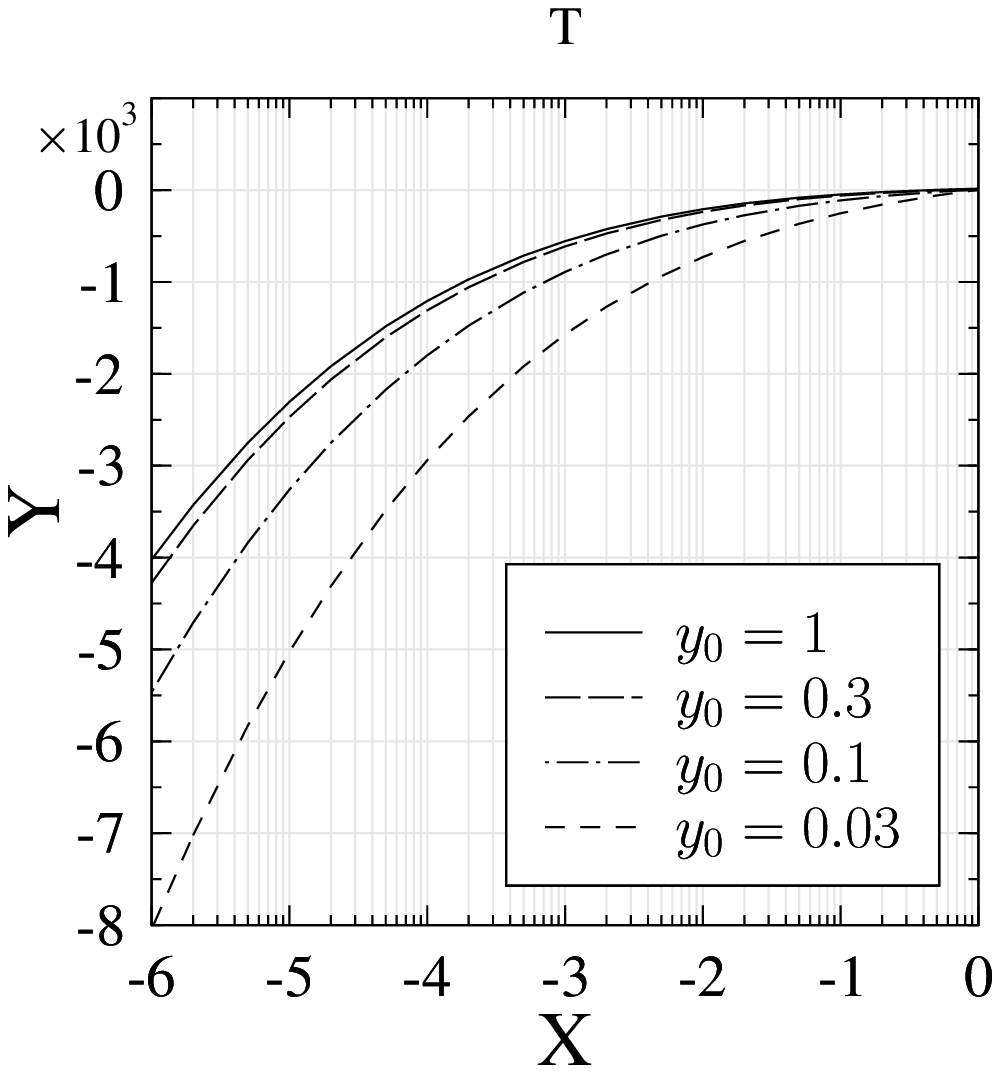}}
}
\caption{Expansion coefficients of the function $\TcS{ik}{(0)}(Y;\eps)$
with $d'_0=3-3\eps$.}
}

%%%
%%% Integrals of the real-virtual counterterms
%%%

\section{Integrals of the real-virtual counterterms}
\label{sec:RVA1}

The approximate cross section that regularizes the singly-unresolved
limits of the real-virtual cross section times the jet function is
\beeq
&&
\dsiga{RV}{1}_{m+1}J_{m} = {\cal N}\sum_{\{m+1\}}\PS{m+1}(\mom{})
\frac{1}{S_{\{m+1\}}}
\nn\\&&\qquad\times
\Bigg\{
\sum_{r} \Bigg[\sum_{i\ne r} 
\frac{1}{2} \cC{ir}{(0,1)}(\mom{})J_{m}(\momti{(ir)})
\nn\\&&\qquad\qquad\qquad
+ \left(\cS{r}{(0,1)}(\mom{})
- \sum_{i\ne r} \cCS{ir}{r}{(0,1)}(\mom{})\right)J_{m}(\momti{(r)})
\Bigg]
\nn\\&&\qquad\quad
+\,\sum_{r} \Bigg[\sum_{i\ne r} 
\frac{1}{2} \cC{ir}{(1,0)}(\mom{})J_{m}(\momti{(ir)})
\nn\\&&\qquad\qquad\qquad
+ \left(\cS{r}{(1,0)}(\mom{})
- \sum_{i\ne r} \cCS{ir}{r}{(1,0)}(\mom{})\right)J_{m}(\momti{(r)})
\Bigg]
\Bigg\}\,.
\label{eq:dsigRVA1}
\eeeq
The subtraction terms $\cC{ir}{(k,l)}$, $\cS{r}{(k,l)}$ and 
$\cCS{ir}{r}{(k,l)}$ (with $(k,l)=(0,1)$ or $(1,0)$) that appear in 
\eqn{eq:dsigRVA1} above were explicitly given in \Ref{Somogyi:2006db}. 
Here we compute the integral of the approximate cross section over the 
factorized one-particle phase space.

%
% Integrals of the collinear counterterms
%

\subsection{Integrals of the collinear counterterms}
\label{ssec:intC01C10}

The collinear counterterms $\cC{ir}{(0,1)}(\mom{})$ and
$\cC{ir}{(1,0)}(\mom{})$ are defined as
\bal
\cC{ir}{(0,1)}(\mom{}) &=
8\pi\as\mu^{2\ep}\frac{1}{s_{ir}}
2\Real\bra{m}{(0)}{(\momti{(ir)})}
\hP_{f_i f_r}^{(0)}(\tzz{i}{r},\tzz{r}{i},\kTt{i,r};\ep)
\ket{m}{(1)}{(\momti{(ir)})}
\nt\\[2mm] &\times
(1-\alpha_{ir})^{2d_{0}-2(m-1)(1-\ep)}\Theta(\alpha_{0}-\alpha_{ir})\,,
\label{eq:Cir01}
\eal
\bal
\cC{ir}{(1,0)}(\mom{}) &=
(8\pi\as\mu^{2\ep})^2\frac{1}{s_{ir}^{1+\ep}} c_{\Gamma}\cos(\pi\ep)
\nt\\[2mm] &\times
\bra{m}{(0)}{(\momti{(ir)})}
\hP_{f_i f_r}^{(1)}(\tzz{i}{r},\tzz{r}{i},\kTt{i,r};\ep)
\ket{m}{(0)}{(\momti{(ir)})}
\nt\\[2mm] &\times
(1-\alpha_{ir})^{2d_{0}-2(m-1)(1-\ep)}\Theta(\alpha_{0}-\alpha_{ir})\,.
\label{eq:Cir10}
\eal
The $\hP_{f_i f_r}^{(0)}(\tzz{i}{r},\tzz{r}{i},\kTt{i,r};\ep)$
kernels are the same tree-level Altarelli--Parisi splitting
functions that appear in \eqn{eq:Cir00}, while the
$\hP_{f_i f_r}^{(1)}(\tzz{i}{r},\tzz{r}{i},\kTt{i,r};\ep)$
kernels are the one-loop generalizations of the Altarelli--Parisi
splitting functions as spelled out explicitly in \Ref{Somogyi:2006db}.
The $m$ momenta, $\momti{(ir)}$, entering the matrix elements on the
right hand sides of \eqns{eq:Cir01}{eq:Cir10} are obtained from the
original $m+1$ momenta, $\mom{}$, by the specific momentum mapping of
\eqn{eq:cmap}
\beq
\mom{}\cmap{ir}\momti{(ir)}\,.
\eeq
This mapping leads to an exact factorization of the $(m+1)$-particle
phase space. The form of this factorization is identical to the
one discussed in \sect{ssec:intC00}.  In particular,
\eqns{eq:PSfact_Cir}{eq:dp1Cir} remain valid after making the
replacement $m \to m-1$.

We write the integrals of the collinear subtraction terms over
the factorized phase space as
\bal
\int_{1}\cC{ir}{(0,1)}(\mom{}) &=
\aeps
\IcC{ir}{(0)}(y_{\wti{ir}Q};\ep)\bT_{ir}^{2}\,
2\Real\bra{m}{(0)}{(\momti{(ir)})}\ket{m}{(1)}{(\momti{(ir)})}
\label{eq:I1Cir01}
\intertext{and}
\int_{1}\cC{ir}{(1,0)}(\mom{}) &=
\left[\aeps\right]^2 \frac{(4\pi)^2}{S_\ep}
c_{\Gamma}\cos(\pi\ep)
\IcC{ir}{(1)}(y_{\wti{ir}Q};\ep)\bT_{ir}^{2}\,
\SME{m}{0}{\momti{(ir)}}\,,
\label{eq:I1Cir10}
\eal
with
\beq
c_{\Gamma}=\frac{1}{(4\pi)^{2-\ep}}
\frac{\Gamma(1+\ep)\Gamma^2(1-\ep)}{\Gamma(1-2\ep)}\,.
\eeq
Note that
\beq
\frac{(4\pi)^2}{S_\ep} c_{\Gamma}\cos(\pi\ep) =
\frac{\Gamma^2(1+\ep)\Gamma^4(1-\ep)}{\Gamma(1+2\ep) \Gamma^2(1-2\ep)} =
1 - \frac{\pi^2}{2} \ep^2 - 2 \zeta(3) \ep^3 + \frac{\pi^4}{120} \ep^4 +
\Oe{5}\,.
\label{eq:gammafactors}
\eeq
In writing \eqns{eq:I1Cir01}{eq:I1Cir10} we used that by the usual
arguments, the spin correlations generally present in
\eqns{eq:Cir01}{eq:Cir10} vanish after azimuthal integration so that we
can replace the splitting functions by their azimuthally averaged
counterparts when computing the integrals.  The
$\IcC{ir}{(0)}(y_{\wti{ir}Q};\ep)$ function appearing
in \eqn{eq:I1Cir01} was computed in \sect{ssec:intC00} while
$\IcC{ir}{(1)}(y_{\wti{ir}Q};\ep)$ in \eqn{eq:I1Cir10}
is given by
\beq
\bsp
&
\IcC{ir}{(1)}(y_{\wti{ir}Q};\ep) =
\frac{(4\pi)^{2}}{S_{\ep}}(Q^{2})^{2\ep}\\
&\qquad\times
\int_{0}^{\alpha_{0}}\!\rd\alpha\,(1-\alpha)^{2d_{0}-1}
\frac{s_{\wti{ir}Q}}{2\pi}\PS{2}(p_{i},p_{r};p_{(ir)})
\frac{1}{s_{ir}^{1+\ep}}P_{f_{i}f_{r}}^{(1)}(\tzz{i}{r},\tzz{r}{i};\ep)
\frac{1}{\bT_{ir}^{2}}\,.
\esp
\label{eq:ICir1}
\eeq
We note that $\IcC{ir}{(1)}(y_{\wti{ir}Q};\eps)$ is independent of $m$ 
as the arguments of the function indicate and its explicit $\alpha_0$ 
and $d_0$ dependence is suppressed in the notation as usual.

Inspecting the actual form of the one-loop Altarelli--Parisi splitting
functions as recalled in \appx{app:APfcns} and using the symmetry
property of the factorized phase space under the interchange
$i\leftrightarrow r$,  we find
that $\IcC{ir}{(1)}(y_{\wti{ir}Q};\ep)$ can be expressed
as a linear combination of the integrals
\beq
\frac{(4\pi)^2}{S_\ep}(Q^2)^{2\ep}\,
\int_0^{\alpha_0}\!\rd \alpha_{ir}\,(1-\alpha_{ir})^{2d_0-1}
\,\frac{s_{\wti{ir}Q}}{2\pi}
\int\!\PS{2}(p_i,p_r; p_{(ir)})
\, \frac{ \tzz{r}{i}^{k+\delta\ep}}{s_{ir}^{1+\ep}}
\, g_I^{(\pm)}(\tzz{r}{i})
\,,
\label{eq:I1k}
\eeq
for $k=-1,0,1,2$ and the values of $\delta$ and $g_I^{(\pm)}$ as given
in \tab{tab:I1ints}.
%\begin{table}
\TABLE{
\label{tab:I1ints}
%\begin{center}
\qquad~
\begin{tabular}{|c|c|c|}
\hline
\hline
$\delta$ & Function & $g_I^{(\pm)}(z)$ \\
\hline
  & & \\[-4mm]
$0$ & $g_A$ & $1$ \\[2mm]
$\mp 1$ & $g_B^{(\pm)}$ & $(1-z)^{\pm\ep}$ \\[2mm]
$0$ & $g_C^{(\pm)}$ & $(1-z)^{\pm\ep}{}_2F_1(\pm\ep,\pm\ep,1\pm\ep,z)$\\[2mm]
$\pm 1$ & $g_D^{(\pm)}$ & ${}_2F_1(\pm\ep,\pm\ep,1\pm\ep,1-z)$ \\[2mm]
\hline
\hline
\end{tabular}
\caption{The values of $\delta$ and $g_I^{(\pm)}(z_r)$ at which
\eqn{eq:I1k} needs to be evaluated.}
%\end{center}
}
%\end{table}
Here we compute the $\eps$-expansion of these integrals using sector
decomposition and residuum subtraction. Analytic results are presented 
in \Refs{Aglietti:2008,Bolzoni:2008}.

%
% Integrals of the soft-type counterterms
%

\subsection{Integrals of the soft-type counterterms}
\label{ssec:intS01S10}

The soft and soft-collinear counterterms appearing in \eqn{eq:dsigRVA1} 
are
\bal
\cS{r}{(0,1)}(\mom{}) &=
-8\pi\as\mu^{2\ep}\sum_{i}\sum_{k\ne i}\frac{1}{2}\calS_{ik}(r)
2\Real\bra{m}{(0)}{(\momti{(r)})}\bT_{i}\ldot\bT_{k}
\ket{m}{(1)}{(\momti{(r)})}
\nt\\[2mm] &\times
(1-y_{rQ})^{d'_{0}-(m-1)(1-\ep)}\Theta(y_{0}-y_{rQ})\,,
\label{eq:Sr01}\\[2mm]
\cC{ir}{}\cS{r}{(0,1)} &=
8\pi\as\mu^{2\ep}\frac{1}{s_{ir}}\frac{2\tzz{i}{r}}{\tzz{r}{i}}
\bT_{i}^{2}\,
2\Real\bra{m}{(0)}{(\momti{(r)})}
\ket{m}{(1)}{(\momti{(r)})}
\nt\\[2mm] &\times
(1-y_{rQ})^{d'_{0}-(m-1)(1-\ep)}\Theta(y_{0}-y_{rQ})\,,
\label{eq:CirSr01}
\eal
\bal
\cS{r}{(1,0)}(\mom{}) &=
(8\pi\as\mu^{2\ep})^{2}c_{\Gamma}\cos(\pi\ep)\sum_{i}\sum_{k\ne i}
\frac{1}{2}\calS_{ik}(r)
\nt\\[2mm] &\times
\Bigg\{\Bigg[\CA\frac{1}{\ep^{2}}\frac{\pi\ep}{\sin(\pi\ep)}
\left(\frac{1}{2}\calS_{ik}(r)\right)^{\ep}
+\frac{\beta_{0}}{2\ep}\frac{S_{\ep}}{(4\pi)^{2}c_{\Gamma}}
%\left[\mu^{2\ep}\cos(\pi\ep)\right]^{-1}
\frac{1}{\mu^{2\ep}\cos(\pi\ep)}
\Bigg]\SME{m;(i,k)}{0}{\momti{(r)}}
\nt\\[2mm] &\qquad
+2\frac{\pi}{\ep}\frac{1}{\cos(\pi\ep)}
\sum_{l\ne i,k}\left(\frac{1}{2}\calS_{il}(r)\right)^{\ep}
\SME{m;(i,k,l)}{0}{\momti{(r)}}
\Bigg\}
\nt\\[2mm] &\times
(1-y_{rQ})^{d'_{0}-(m-1)(1-\ep)}\Theta(y_{0}-y_{rQ})\,,
\label{eq:Sr10}
\eal
\bal
\cC{ir}{}\cS{r}{(1,0)} &=
-(8\pi\as\mu^{2\ep})^{2}c_{\Gamma}\cos(\pi\ep)
\frac{1}{s_{ir}}\frac{2\tzz{i}{r}}{\tzz{r}{i}}\bT_{i}^{2}
\nt\\[2mm] &\times
\Bigg[
\CA\frac{1}{\ep^{2}}\frac{\pi\ep}{\sin(\pi\ep)}
\left(\frac{1}{s_{ir}}\frac{\tzz{i}{r}}{\tzz{r}{i}}\right)^{\ep}
+\frac{\beta_{0}}{2\ep}\frac{S_{\ep}}{(4\pi)^{2}c_{\Gamma}}
%\left[\mu^{2\ep}\cos(\pi\ep)\right]^{-1}
\frac{1}{\mu^{2\ep}\cos(\pi\ep)}
\Bigg]\SME{m}{0}{\momti{(r)}}
\nt\\[2mm] &\times
(1-y_{rQ})^{d'_{0}-(m-1)(1-\ep)}\Theta(y_{0}-y_{rQ})\,,
\label{eq:CirSr10}
\eal
with
\beq
\beta_0 = \frac{11}{3}\CA-\frac{4}{3}\TR\Nf-\frac{2}{3}\TR\Ns\,.
\label{eq:beta0}
\eeq
In QCD, for the number of scalars we have $\Ns=0$.
The three-parton colour-correlated squared  matrix element 
$\M{m;(i,k,l)}{(0)}$ appearing in \eqn{eq:Sr10} is defined in \eqn{eq:colam3}.
The $m$ momenta $\momti{(r)}$ that enter the squared matrix 
elements on the right hand sides of \eqnss{eq:Sr01}{eq:CirSr10}
are obtained by applying the momentum mapping of \eqn{eq:smap} to the 
original set of $m+1$ momenta $\mom{}$,
\beq
\mom{}\smap{r}\momti{(r)}\,.
\eeq
As discussed in \sect{ssec:intS00}, this mapping leads to an exact 
factorization of the $(m+1)$-particle phase space. Indeed, 
\eqns{eq:PSfact_Sr}{eq:dp1Sr} remain valid after making the 
replacement $m\to m-1$.

Integrating the soft-type subtraction
terms of \eqnss{eq:Sr01}{eq:CirSr10} over the factorized phase space,
we can write the results as
\bal
\int_{1}\cS{r}{(0,1)}(\mom{}) &=
\aeps
\label{eq:I1Sr01}\\
&\times
\sum_{i}\sum_{k\ne i}\IcS{ik}{(0)}(\Y{i}{k};\ep)
\,2\Real\bra{m}{(0)}{(\momti{(r)})}\bT_{i}\ldot\bT_{k}
\ket{m}{(1)}{(\momti{(r)})}\,,
\nt\\[2mm]
\int_{1}\cC{ir}{}\cS{r}{(0,1)}(\mom{}) &=
\aeps
\IcCS{(0)}(\ep)\,\bT_{i}^{2}\,
2\Real\bra{m}{(0)}{(\momti{(r)})}
\ket{m}{(1)}{(\momti{(r)})}\,,
\label{eq:I1CirSr01}
\eal
\bal
\int_{1}\cS{r}{(1,0)}(\mom{}) &=
\left[\aeps\right]^2 \frac{(4\pi)^2}{S_\ep}
c_{\Gamma}\cos(\pi\ep)
\nt\\[2mm]
&\times\sum_{i}\sum_{k\ne i}\bigg[
\IcS{ik}{(1)}(\Y{i}{k};\ep)\SME{m;(i,k)}{0}{\momti{(r)}}
\label{eq:I1Sr10}\\
&\qquad\qquad +
\sum_{l\ne i,k}\IcS{ikl}{(1)}(\Y{i}{k},\Y{i}{l},\Y{k}{l};\ep)
\SME{m;(i,k,l)}{0}{\momti{(r)}}\bigg]
\,,
\nt
\\[2mm]
\int_{1}\cC{ir}{}\cS{r}{(1,0)}(\mom{}) &=
\left[\aeps\right]^2 \frac{(4\pi)^2}{S_\ep}
c_{\Gamma}\cos(\pi\ep)
\IcCS{(1)}(\ep)\,\bT_{i}^{2}
\SME{m}{0}{\momti{(r)}}\,.
\label{eq:I1CirSr10}
\eal
The notation anticipates that  $\IcS{ik}{(1)}$, 
$\IcS{ikl}{(1)}$ and $\IcCS{(1)}$ are $m$-independent. (As usual
the explicit $y_0$ and $d'_0$ dependences are not indicated.)
The tree-level
functions $\IcS{ik}{(0)}(\Y{i}{k};\ep)$ and 
$\IcCS{(0)}(\ep)$ are given in 
\eqns{eq:ISik0}{eq:ICS0} respectively, while their one-loop
counterparts read
\beeq
&&
\IcS{ik}{(1)}(\Y{i}{k};\ep) =
\nn\\[2mm] &&\qquad
\CA\frac{1}{\ep^{2}}\frac{\pi\ep}{\sin(\pi\ep)}
\frac{(4\pi)^{2}}{S_{\ep}}(Q^{2})^{2\ep}
\int_{0}^{y_{0}}\!\rd y\,(1-y)^{d'_{0}-1}\frac{Q^{2}}{2\pi}
\PS{2}(p_{r},K;Q)
\left(\frac{1}{2}\calS_{ik}(r)\right)^{1+\ep}
\nn\\[2mm]
&&\qquad
- \frac{\beta_{0}}{2\ep}\frac{S_{\ep}}{(4\pi)^{2}c_{\Gamma}}
\left[\left(\frac{\mu^{2}}{Q^{2}}\right)^{\ep}\cos(\pi\ep)\right]^{-1}
\IcS{ik}{(0)}(\Y{i}{k};\ep)\,,
\label{eq:ISik1}
\eeeq
\beeq &&
\IcS{ikl}{(1)}(\Y{i}{k},\Y{i}{l},\Y{k}{l};\ep) =
\\[2mm] &&\qquad
\nn
2\frac{\pi}{\ep}\frac{1}{\cos(\pi\ep)}
\frac{(4\pi)^{2}}{S_{\ep}}(Q^{2})^{2\ep}
\int_{0}^{y_{0}}\!\rd y\,(1-y)^{d'_{0}-1}\frac{Q^{2}}{2\pi}
\PS{2}(p_{r},K;Q)
\frac{1}{2}\calS_{ik}(r)\left(\frac{1}{2}\calS_{kl}(r)\right)^{\ep}
\,,
\label{eq:ISikl1}
\\[2mm] &&
\IcCS{(1)}(\ep) =
\nn\\[2mm] &&\qquad
- \CA\frac{1}{\ep^{2}}\frac{\pi\ep}{\sin(\pi\ep)}
\frac{(4\pi)^{2}}{S_{\ep}}(Q^{2})^{2\ep}
\int_{0}^{y_{0}}\!\rd y\,(1-y)^{d'_{0}-1}\frac{Q^{2}}{2\pi}
\PS{2}(p_{r},K;Q)
2\left(\frac{1}{s_{ir}}\frac{\tzz{i}{r}}{\tzz{r}{i}}\right)^{1+\ep}
\nn\\ &&\qquad
-\frac{\beta_{0}}{2\ep}\frac{S_{\ep}}{(4\pi)^{2}c_{\Gamma}}
\left[\left(\frac{\mu^{2}}{Q^{2}}\right)^{\ep}\cos(\pi\ep)\right]^{-1}
\IcCS{(0)}(\ep)\,.
\label{eq:ICS1}
\eeeq
These integrals can also be evaluated using energy and angle variables 
The explicit computation is performed in \Ref{Aglietti:2008}.

%
% The integrated approximate cross section
%

\subsection{The integrated approximate cross section}
\label{ssec:intRVA1}

The computation of $\int_1\dsiga{RV}{1}_{m+1}$ (including the counting 
of symmetry factors) proceeds along the same lines as that in 
\sect{ssec:intRRA1}.
The final result for the integral of the real-virtual singly-unresolved
approximate cross section can be written as (cf.~with \eqn{eq:I1dsigRRA1})
\beq
\int_{1}\dsiga{RV}{1}_{m+1} =
  \dsig{V}_{m}\otimes \bI^{(0)}(\mom{}_{m};\ep)
+ \dsig{B}_{m}\otimes \bI^{(1)}(\mom{}_{m};\ep)
\,,
\label{eq:I1dsigRVA1}
\eeq
where the insertion operator $\bI^{(0)}$ is given in \eqnss{eq:I0}{eq:ICi0},
while $\bI^{(1)}$ reads
\beq
\bsp
&
\bI^{(1)}(\mom{}_{m};\ep) =
\left[\aeps\right]^2
\frac{\Gamma^2(1+\ep) \Gamma^4(1-\ep)}{\Gamma(1+2\ep) \Gamma^2(1-2\ep)}
\\ &\qquad\times
\sum_{i}\bigg[
\IcC{i}{(1)}(y_{iQ};\ep)\,\bT_{i}^{2}
+\sum_{k\ne i} \TcS{ik}{(1)}(\Y{i}{k};\ep)\,\bT_{i}\ldot\bT_{k}
\\ &\qquad\qquad\quad
+\sum_{k\ne i} \sum_{l\ne i,k}
\IcS{ikl}{(1)}(\Y{i}{k},\Y{i}{l},\Y{k}{l};\ep)
\sum_{a,b,c}f_{abc}T_i^a T_k^b T_l^c
\bigg]\,.
\esp
\label{eq:I1}
\eeq
In \eqn{eq:I1} we introduced the functions
\beq
\IcC{q}{(1)} = 
\IcC{qg}{(1)}\,,\qquad
\IcC{g}{(1)} = 
\frac{1}{2}\IcC{gg}{(1)} + \Nf \IcC{q\qb}{(1)}\,,\qquad
\TcS{ik}{(1)} = \IcS{ik}{(1)} + \IcCS{(1)}\,,
\label{eq:ICi1}
\eeq
with $\IcC{ir}{(1)}$, $\IcS{ik}{(1)}$ and $\IcCS{(1)}$ defined in
Eqs.~(\ref{eq:ICir1}), (\ref{eq:ISik1}) and (\ref{eq:ICS1}),
respectively.  The terms proportional to $\beta_0$ in the one-loop
functions (superscript (1)) are proportional to the corresponding
tree-level contributions (same function with superscript (0)), with
proportionality factor
\beq
-\frac{\beta_0}{2\ep}
\left[\left(\frac{\mu^2}{Q^2}\right)^\ep \cos(\pi\ep)\right]^{-1}
\,,
\eeq
and represent the effect of one-loop UV renormalization. In the
following we present results for the {\em unrenormalized} functions
$\IcC{q,\rm{bare}}{(1)}$, $\IcC{g,\rm{bare}}{(1)}$
and $\TcS{ik,\rm{bare}}{(1)}$, obtained by setting $\beta_0 = 0$.

Expanding these functions in $\eps$, we find that the coefficient of
the two leading poles are independent of the cut parameters $\alpha_0$
and $y_0$ as well as the exponents $d_0$ and $d'_0$:
\beeq
&&
\IcC{q,\rm{bare}}{(1)}(x;\eps) = 
-\frac{1}{4\eps^4}\CA - \frac{1}{\eps^3}\left(\frac{3}{4}-\ln x\right)\CA 
+ \Oe{-2}\,,
\\
&&
\IcC{g,\rm{bare}}{(1)}(x;\eps) = 
-\frac{1}{4\eps^4}\CA - \frac{1}{\eps^3}\left(\frac{11}{12}-\ln x
+\frac{1}{3}\Nf\frac{\TR}{\CA} - \frac{2}{3}\Nf\frac{\TR\CF}{\CA^2}\right)\CA 
+ \Oe{-2}\,,
~\qquad
\\
&&
\TcS{ik,\rm{bare}}{(1)}(Y;\eps) =
-\frac{1}{\eps^3}\CA\frac{\ln Y}{2} + \Oe{-2}\,,
\\
&&
\IcS{ikl}{(1)}(Y_1,Y_2,Y_3;\eps) =
\frac{3\pi}{4\eps^3} + \Oe{-2}\,.
\eeeq 
For the remaining coefficients we obtain integral representations which
can be evaluated numerically. The results for four values of $\alpha_0$
or $y_0=1$, $0.3$, $0.1$ and $0.03$ with fixed values of
$d_0=d'_0=3-3\eps$ are presented in
\figss{fig:IcCg1}{fig:cSikl3}.
As in the case of $\IcC{i}{(0)}$ functions, the dependence on
$\alpha_0$ in the collinear functions is hardly visible and the expansion
coefficients of $\IcC{q,\rm{bare}}{(1)}(x;\eps)$ and
$\IcC{g,\rm{bare}}{(1)}(x;\eps)$ are very similar. The
function $\IcS{ikl}{(1)}$ depends on three variables $Y_1$, $Y_2$ and
$Y_3$. In the plots we  only show representative results obtained by
fixing two variables at  0.1 and varying the third.
%
% Collinear figures: C1q
%
\FIGURE{
\label{fig:TcSik1}
\makebox{
\hspace{-1.5em}
\psfrag{X}[ct]{$\log_{10} x$}
\psfrag{T}{}
\makebox{
\psfrag{Y}[cb]{\raisebox{0.5em}{$\IcC{q,\rm{bare}}{(1)}(x;\eps)$}}
\psfrag{T}[b]{\raisebox{0.5em}{Order: $\eps^{-2}$}}
\includegraphics[scale=0.42]{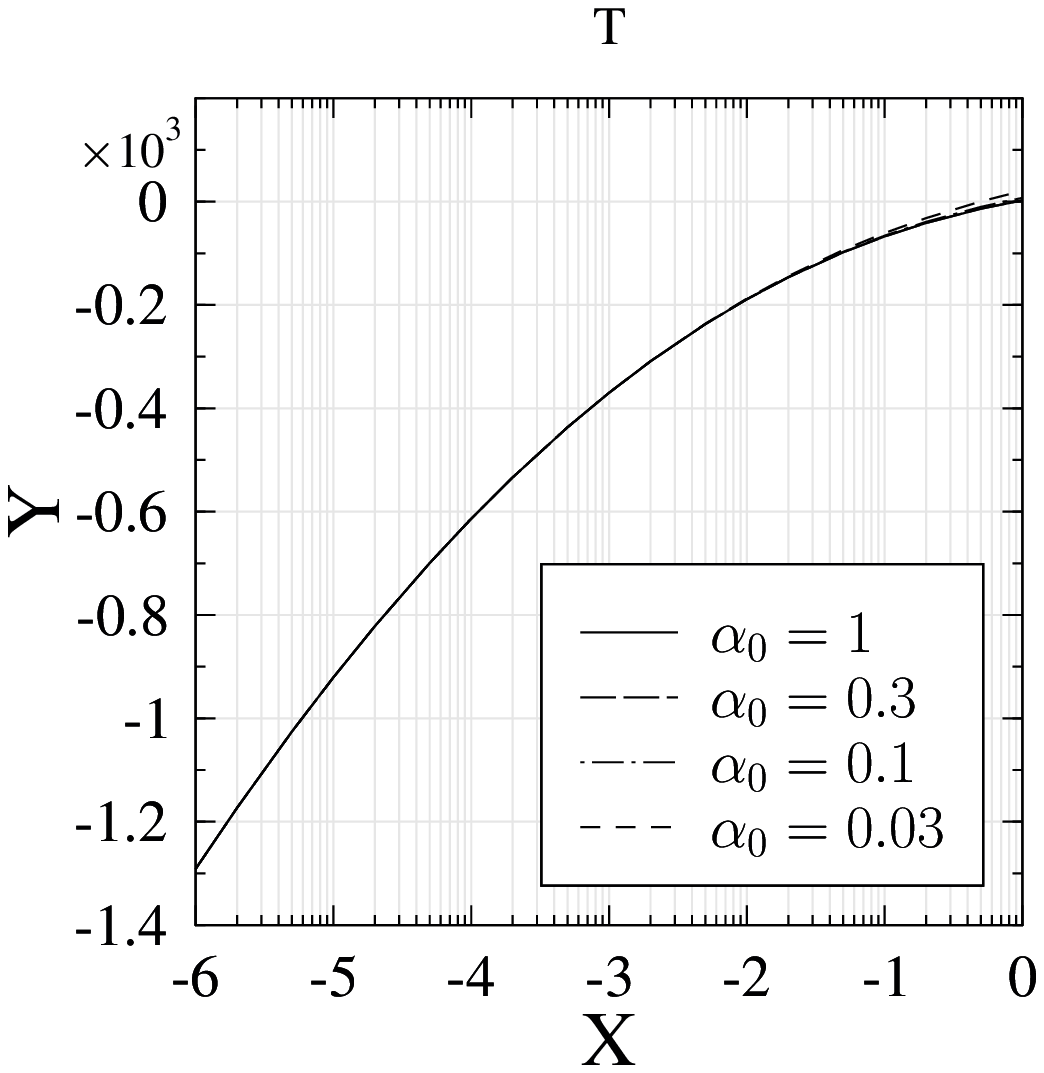}
}
\hspace{-1.5em}
\psfrag{Y}{}
\makebox{
\psfrag{T}[b]{\raisebox{0.5em}{Order: $\eps^{-1}$}}
\includegraphics[scale=0.42]{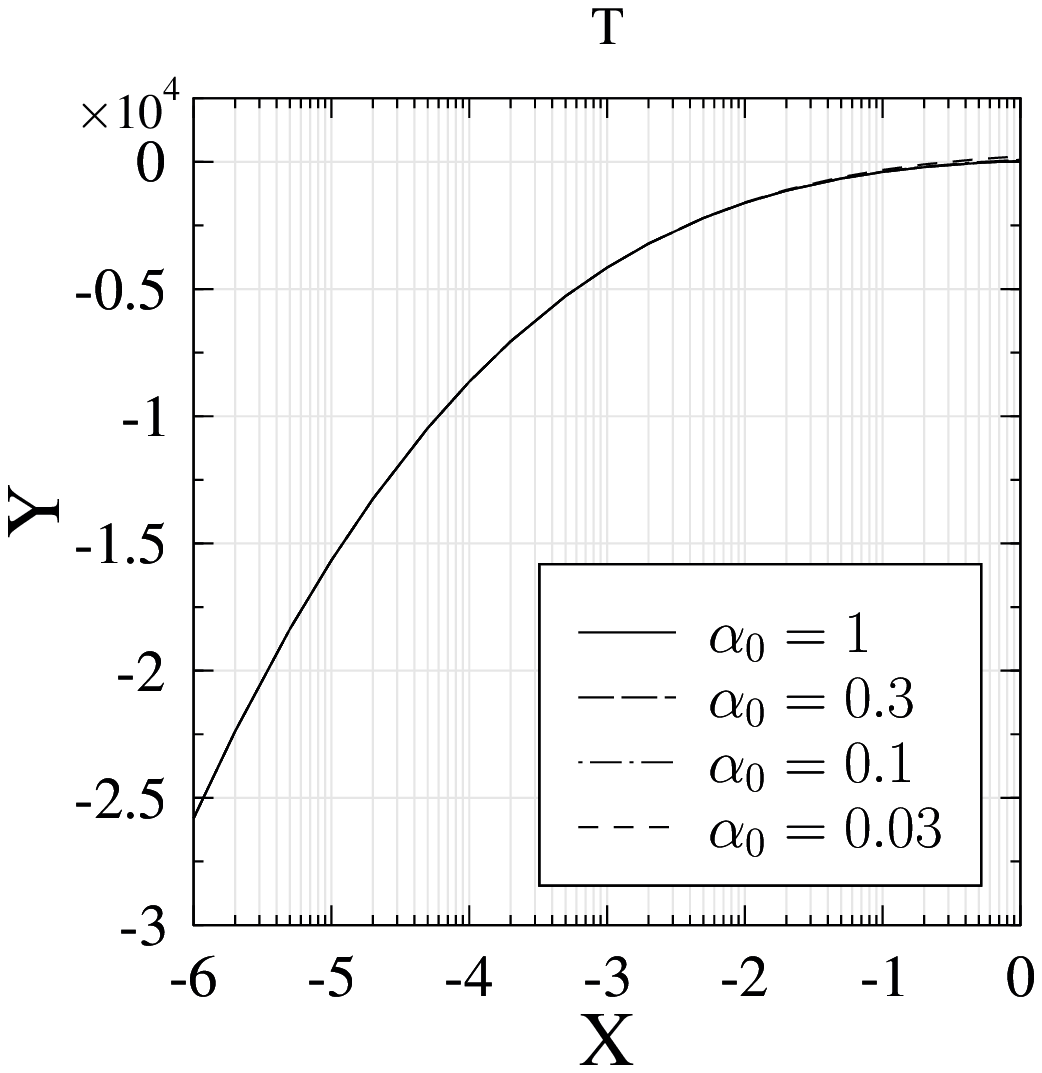}
}
\hspace{-1.5em}
\makebox{
\psfrag{T}[b]{\raisebox{0.5em}{Order: $\eps^0$}}
\includegraphics[scale=0.42]{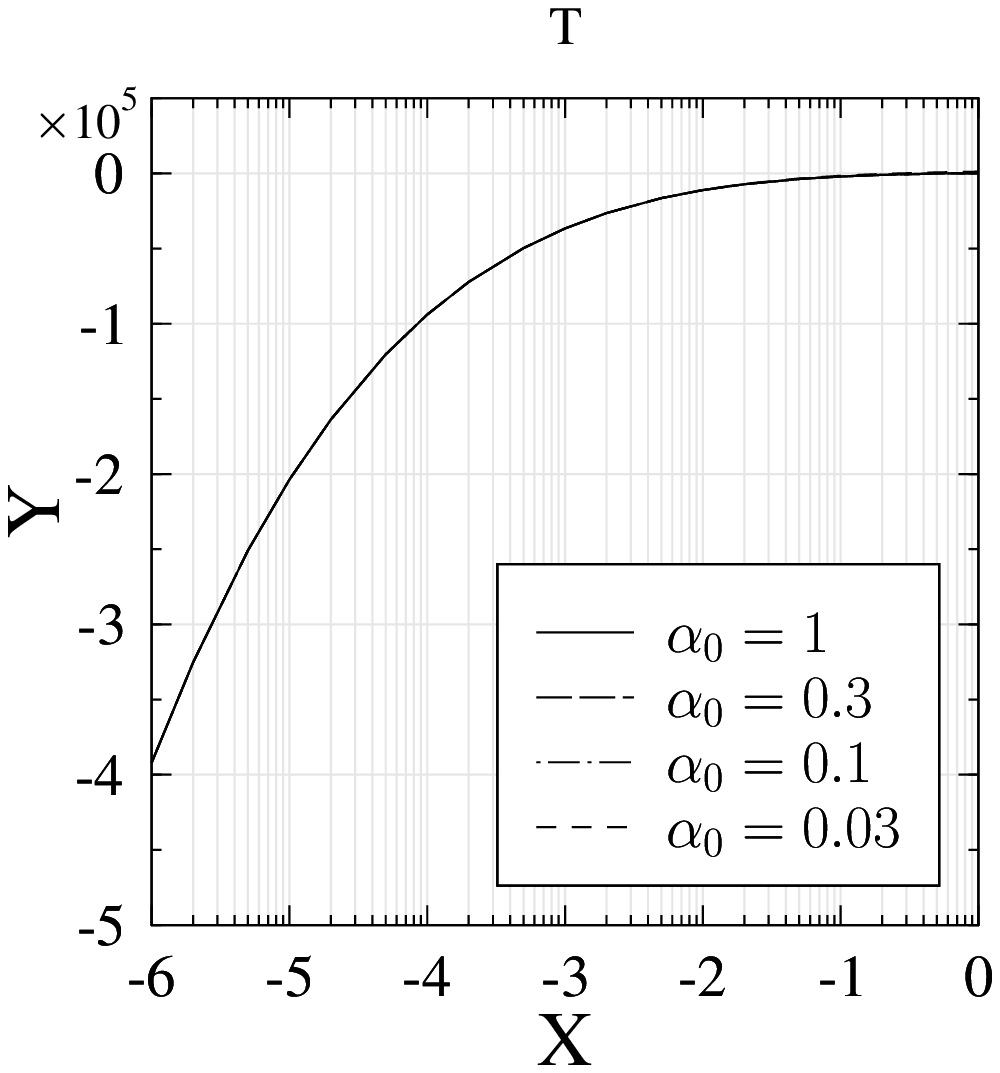}}
}
\makebox{
\hspace{-1.5em}
\psfrag{X}[ct]{$\log_{10} x$}
\psfrag{T}{}
\makebox{
\psfrag{Y}[cb]{\raisebox{0.5em}{$\IcC{g,\rm{bare}}{(1)}(x;\eps)$}}
\includegraphics[scale=0.42]{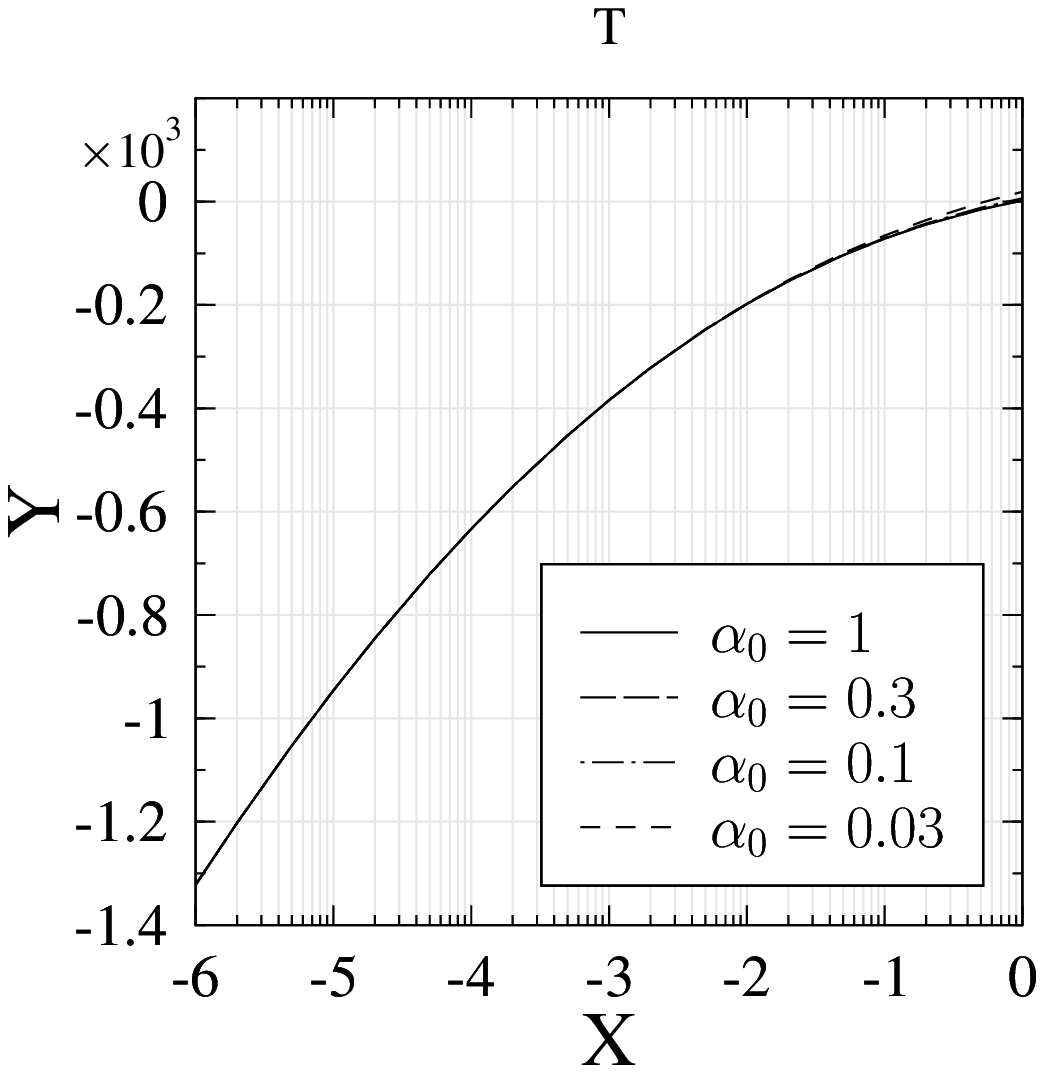}
}
\hspace{-1.5em}
\psfrag{Y}{}
\makebox{
\includegraphics[scale=0.42]{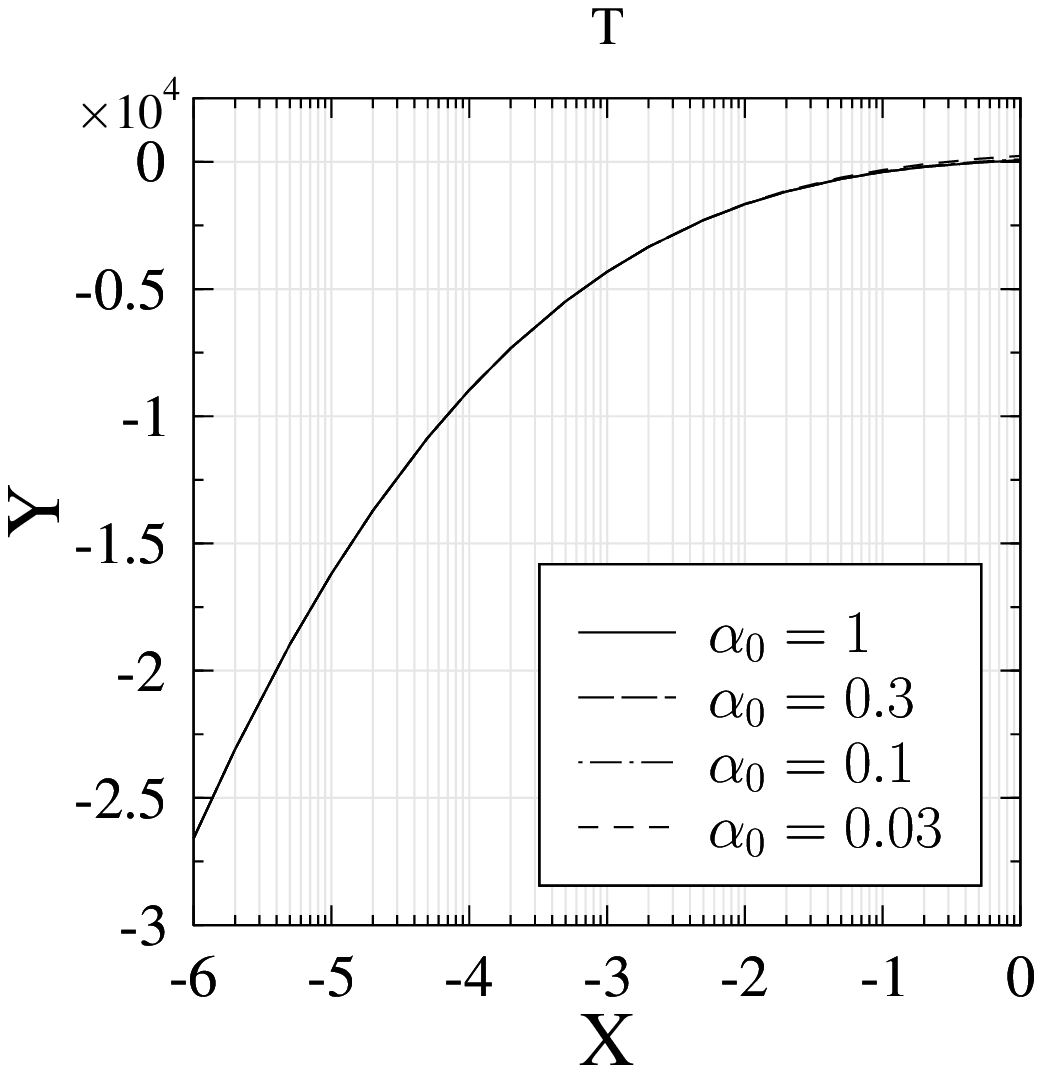}
}
\hspace{-1.5em}
\makebox{
\includegraphics[scale=0.42]{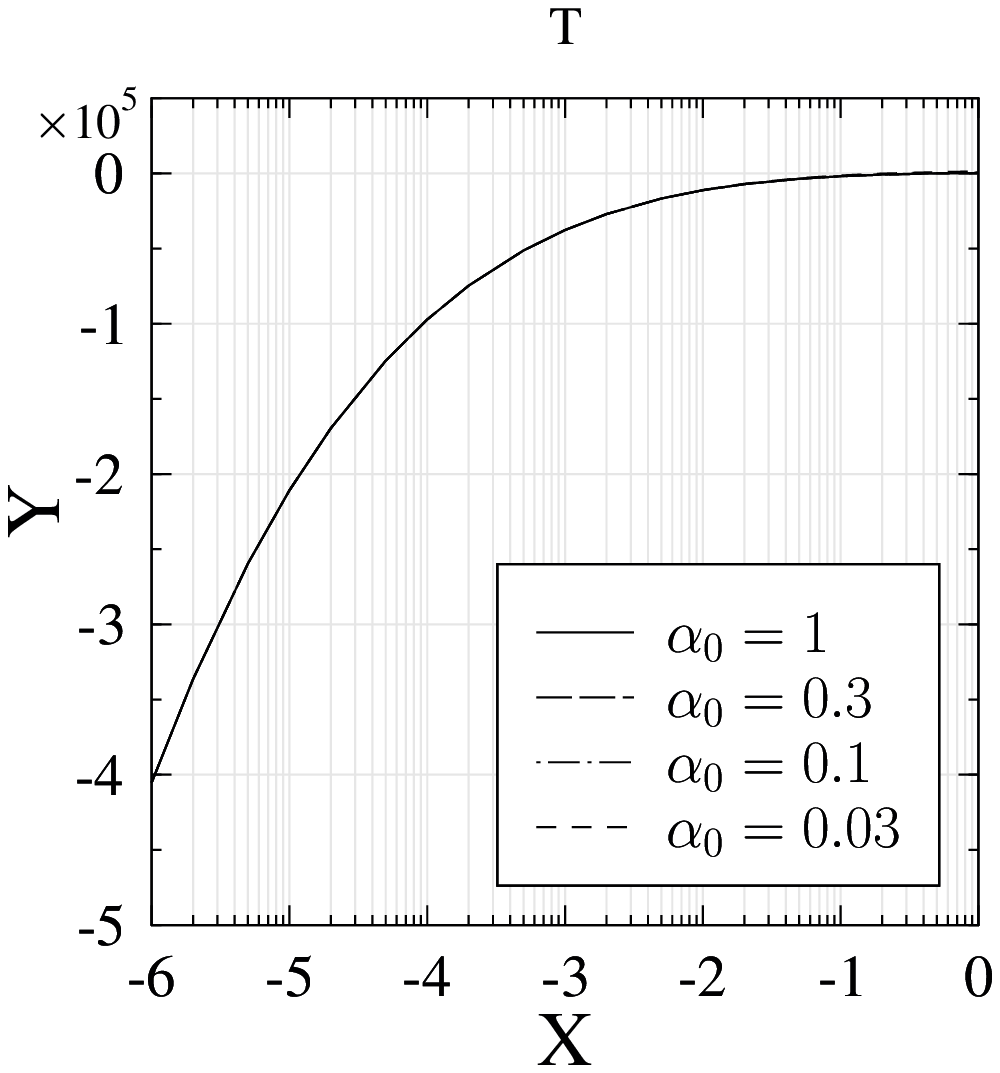}}
}
\caption{Expansion coeffiecients of the functions 
$\IcC{i,\rm{bare}}{(1)}(x;\eps)$ with $d_0=3-3\eps$ and $\Nf=5$.
Upper row: $i = q$, lower row: $i = g$.}
\label{fig:IcCg1}
}
\FIGURE{
\makebox{
\hspace{-1.5em}
\psfrag{X}[ct]{$\log_{10} Y$}
\makebox{
\psfrag{Y}[cb]{\raisebox{0.5em}{$\TcS{ik,\rm{bare}}{(1)}(Y;\eps)$}}
\psfrag{T}[b]{\raisebox{0.5em}{Order: $\eps^{-2}$}}
\includegraphics[scale=0.42]{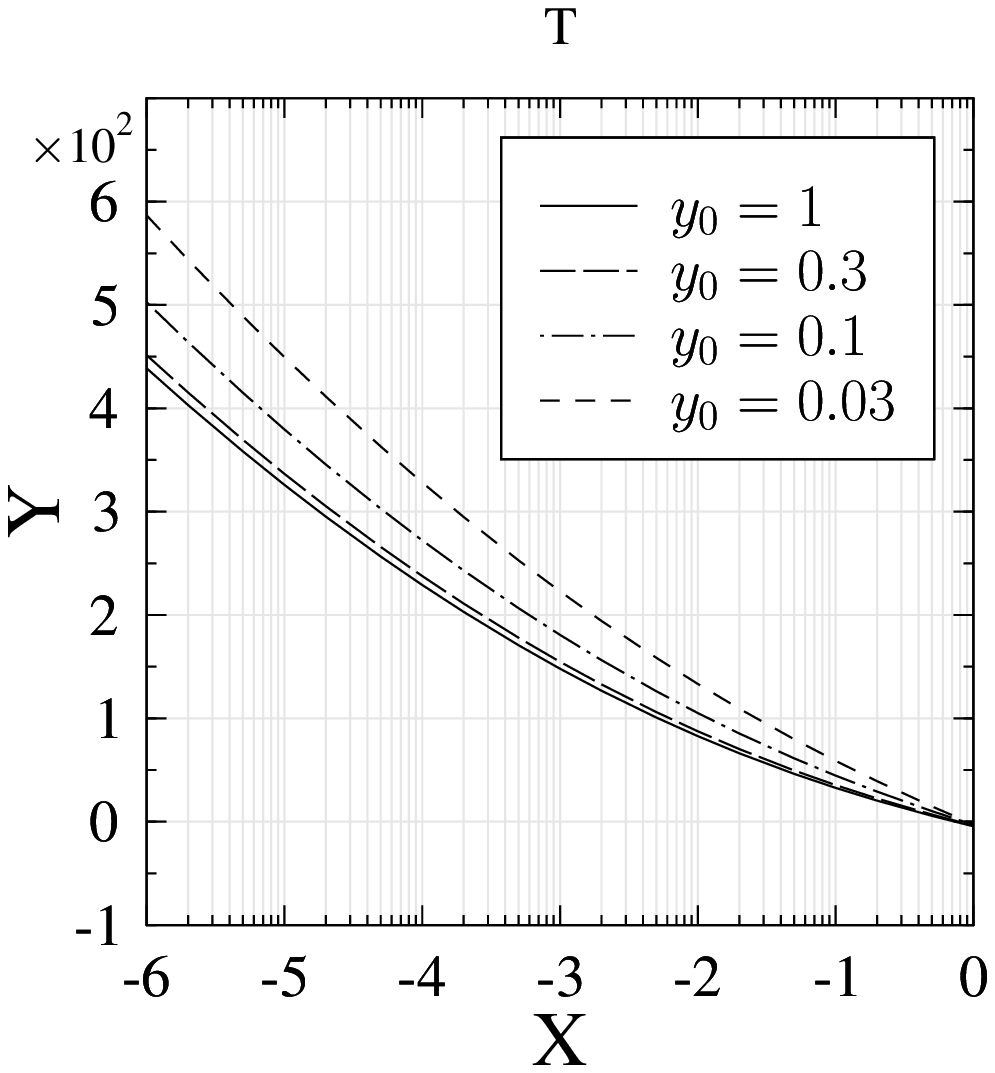}
}
\hspace{-1.5em}
\psfrag{Y}{}
\makebox{
\psfrag{T}[b]{\raisebox{0.5em}{Order: $\eps^{-1}$}}
\includegraphics[scale=0.42]{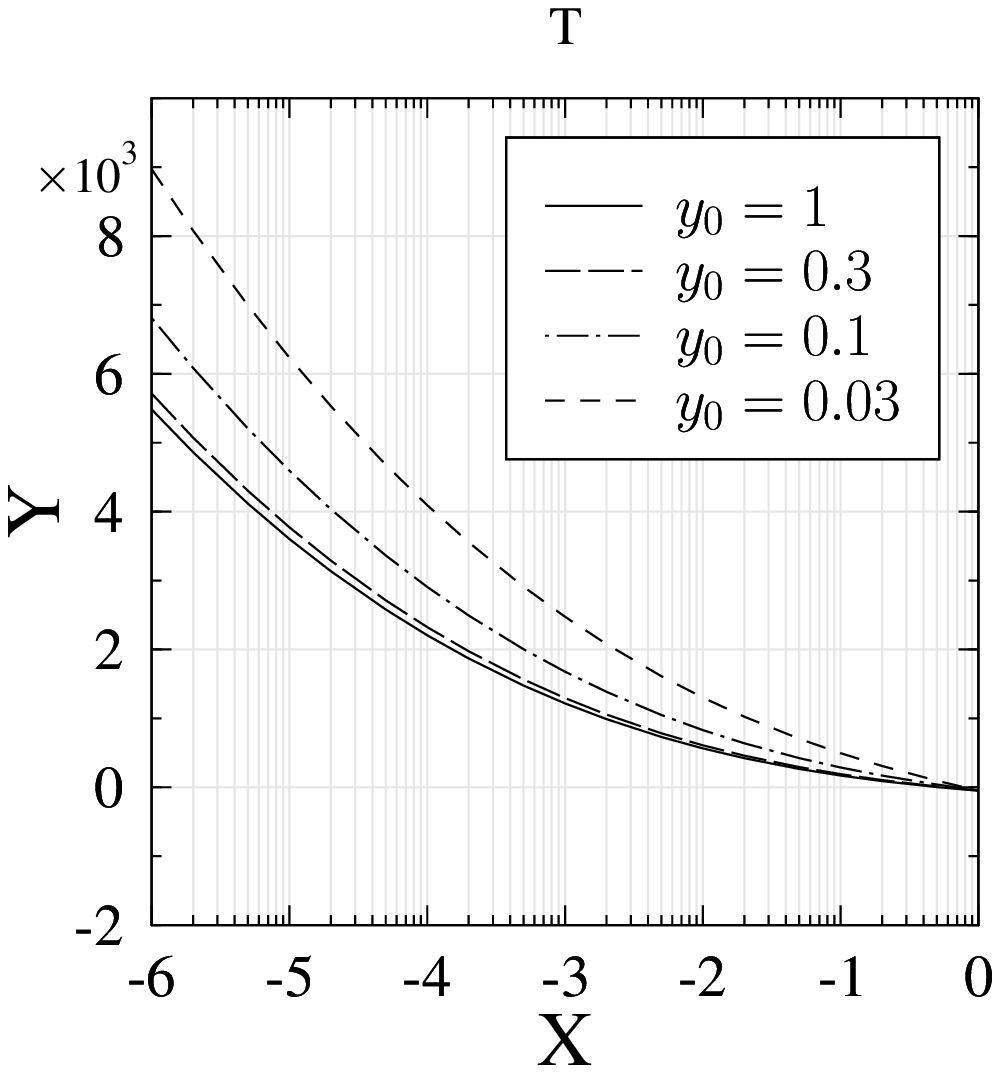}
}
\hspace{-1.5em}
\makebox{
\psfrag{T}[b]{\raisebox{0.5em}{Order: $\eps^0$}}
\includegraphics[scale=0.42]{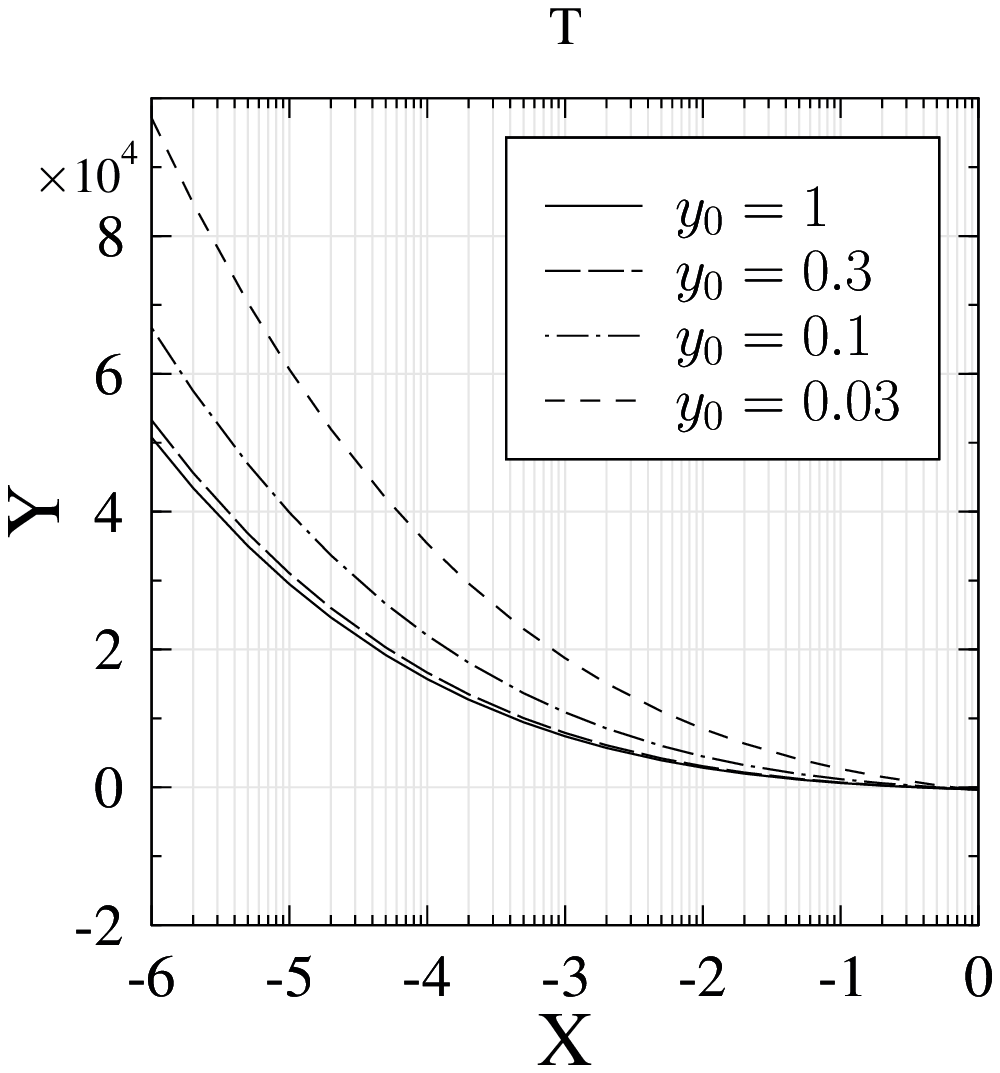}}
}
\caption{Expansion coeffiecients  of the function 
$\TcS{ik,\rm{bare}}{(1)}(Y;\eps)$ with $d'_0=3-3\eps$.}
}

%
% Soft figures: S1ikl
%
\FIGURE{
\makebox{
\hspace{-1.5em}
\psfrag{X}[ct]{$\log_{10} Y$}
\makebox{
\psfrag{Y}[cb]{\raisebox{0.5em}{$\IcS{ikl}{(1)}(Y,0.1,0.1;\eps)$}}
\psfrag{T}[b]{\raisebox{0.5em}{Order: $\eps^{-2}$}}
\includegraphics[scale=0.42]{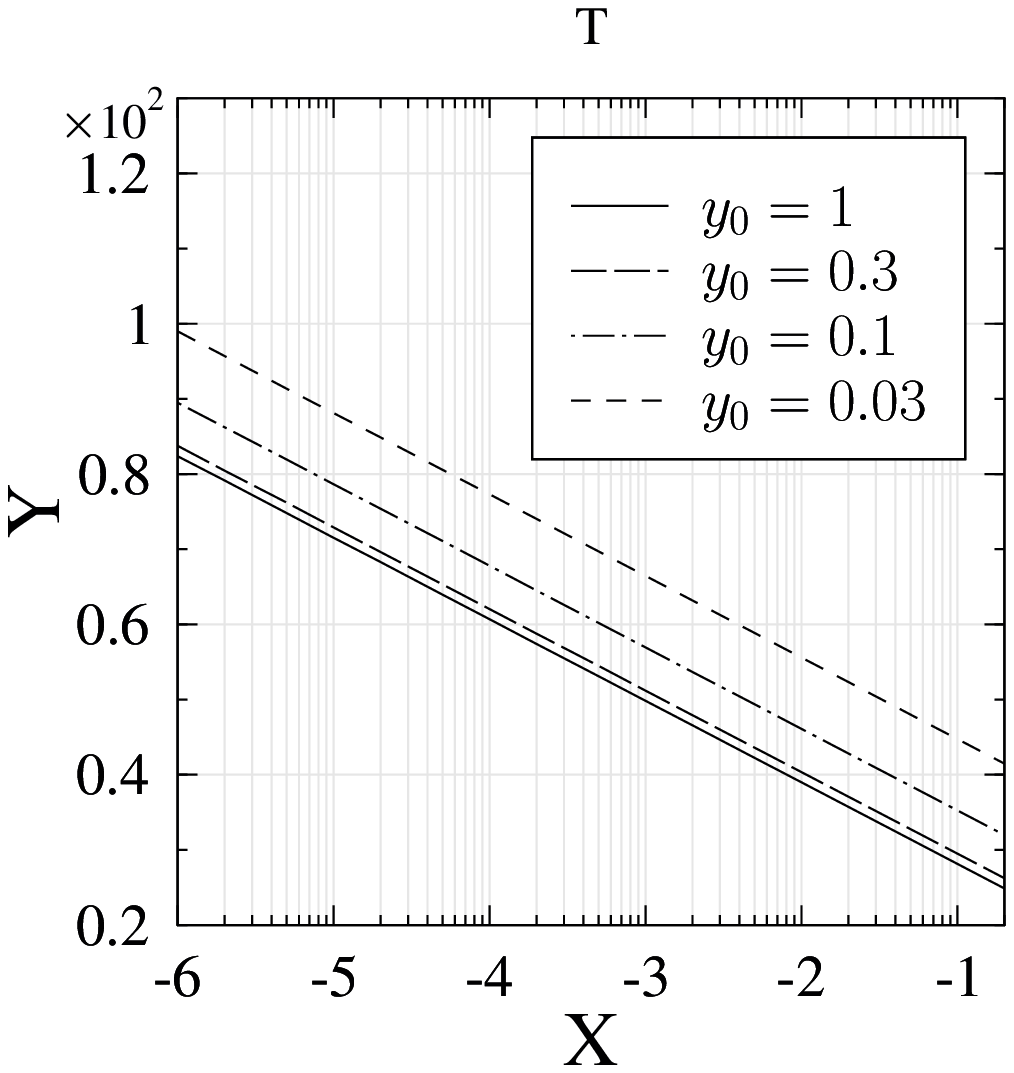}
}
\hspace{-1.5em}
\psfrag{Y}{}
\makebox{
\psfrag{T}[b]{\raisebox{0.5em}{Order: $\eps^{-1}$}}
\includegraphics[scale=0.42]{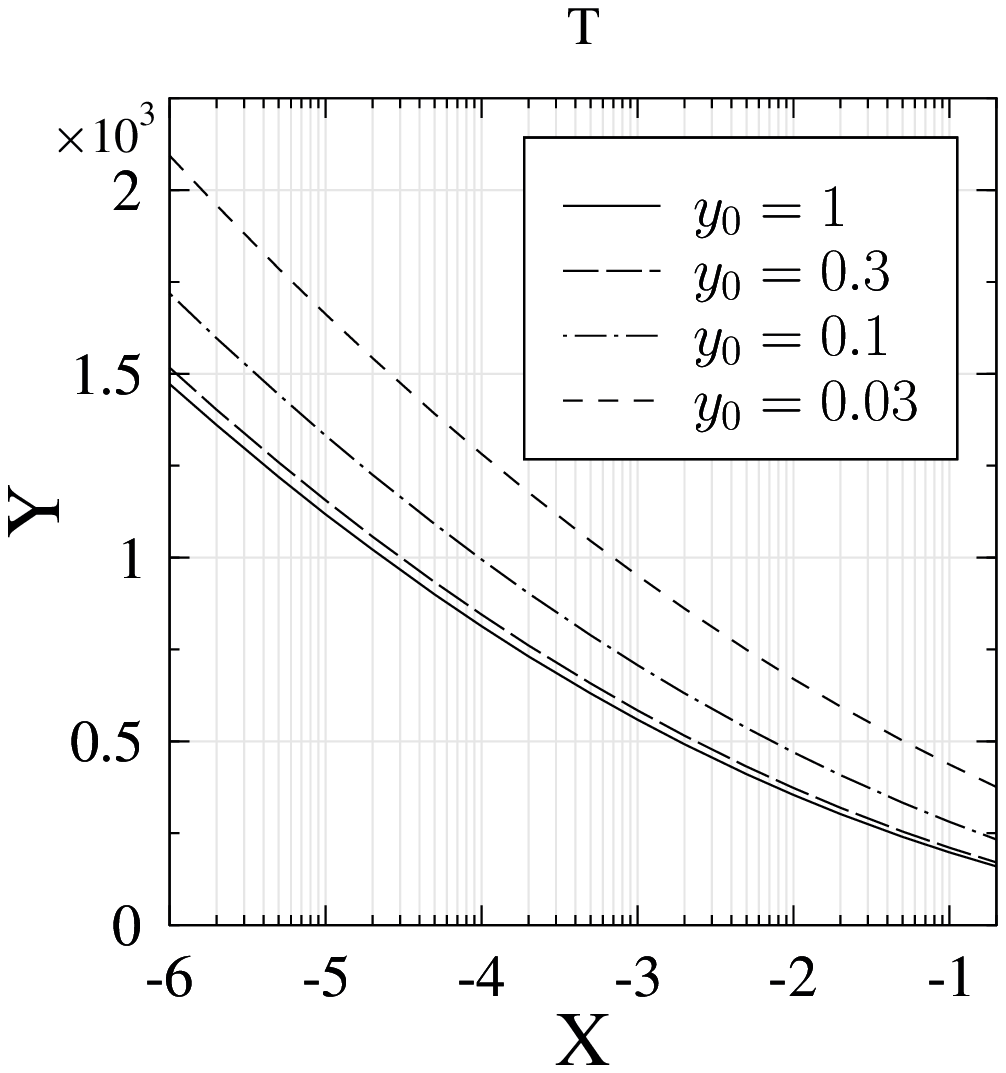}
}
\hspace{-1.5em}
\makebox{
\psfrag{T}[b]{\raisebox{0.5em}{Order: $\eps^0$}}
\includegraphics[scale=0.42]{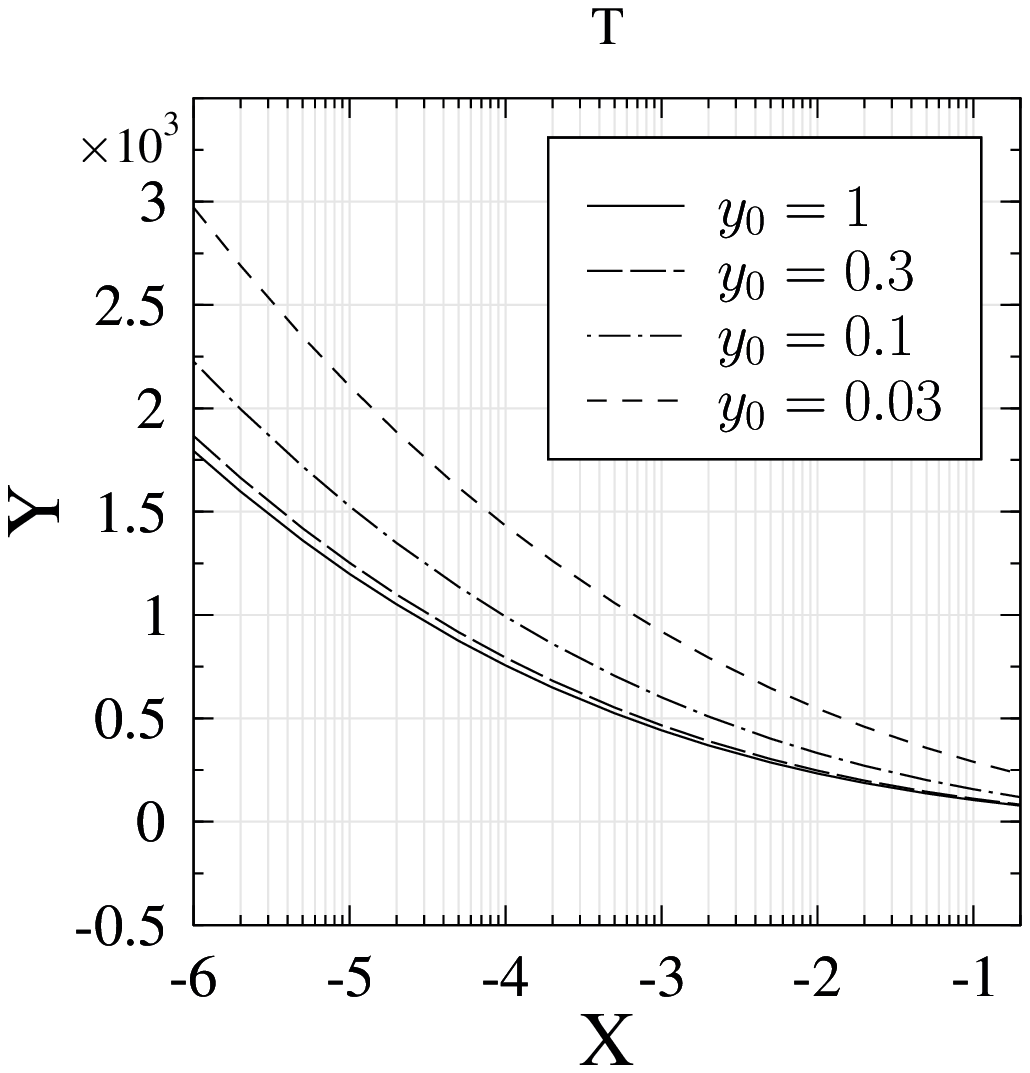}}
}
\vspace{-1em}
\caption{Expansion coeffiecients of the function 
$\IcS{ikl}{(1)}(Y,0.1,0.1;\eps)$ with $d'_0=3-3\eps$.}
}
\FIGURE{
\makebox{
\hspace{-1.5em}
\psfrag{X}[ct]{$\log_{10} Y$}
\makebox{
\psfrag{Y}[cb]{\raisebox{1.0em}{$\IcS{ikl}{(1)}(0.1,Y,0.1;\eps)$}}
\psfrag{T}[b]{\raisebox{0.5em}{Order: $\eps^{-2}$}}
\includegraphics[scale=0.42]{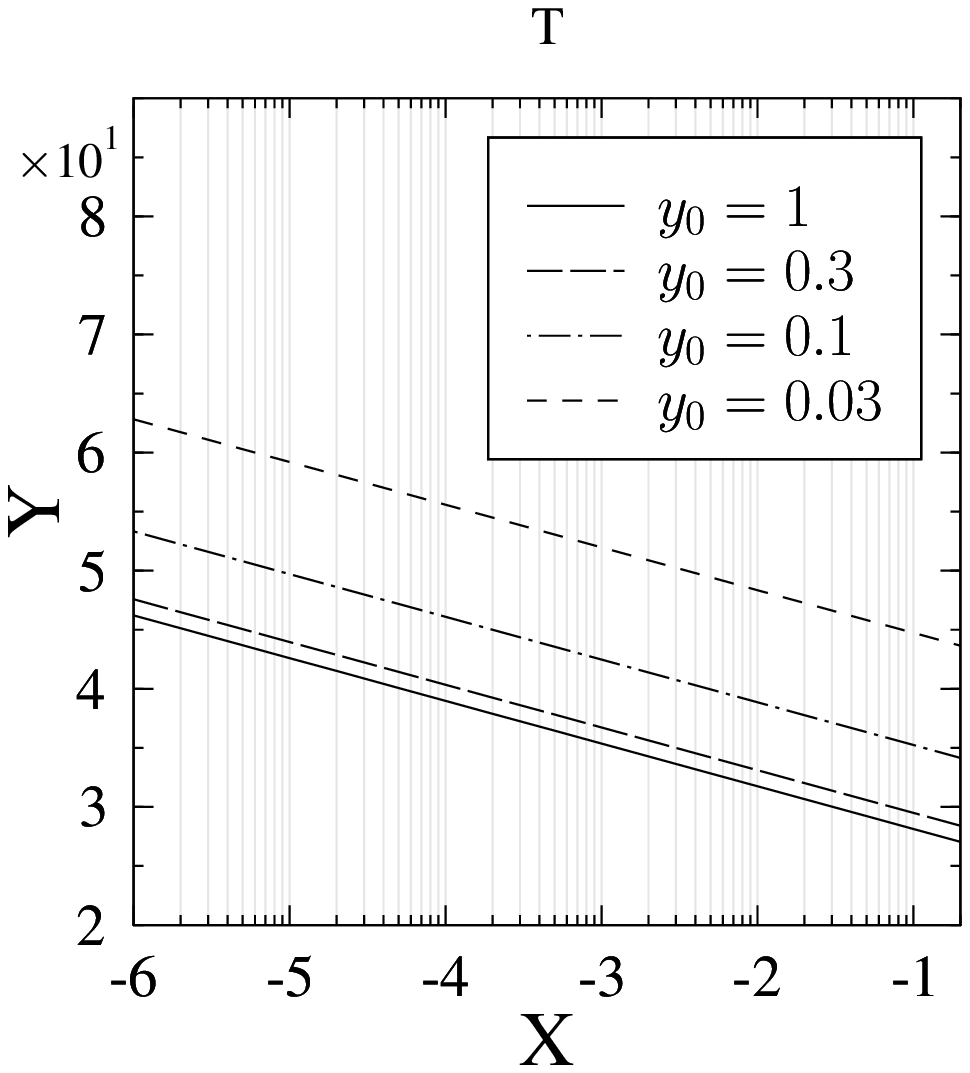}
}
\hspace{-1.5em}
\psfrag{Y}{}
\makebox{
\psfrag{T}[b]{\raisebox{0.5em}{Order: $\eps^{-1}$}}
\includegraphics[scale=0.42]{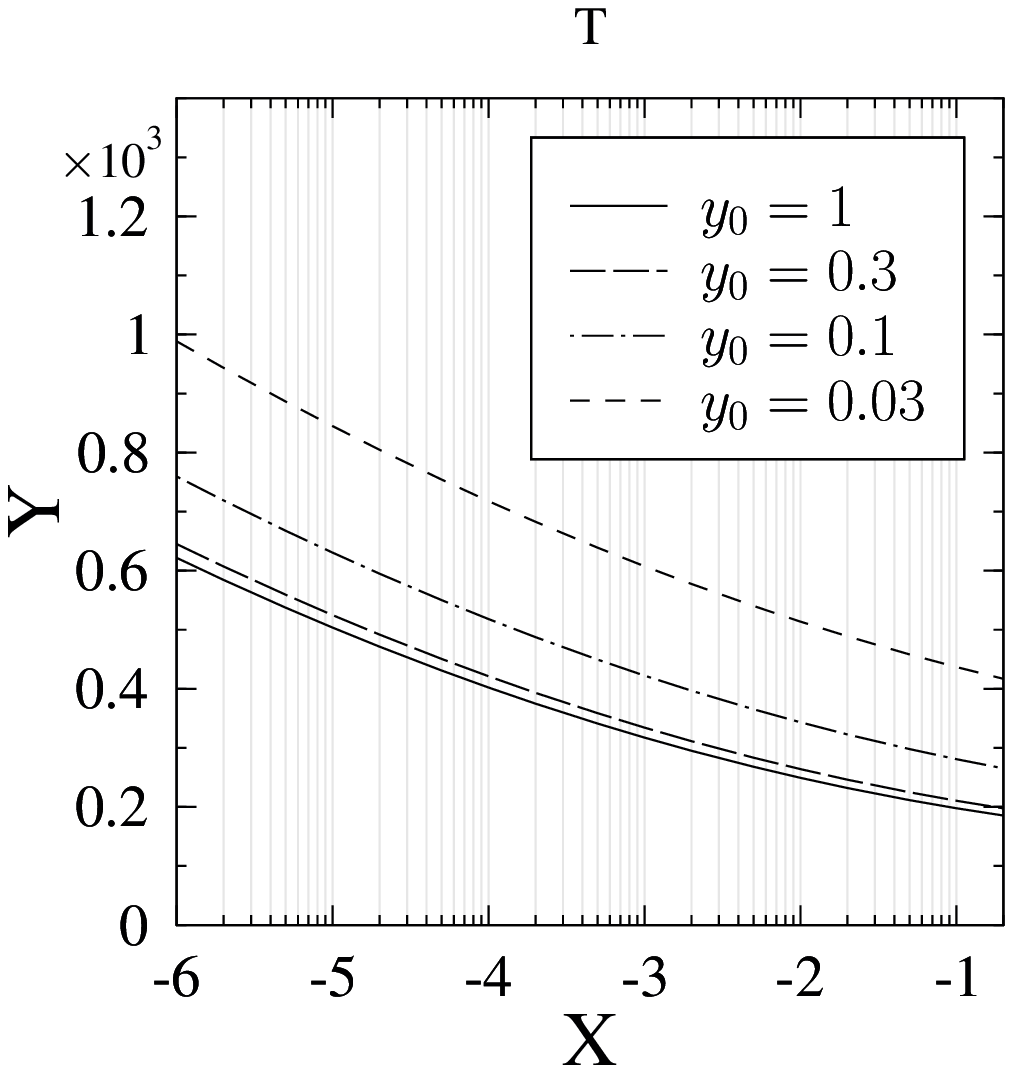}
}
\hspace{-1.5em}
\makebox{
\psfrag{T}[b]{\raisebox{0.5em}{Order: $\eps^0$}}
\includegraphics[scale=0.42]{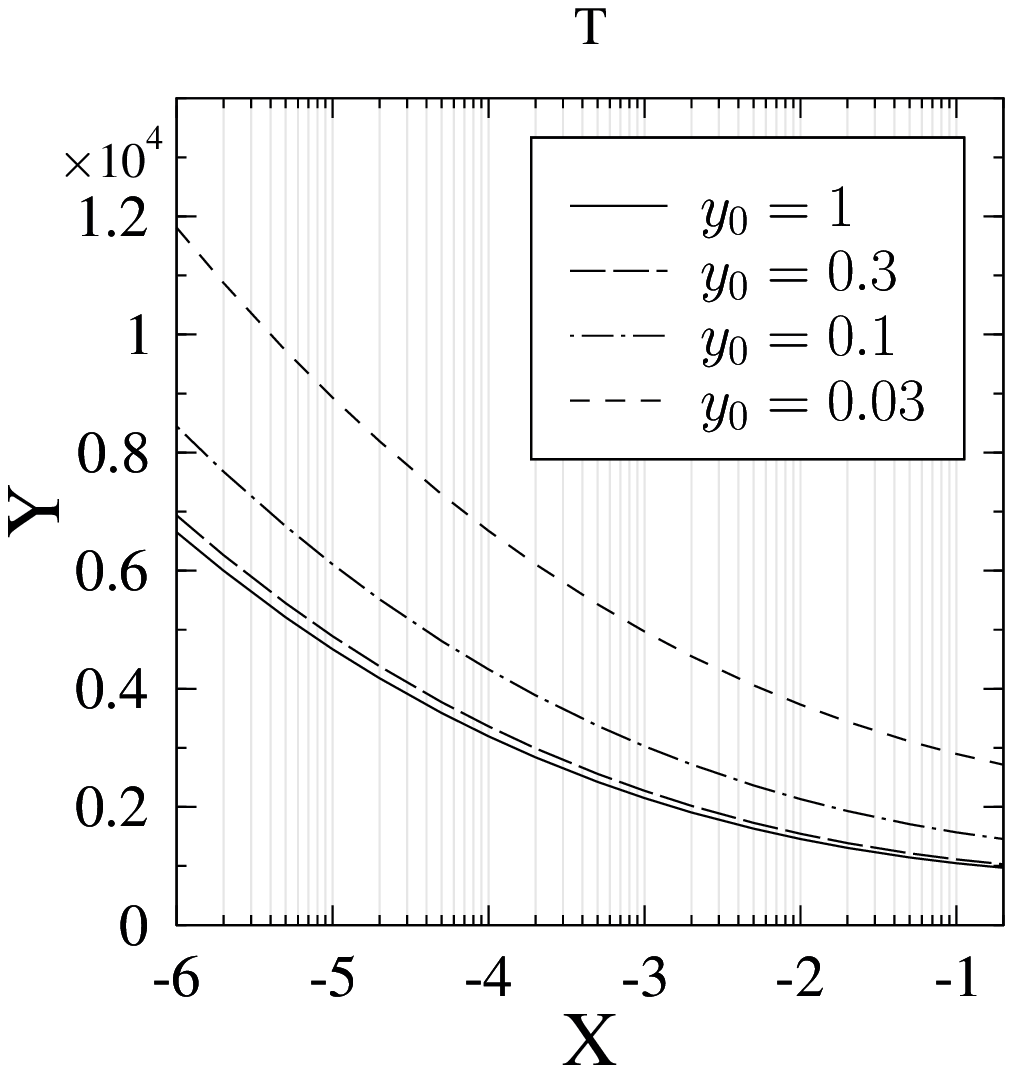}}
}
\vspace{-1em}
\caption{Expansion coeffiecients of the function 
$\IcS{ikl}{(1)}(0.1,Y,0.1,;\eps)$ with $d'_0=3-3\eps$.}
}
\FIGURE{
\label{fig:cSikl3}
\makebox{
\hspace{-1.5em}
\psfrag{X}[ct]{$\log_{10} Y$}
\makebox{
\psfrag{Y}[cb]{\raisebox{1.0em}{$\IcS{ikl}{(1)}(0.1,0.1,Y;\eps)$}}
\psfrag{T}[b]{\raisebox{0.5em}{Order: $\eps^{-2}$}}
\includegraphics[scale=0.42]{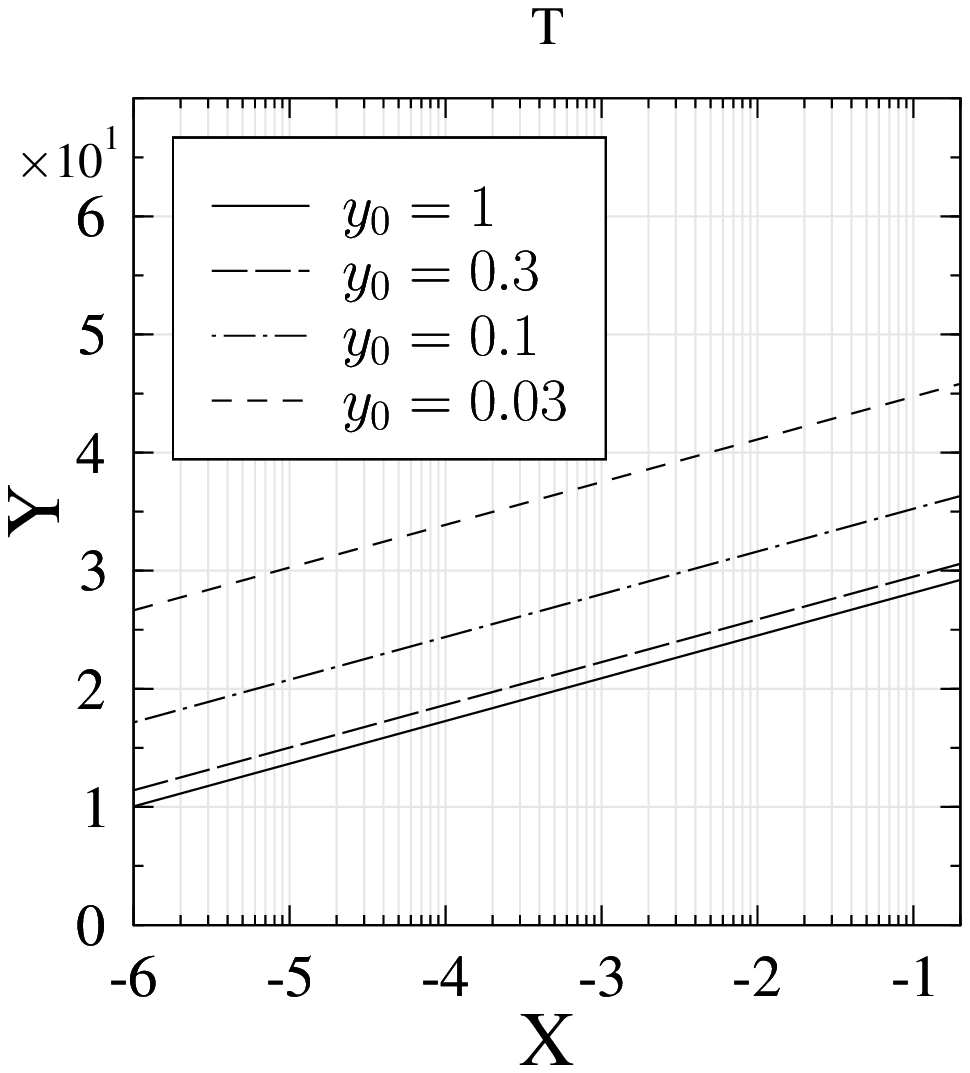}
}
\hspace{-1.5em}
\psfrag{Y}{}
\makebox{
\psfrag{T}[b]{\raisebox{0.5em}{Order: $\eps^{-1}$}}
\includegraphics[scale=0.42]{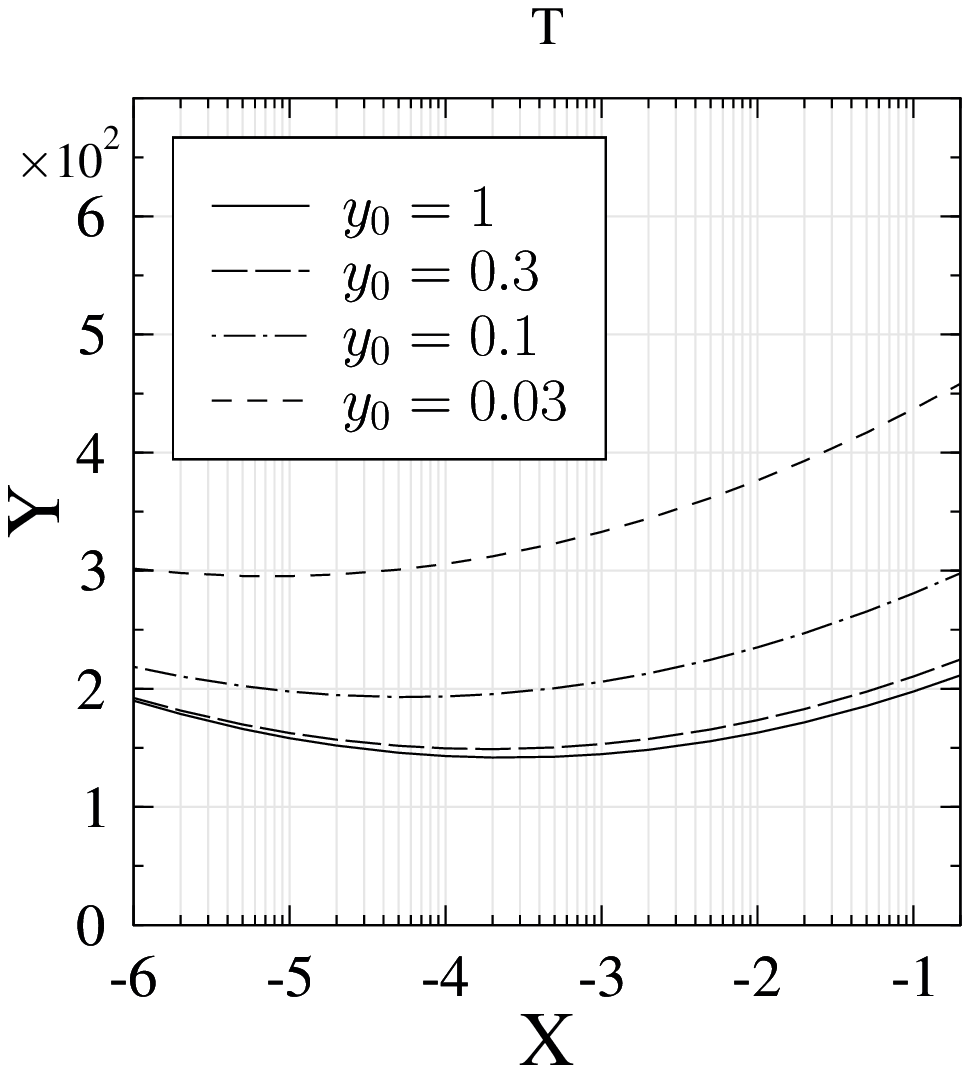}
}
\hspace{-1.5em}
\makebox{
\psfrag{T}[b]{\raisebox{0.5em}{Order: $\eps^0$}}
\includegraphics[scale=0.42]{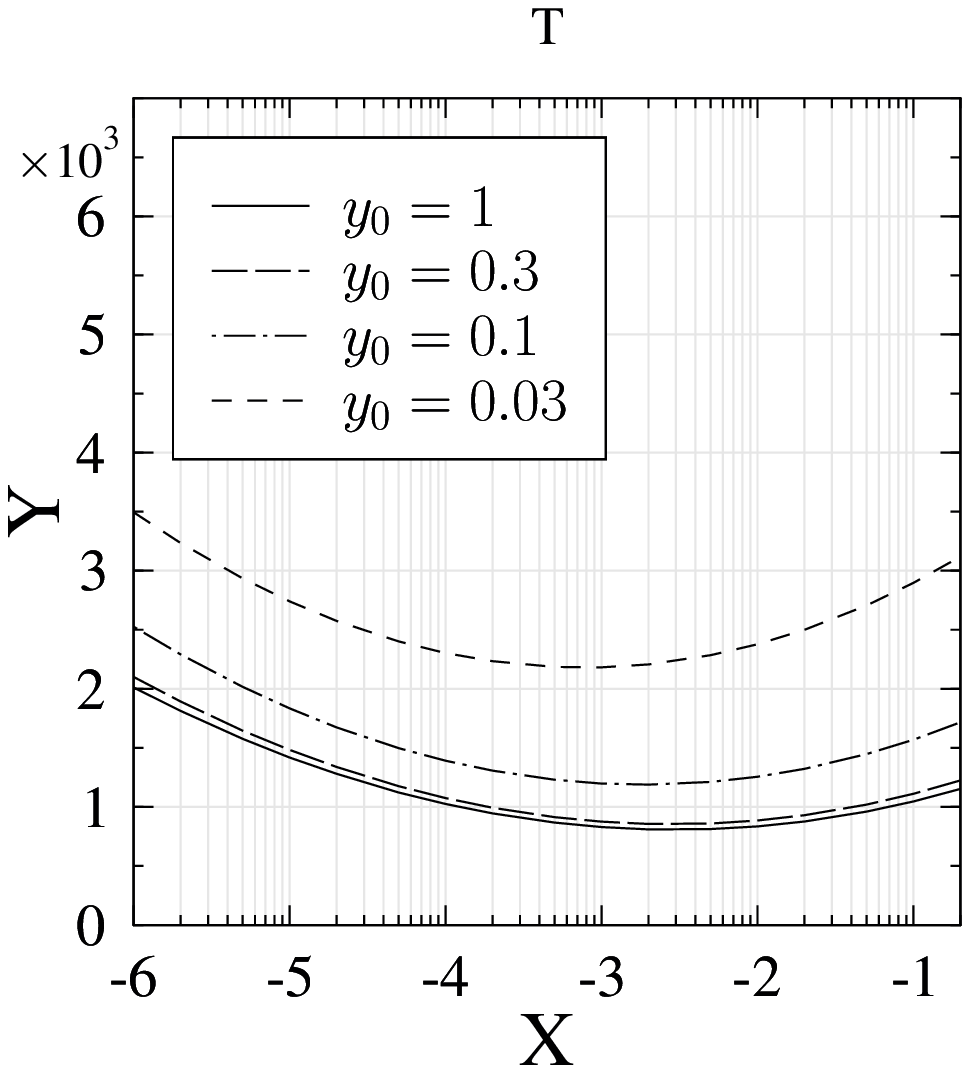}}
}
\caption{Expansion coeffiecients of the function 
$\IcS{ikl}{(1)}(0.1,0.1,Y;\eps)$ with $d'_0=3-3\eps$.}
}

%%%
%%% Integrals of the counterterms for the integrated approximate 
%%% cross section
%%%
\clearpage

\section{Integrals of the counterterms for the integrated approximate 
cross section}
\label{sec:IRRA1_A1}

The singly-unresolved approximate cross section regularizing the 
integrated approximate cross section $\int_{1}\dsiga{RR}{1}_{m+2}$ 
times the jet function reads
\beeq
&&
\Big(\int_1\dsiga{RR}{1}_{m+2}\Big) 
\strut^{{\rm A}_{\scriptscriptstyle 1}} J_m = {\cal N}\sum_{\{m+1\}}
\PS{m+1}(\mom{})\frac{1}{S_{\{m+1\}}}
\nn\\&&\qquad\times
\Bigg\{
\sum_{r}\Bigg[\sum_{i\ne r}
\frac{1}{2}\cC{ir}{(0,0\otimes I)}(\mom{})J_m(\momti{(ir)})
\nn\\&&\qquad\qquad\qquad
+\,\Bigg(
\cS{r}{(0,0\otimes I)}(\mom{})
-\sum_{r\ne i}\cC{ir}{}\cS{r}{(0,0\otimes I)}(\mom{})
\Bigg)J_m(\momti{(r)})
\Bigg]
\nn\\&&\qquad\quad 
+\,\sum_{r}\Bigg[\sum_{i\ne r}
\frac{1}{2}\cC{ir}{R\times(0,0)}(\mom{})J_m(\momti{(ir)})
\nn\\&&\qquad\qquad\qquad
+\,\Bigg(
\cS{r}{R\times(0,0)}(\mom{})
-\sum_{r\ne i}\cC{ir}{}\cS{r}{R\times(0,0)}(\mom{})
\Bigg)J_m(\momti{(r)})
\Bigg]
\Bigg\}\,.
\label{eq:dsigRRA1A1}
\eeeq
The precise meaning of each subtraction term appearing above was 
spelled out explicitly in \Ref{Somogyi:2006db}.

%
% Integral of the collinear counterterms
%

\subsection{Integrals of the collinear counterterms}
\label{ssec:intC00ICR00}

The collinear counterterms are
\bal
\cC{ir}{(0,0\otimes I)}(\mom{}) &=
8\pi\as\mu^{2\ep}\frac{1}{s_{ir}}
\nt\\[2mm] &\times
\bra{m}{(0)}{(\momti{(ir)})}
\bI^{(0)}(\momti{(ir)},\ep)
\hP_{f_{i}f_{r}}^{(0)}(\tzz{i}{r},\tzz{r}{i},\kTt{i,r};\ep)
\ket{m}{(0)}{(\momti{(ir)})}
\nt\\[2mm] &\times
(1-\alpha_{ir})^{2d_{0}-2(m-1)(1-\ep)}\Theta(\alpha_{0}-\alpha_{ir})
\label{eq:Cir00I}
\eal
and
\bal
\cC{ir}{R\times(0,0)}(\mom{}) &=
8\pi\as\mu^{2\ep}\frac{1}{s_{ir}}
\R{ir}(y_{ir},\tzz{i}{r}y_{\wti{ir}Q},\tzz{r}{i}y_{\wti{ir}Q};\ep)
\nt\\[2mm] &\times
\bra{m}{(0)}{(\momti{(ir)})}
\hP_{f_{i}f_{r}}^{(0)}(\tzz{i}{r},\tzz{r}{i},\kTt{i,r};\ep)
\ket{m}{(0)}{(\momti{(ir)})}
\nt\\[2mm] &\times
(1-\alpha_{ir})^{2d_{0}-2(m-1)(1-\ep)}\Theta(\alpha_{0}-\alpha_{ir})\,,
\label{eq:CirR00}
\eal
where $\hP_{f_i f_r}^{(0)}$ are again the tree-level Altarelli--Parisi
kernels and
the momentum mapping used to define the momenta entering the
matrix elements on the right hand sides and the corresponding phase
space factorization are the same as in \sect{ssec:intC01C10}. The function 
$\R{ir}(y_{ir},\tzz{i}{r}y_{\wti{ir}Q},\tzz{r}{i}y_{\wti{ir}Q};\ep)$
reads
\beq
\bsp
&
\R{ir}(y_{ir},\tzz{i}{r}y_{\wti{ir}Q},\tzz{r}{i}y_{\wti{ir}Q};\ep) =
\aeps
\\ &\qquad\times
\Bigg[
\IcC{i}{(0)}(\tzz{i}{r}y_{\wti{ir}Q};\ep)\,\bT_{i}^{2}
+\IcC{r}{(0)}(\tzz{r}{i}y_{\wti{ir}Q};\ep)\,\bT_{r}^{2}
-\IcC{(ir)}{(0)}(y_{\wti{ir}Q};\ep)\,\bT_{ir}^{2}\\
&\qquad\qquad
+(\bT_{ir}^{2}-\bT_{i}^{2}-\bT_{r}^{2})\,
\TcS{ir}{(0)}\left(\frac{y_{ir}}{\tzz{i}{r}\tzz{r}{i}y_{\wti{ir}Q}^{2}},
\ep\right)
\Bigg]\,.
\esp
\label{eq:Rir}
\eeq
Note that in order to lighten the notation, we do not explicitly
indicate the dependence of $\R{ir}$ on $\alpha_{0},d_{0},y_{0}$ and $d'_{0}$.

As usual, we write the integrals of the collinear subtraction terms
over the factorized phase space in the following form
\bal
\int_{1}\cC{ir}{(0,0\otimes I)} &=
\aeps
\IcC{ir}{(0)}(y_{\wti{ir}Q};\ep)\,\bT_{ir}^{2}
\nt\\[2mm] &\times
\bra{m}{(0)}{(\momti{(ir)})}
\bI^{(0)}(\momti{(ir)};\ep)
\ket{m}{(0)}{(\momti{(ir)})}
\label{eq:I1Cir00I}
\intertext{and}
\int_{1}\cC{ir}{R\times(0,0)} &=
\left[\aeps\right]^2 \frac{(4\pi)^2}{S_\ep}
\IcC{ir}{R\times(0)}(y_{\wti{ir}Q};\ep)\,
\bT_{ir}^{2}\,\SME{m}{0}{\momti{(ir)}}\,.
\label{eq:I1CirR00}
\eal
The functions $\IcC{ir}{(0)}$ and $\bI^{(0)}$ in \eqn{eq:I1Cir00I}
are defined in \eqn{eq:ICir0} and \eqnss{eq:I0}{eq:ICi0}, 
while $\IcC{ir}{R\times(0)}$ reads
\beq
\bsp
\IcC{ir}{R\times(0)}(y_{\wti{ir}Q};\ep) &=
(Q^{2})^{\ep}\int_{0}^{\alpha_{0}}\rd \alpha\,(1-\alpha)^{2d_{0}-1}
\frac{s_{\wti{ir}Q}}{2\pi}\PS{2}(p_{i},p_{r};p_{(ir)})\\
&\times
\frac{1}{s_{ir}}
\tiR{ir}(y_{ir},\tzz{i}{r}y_{\wti{ir}Q},\tzz{r}{i}y_{\wti{ir}Q};\ep)
P_{f_{i}f_{r}}^{(0)}(\tzz{i}{r},\tzz{r}{i};\ep)\frac{1}{\bT_{ir}^{2}}\,,
\esp
\label{eq:ICirR00}
\eeq
where
\beq
\tiR{ir} = \left[\aeps\right]^{-1} \R{ir}\,.
\label{eq:tiRir}
\eeq
As the rest of the integrated collinear functions, $\IcC{ir}{R\times(0)}$
too is independent of $m$ as the notation suggests. Also in keeping with
our conventions, we do not indicate the explicit dependence on the cut
parameters $\alpha_0$ and $y_0$ or the exponents $d_0$ and $d'_0$.
The analytic evaluation of the integrals is performed in
\Ref{Bolzoni:2008}.

%
% Integral of the soft-type counterterms
%

\subsection{Integrals of the soft-type counterterms}
\label{ssec:intS00ISR00}

The soft-type counterterms read
\bal
\cS{r}{(0,0\otimes I)}(\mom{}) &=
-8\pi\as\mu^{2\ep}\sum_{i}\sum_{k\ne i}\frac{1}{4}\calS_{ik}(r)
\nt\\[2mm] &\times
\bra{m}{(0)}{(\momti{(r)})}
\{\bI(\momti{(r)};\ep),
\bT_{i}\ldot\bT_{k}\}
\ket{m}{(0)}{(\momti{(r)})}
\nt\\[2mm] &\times
(1-y_{rQ})^{d'_{0}-(m-1)(1-\ep)}\Theta(y_{0}-y_{rQ})\,,
\label{eq:Sr00I}\\[2mm]
\cC{ir}{}\cS{r}{(0,0\otimes I)}(\mom{}) &=
8\pi\as\mu^{2\ep}\frac{1}{s_{ir}}\frac{2\tzz{i}{r}}{\tzz{r}{i}}
\,\bT_{i}^{2}\,
\bra{m}{(0)}{(\momti{(r)})}
\bI(\momti{(r)};\ep)
\ket{m}{(0)}{(\momti{(r)})}
\nt\\[2mm] &\times
(1-y_{rQ})^{d'_{0}-(m-1)(1-\ep)}\Theta(y_{0}-y_{rQ})\,,
\label{eq:CirSr00I}
\eal
and
\bal
\cS{r}{R\times(0,0)}(\mom{}) &=
-8\pi\as\mu^{2\ep}\sum_{i}\sum_{k\ne i}\frac{1}{2}\calS_{ik}(r)
\R{ik,r}(y_{ik},y_{ir},y_{kr},y_{iQ},y_{kQ},y_{rQ};\ep)
\nt\\[2mm] &\times
\SME{m;(i,k)}{0}{\momti{(r)}}
(1-y_{rQ})^{d'_{0}-(m-1)(1-\ep)}\Theta(y_{0}-y_{rQ})
\label{eq:SrR00}
\,,\\
\cC{ir}{}\cS{r}{R\times(0,0)}(\mom{}) &=
8\pi\as\mu^{2\ep}\frac{1}{s_{ir}}\frac{2\tzz{i}{r}}{\tzz{r}{i}}
\bT_{i}^{2}\,
\CA\frac{\as}{2\pi}S_{\ep}\left(\frac{\mu^{2}}{Q^{2}}\right)^{\ep}
\left[
\IcC{g}{(0)}(y_{rQ};\ep)
-\TcS{ir}{(0)}\left(\frac{y_{ir}}{y_{iQ}y_{rQ}};\ep\right)
\right]
\nt\\[2mm] &\times
\SME{m}{0}{\momti{(r)}}
(1-y_{rQ})^{d'_{0}-(m-1)(1-\ep)}\Theta(y_{0}-y_{rQ})\,.
\label{eq:CirSrR00}
\eal
The momenta which enter the matrix elements on the right hand sides and
the corresponding phase space factorization are the same as in 
\sect{ssec:intS01S10}. In \eqn{eq:SrR00} the function 
$\R{ik,r}(y_{ik},y_{ir},y_{kr},y_{iQ},y_{kQ},y_{rQ};\ep)$ is
\beq
\bsp
&
\R{ik,r}(y_{ik},y_{ir},y_{kr},y_{iQ},y_{kQ},y_{rQ};\ep) =
\CA\frac{\as}{2\pi}S_{\ep}\left(\frac{\mu^{2}}{Q^{2}}\right)^{\ep}\\
&\qquad\times
\left[
\IcC{g}{(0)}(y_{rQ};\ep)
+\TcS{ik}{(0)}\left(\frac{y_{ik}}{y_{iQ}y_{kQ}};\ep\right)
-\TcS{ir}{(0)}\left(\frac{y_{ir}}{y_{iQ}y_{rQ}};\ep\right)
-\TcS{kr}{(0)}\left(\frac{y_{kr}}{y_{kQ}y_{rQ}};\ep\right)
\right]\,,
\esp
\label{eq:Rikr}
\eeq
where again we do not indicate the dependence of $\R{ik,r}$ on 
$\alpha_{0},d_{0},y_{0}$ and $d'_{0}$ explicitly.

After integrating the soft-type subtraction terms over the unresolved
phase space we can write the results as
\bal
\int_{1}\cS{r}{(0,0\otimes I)}(\mom{}) &=
\aeps
\sum_{i}\sum_{k\ne i}\IcS{ik}{(0)}(\Y{i}{k};\ep)
\nt\\[2mm] &\times
\frac{1}{2}
\bra{m}{(0)}{(\momti{(r)})}
\{\bI(\momti{(r)};\ep),
\bT_{i}\ldot \bT_{k}\}
\ket{m}{(0)}{(\momti{(r)})}\,,
\label{eq:I1Sr00I}\\[2mm]
\int_{1}\cC{ir}{}\cS{r}{(0,0\otimes I)}(\mom{}) &=
\aeps
\IcCS{(0)}(\ep)\,\bT_{i}^{2}
\nt\\[2mm] &\times
\bra{m}{(0)}{(\momti{(r)})}
\bI(\momti{(r)};\ep)
\ket{m}{(0)}{(\momti{(r)})}\,,
\label{eq:I1CirSr00I}
\eal
\bal
\int_{1}\cS{r}{R\times(0,0)}(\mom{}) &=
\left[\aeps\right]^2 \frac{(4\pi)^2}{S_\ep}
\sum_{i}\sum_{k\ne i}
\IcS{ik}{R\times(0)}(\Y{i}{k};\ep)
\SME{m;(i,k)}{0}{\momti{(r)}}
\label{eq:I1SrR00}
\,,\\
\int_{1}\cC{ir}{}\cS{r}{R\times(0,0)}(\mom{}) &=
\left[\aeps\right]^2 \frac{(4\pi)^2}{S_\ep}
\IcCS{R\times(0)}(\Y{i}{k};\ep)\,
\bT_{i}^{2}\,\SME{m}{0}{\momti{(r)}}\,.
\label{eq:I1CirSrR00}
\eal
The $\IcS{ik}{(0)}$ and $\IcCS{(0)}$ functions are defined in 
\eqns{eq:ISik0}{eq:ICS0} respectively, while for $\IcS{ik}{R\times(0)}$ 
and $\IcCS{R\times(0)}$ we need to compute the integrals
\bal
\IcS{ik}{R\times(0)}(\Y{i}{k};\ep) &=
-(Q^{2})^{\ep}\int_{0}^{y_{0}}\!\rd y\,(1-y)^{d'_{0}-1}
\frac{Q^{2}}{2\pi}\PS{2}(p_{r},K;Q)
 \frac{1}{2}\calS_{ik}(r)
\nt\\[2mm] &\qquad\qquad\quad\times
 \CA\,\tiR{ik,r}(y_{ik},y_{ir},y_{kr},y_{iQ},y_{kQ},y_{rQ};\ep)
 \label{eq:ISikR00}
 \intertext{and}
 \IcCS{R\times(0)}(\ep) &=
 (Q^{2})^{\ep}\int_{0}^{y_{0}}\!\rd y\,(1-y)^{d'_{0}-1}
\frac{Q^{2}}{2\pi}\PS{2}(p_{r},K;Q)
\frac{1}{s_{ir}}\frac{2\tzz{i}{r}}{\tzz{r}{i}}
\nt\\[2mm] &\qquad\qquad\times
\CA
\left[
\IcC{g}{(0)}(y_{rQ};\ep)
-\TcS{ir}{(0)}\left(\frac{y_{ir}}{y_{iQ}y_{rQ}};\ep\right)
\right]\,.
\label{eq:ICSR00}
\eal
In \eqn{eq:ISikR00}
\beq
\tiR{ik,r} = \left[\CA \frac{\as}{2\pi}S_{\ep}
\left(\frac{\mu^{2}}{Q^{2}}\right)^{\ep}\right]^{-1}
\R{ik,r}\,.
\label{eq:tiRik,r}
\eeq
The soft functions $\IcS{ik}{R\times(0)}$ and $\IcCS{R\times(0)}$ are
again independent of $m$ as the notation suggests but do depend on the 
cut parameters $\alpha_0$ and $y_0$ as well as the exponents $d_0$ and 
$d'_0$. The dependence on the later four parameters is suppressed in the
notation as usual.
The analytic evaluation of the integrals is performed in
\Ref{Bolzoni:2008}.

%
% The integrated approximate cross section
%

\subsection{The integrated approximate cross section}
\label{ssec:intRRA1:A1}

The computation of
$\int_1\left(\int_1\dsiga{RR}{1}_{m+2}\right)
\strut^{{\rm A}_{\scriptscriptstyle 1}}$
(including the counting of symmetry factors) proceeds along the same
lines as that in \sect{ssec:intRRA1}.
The final result for the integral of the iterated singly-unresolved
approximate cross section can be written as (cf.~with \eqn{eq:I1dsigRRA1})
\beq
\int_1\Big(\int_1\dsiga{RR}{1}_{m+2}\Big) 
\strut^{{\rm A}_{\scriptscriptstyle 1}} = 
\dsig{B}_{m}\otimes \left[
\frac12\Big\{
\bI^{(0)}(\mom{}_{m};\ep),
\bI^{(0)}(\mom{}_{m};\ep)
\Big\}
+ \bI^{R\times (0)}(\mom{}_{m};\ep)
\right]
\,,
\label{eq:I1dsigRA1_A1}
\eeq
where the insertion operator $\bI^{(0)}$ is given in \eqnss{eq:I0}{eq:ICi0},
while $\bI^{R\times(0)}$ reads
\beq
\bI^{R\times (0)}(\mom{}_{m};\ep) =
\left[\aeps\right]^2
\sum_{i}\bigg[
\IcC{i}{R\times (0)}(y_{iQ};\ep)\,\bT_{i}^{2}
+\sum_{k\ne i}
\TcS{ik}{R\times (0)}(\Y{i}{k};\ep)\,\bT_{i}\ldot\bT_{k}
\bigg]\,.
\label{eq:IRx0}
\eeq
In \eqn{eq:IRx0} we introduced the functions
\beeq
\IcC{q}{R\times (0)} &=& 
\frac{(4\pi)^{2}}{S_{\ep}}\,\IcC{qg}{R\times (0)}\,,\qquad
\IcC{g}{R\times (0)} = 
\frac{(4\pi)^{2}}{S_{\ep}}
\,\left(\frac{1}{2}\IcC{gg}{R\times (0)} + \Nf \IcC{q\qb}{R\times (0)}\right)
\,,
\nn\\[2mm]
\TcS{ik}{R\times (0)} &=&
\frac{(4\pi)^{2}}{S_{\ep}}
\,\left( \IcS{ik}{R\times (0)} + \IcCS{R\times (0)} \right)\,.
\label{eq:ICiRx0}
\eeeq

In this paper we evaluate the $\eps$-expansion of these functions using 
sector decomposition. We are able to find the coefficients of the two 
leading poles analytically (the coefficient of the $1/\eps^4$ pole
is zero in each case): 
\beeq
\IcC{q}{R\times(0)}(x;\eps) &=&
\frac{1}{\eps^3}
\Bigg[\frac{5}{3}-\ln x
- \frac{1}{3}\Nf\frac\TR\CA + \Sigma(y_0,d'_0)
\Bigg]\CA + \Oe{-2}\,,
\label{eq:IcCqR0}
\\
\IcC{g}{R\times(0)}(x;\eps) &=&
\frac{1}{\eps^3}
\Bigg[ \frac{11}{6}-\ln x
- \frac{2}{3} \Nf\frac\TR\CA\frac\CF\CA + \Sigma(y_0,d'_0)
\Bigg]\CA + \Oe{-2}\,,\qquad~
\label{eq:IcCgR0}
\eeeq
and
\beq
\TcS{ik}{R\times(0)}(Y;\eps) = 
\frac{1}{\eps^3}\CA\frac{\ln Y}{2} + \Oe{-2}\,.
\eeq
In \eqns{eq:IcCqR0}{eq:IcCgR0}, the function $\Sigma(y_0,d'_0)$
depends on the cut parameter $y_0$ and on the exponent $d'_0$.
If we set $d'_0=D'_0+d'_1\eps$, where $D'_0$ is an integer (greater than 2, 
see \appx{app:ddprimechoice}) and $d'_1$ is real, then we find
\beq
\Sigma(y_0,d'_0) = \ln y_0 - \sum_{k=1}^{D'_0}\frac{1-(1-y_0)^k}{k}\,.
\eeq
For the remaining coefficients we obtain integral representations which 
we can evaluate numerically. The results for $\alpha_0=y_0=1$, $0.3$, $0.1$ 
and $0.03$ with fixed values of $d_0=d'_0=3-3\eps$ are presented in 
\figs{fig:IcCgR}{fig:TcSikR}.

%
% Collinear figures: CRq
%
\FIGURE{
\label{fig:IcCgR}
\makebox{
\hspace{-1.5em}
\psfrag{X}[ct]{$\log_{10} x$}
\psfrag{T}{}
\makebox{
\psfrag{Y}[cb]{\raisebox{0.5em}{$\IcC{q}{R\times(0)}(x;\eps)$}}
\psfrag{T}[b]{\raisebox{0.5em}{Order: $\eps^{-2}$}}
\includegraphics[scale=0.42]{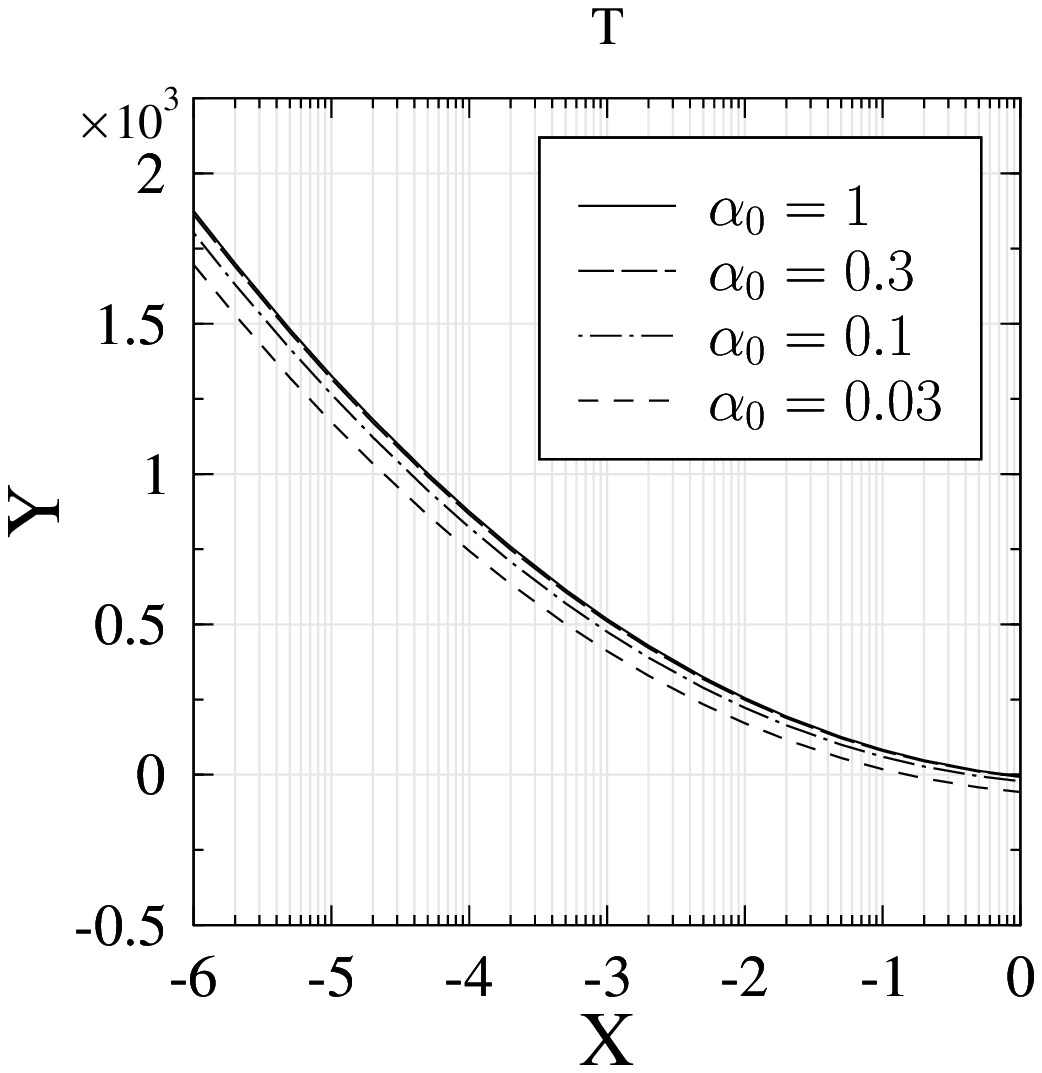}
}
\hspace{-1.5em}
\psfrag{Y}{}
\makebox{
\psfrag{T}[b]{\raisebox{0.5em}{Order: $\eps^{-1}$}}
\includegraphics[scale=0.42]{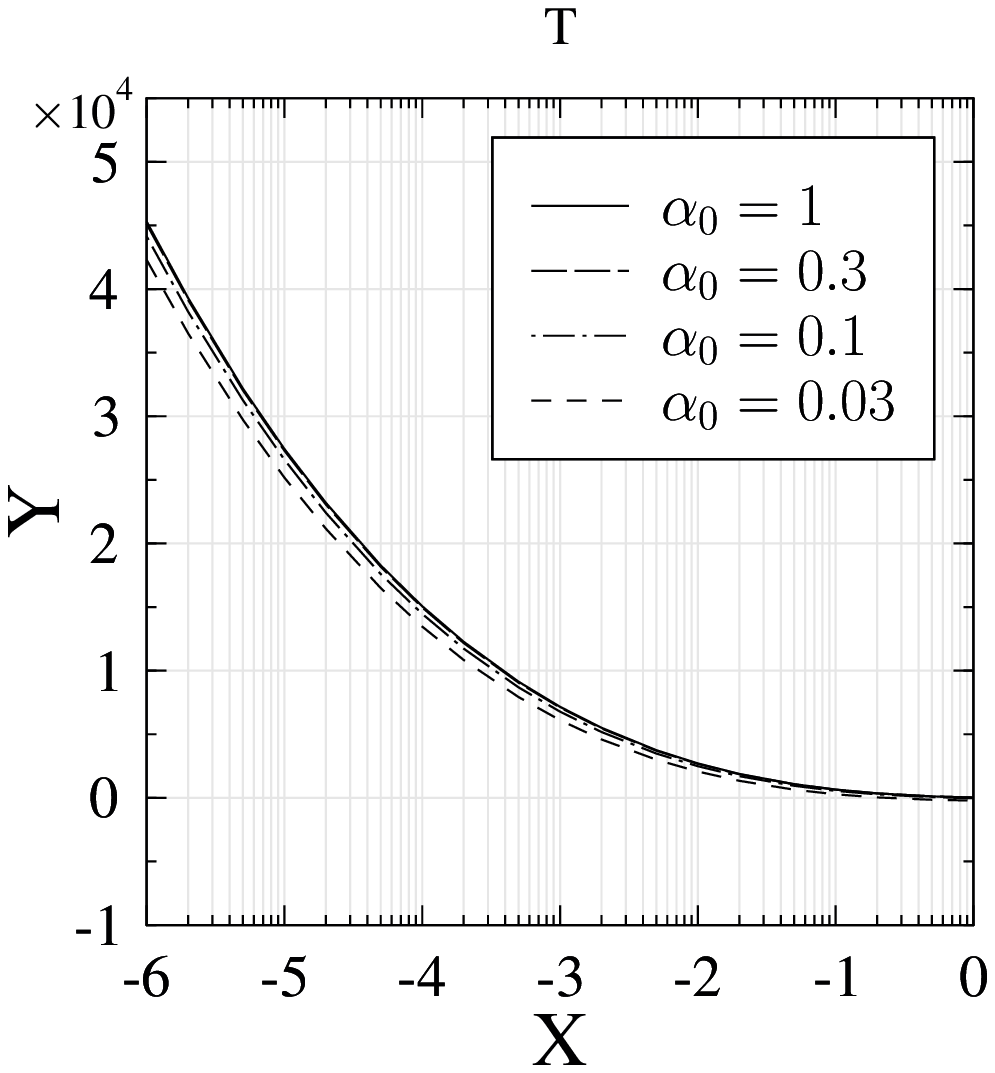}
}
\hspace{-1.5em}
\makebox{
\psfrag{T}[b]{\raisebox{0.5em}{Order: $\eps^0$}}
\includegraphics[scale=0.42]{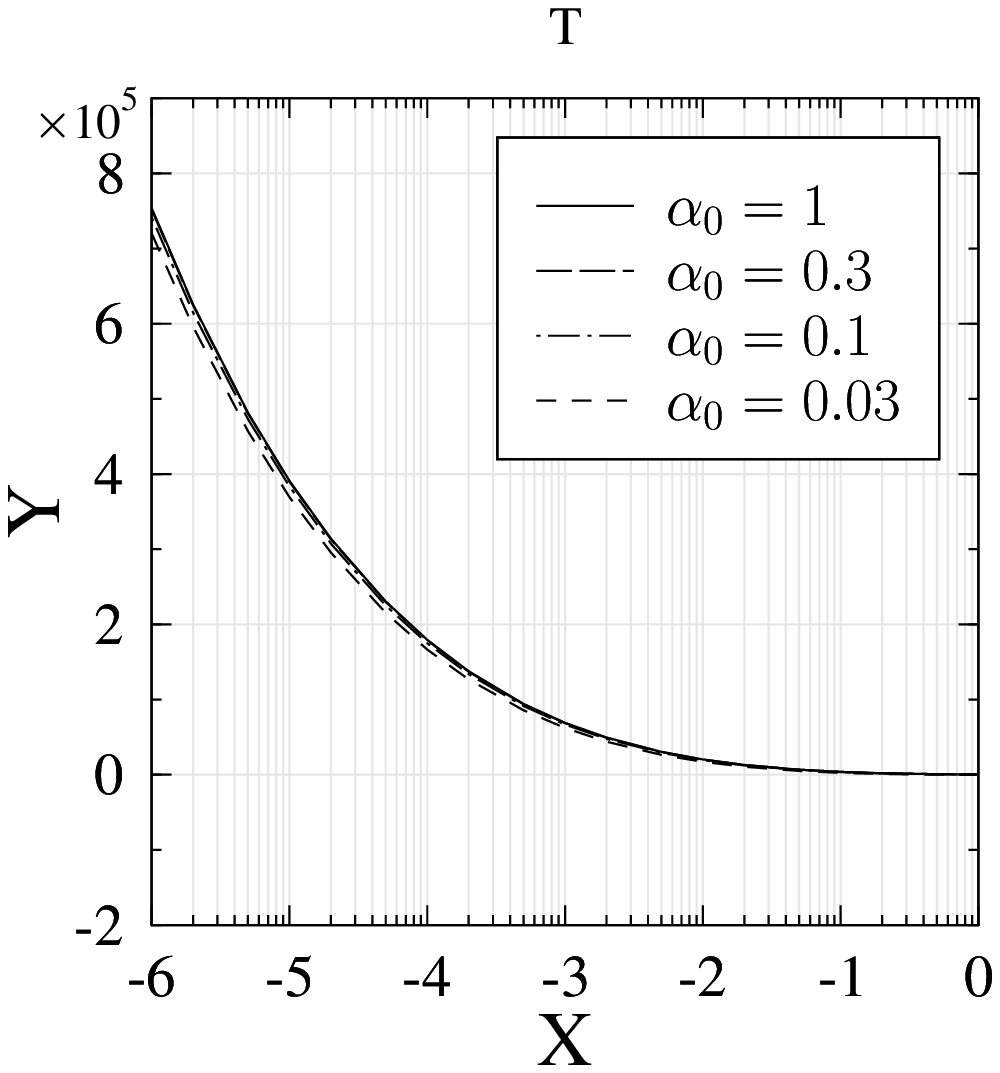}}
}
\makebox{
\hspace{-1.5em}
\psfrag{X}[ct]{$\log_{10} x$}
\psfrag{T}{}
\makebox{
\psfrag{Y}[cb]{\raisebox{0.5em}{$\IcC{g}{R\times(0)}(x;\eps)$}}
\includegraphics[scale=0.42]{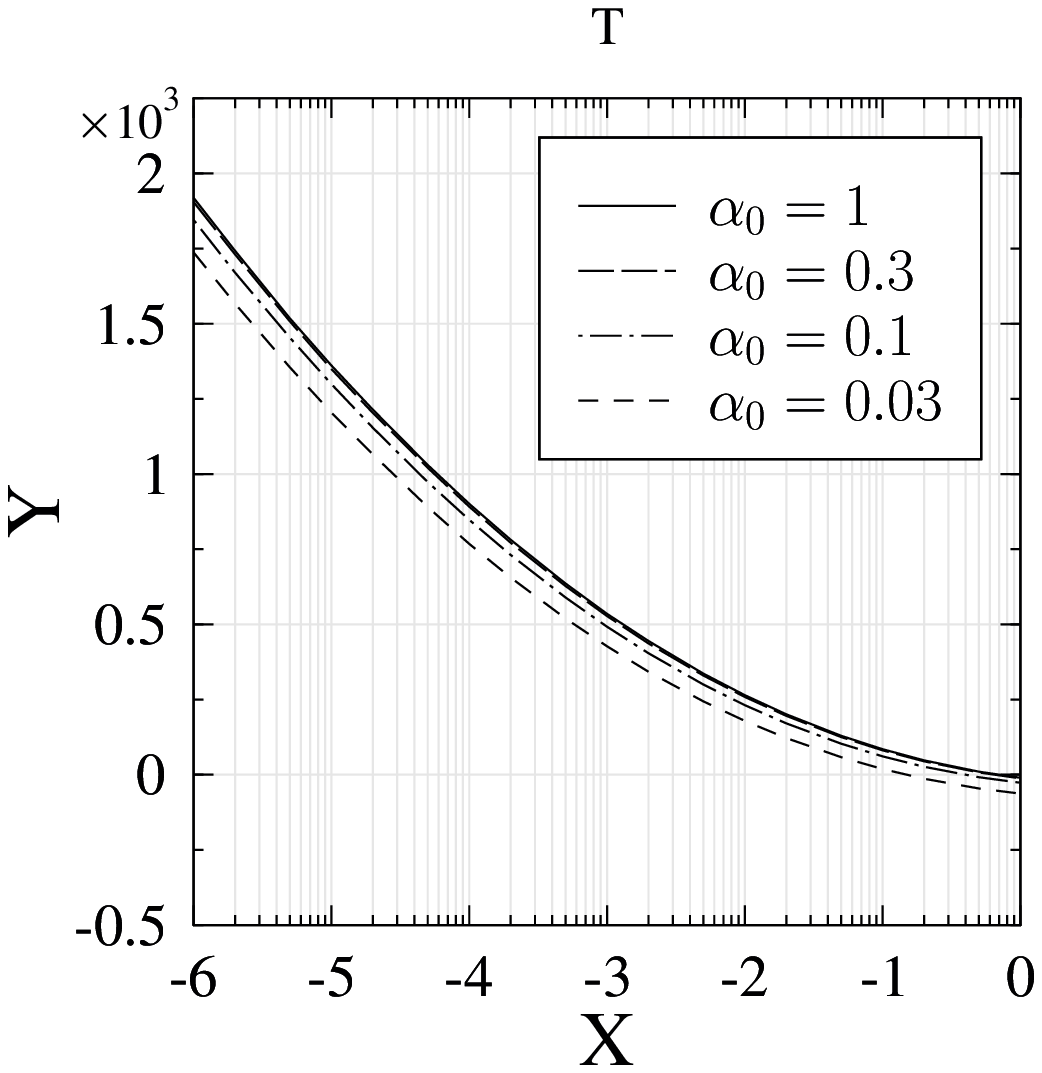}
}
\hspace{-1.5em}
\psfrag{Y}{}
\makebox{
\includegraphics[scale=0.42]{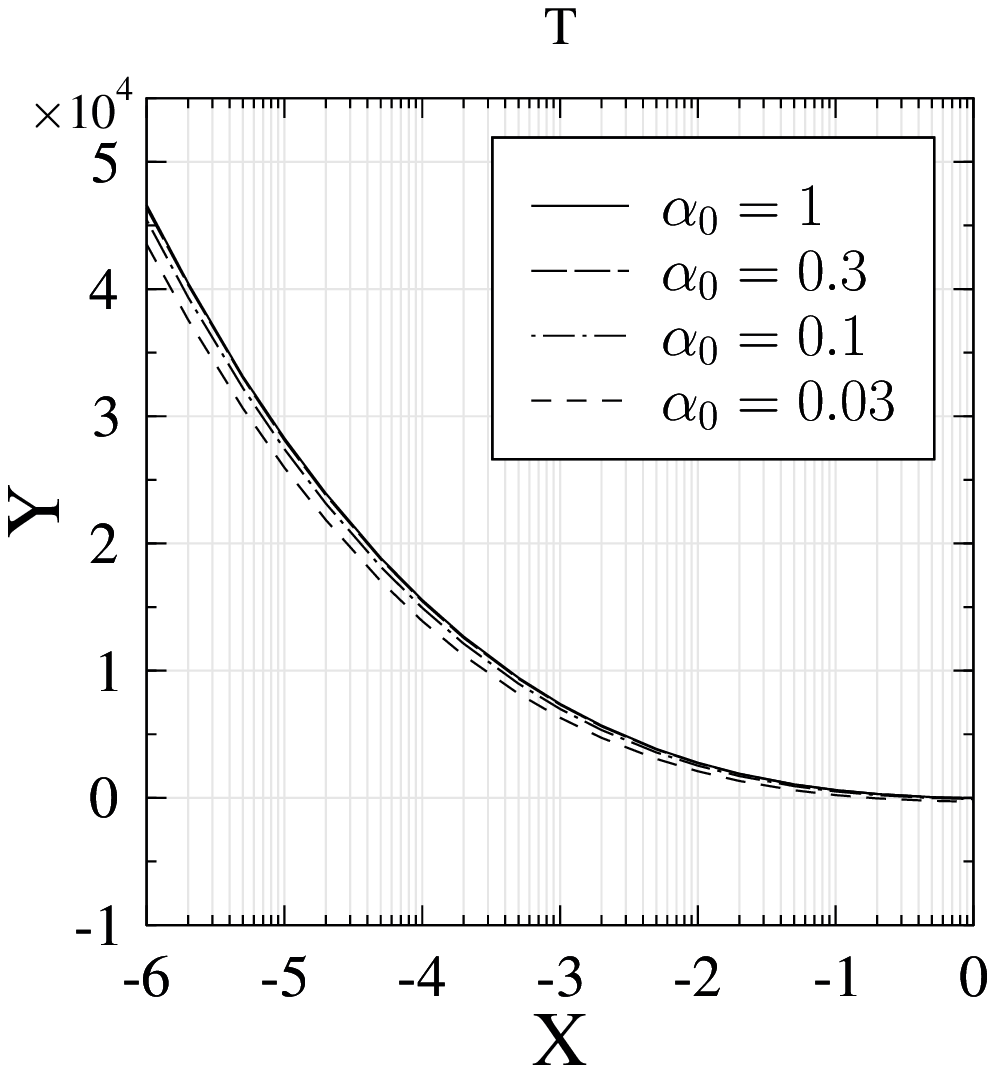}
}
\hspace{-1.5em}
\makebox{
\includegraphics[scale=0.42]{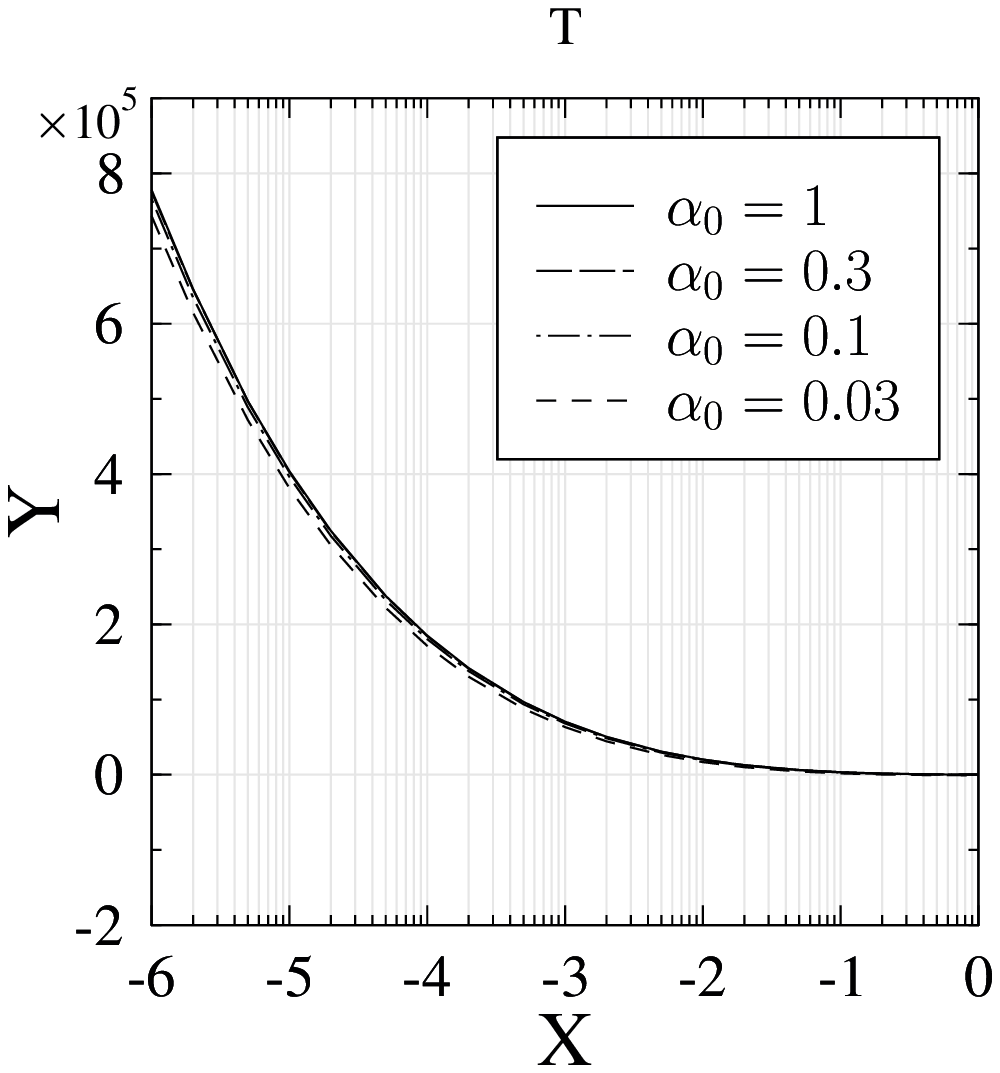}}
}
\caption{Expansion coeffiecients of the functions 
$\IcC{i}{R\times(0)}(x;\eps)$ with $d_0=d'_0=3-3\eps$ and $\Nf=5$.
Upper row: $i = q$, lower row: $i = g$.}
}
\FIGURE{
\label{fig:TcSikR}
\makebox{
\hspace{-1.5em}
\psfrag{X}[ct]{$\log_{10} Y$}
\makebox{
\psfrag{Y}[cb]{\raisebox{0.5em}{$\TcS{ik}{R\times(0)}(Y;\eps)$}}
\psfrag{T}[b]{\raisebox{0.5em}{Order: $\eps^{-2}$}}
\includegraphics[scale=0.42]{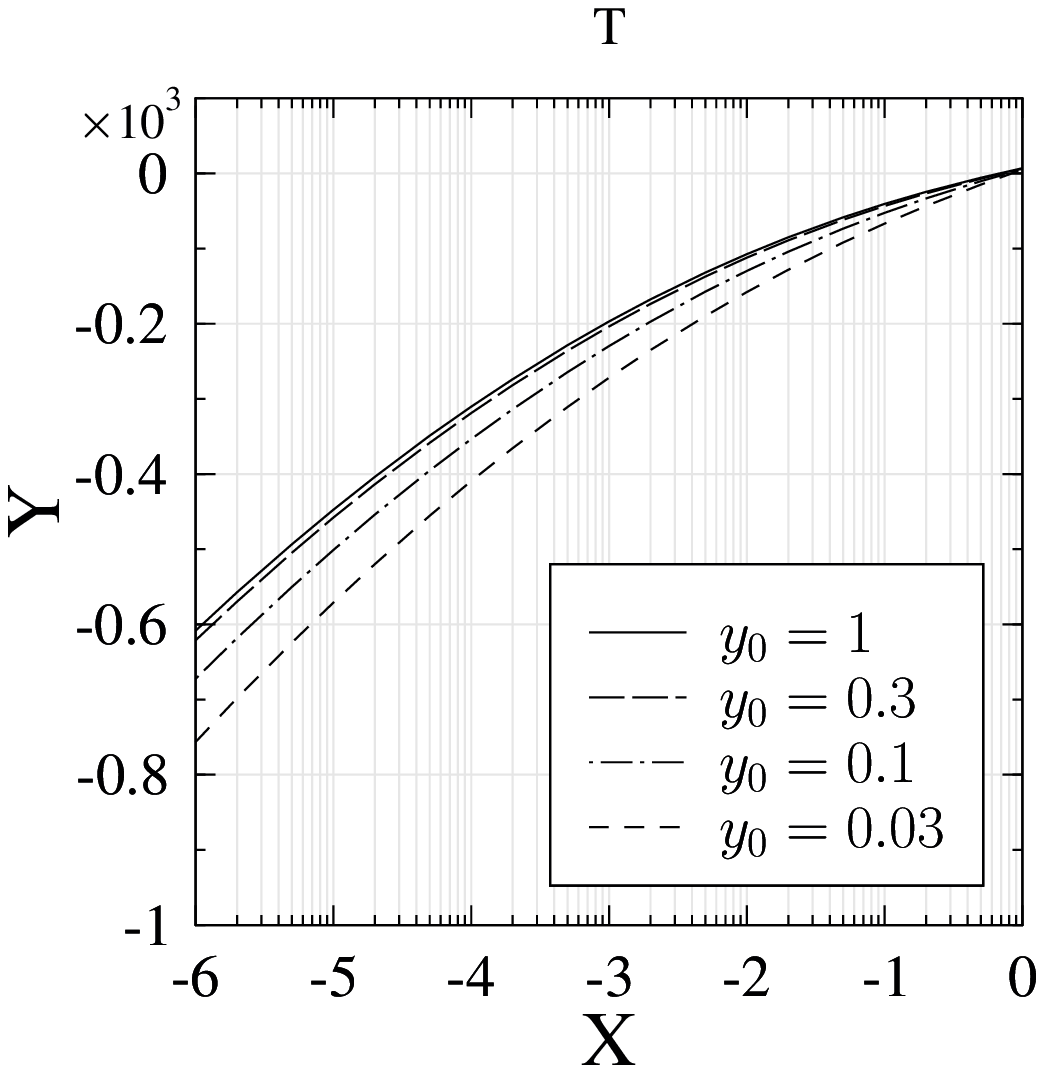}
}
\hspace{-1.5em}
\psfrag{Y}{}
\makebox{
\psfrag{T}[b]{\raisebox{0.5em}{Order: $\eps^{-1}$}}
\includegraphics[scale=0.42]{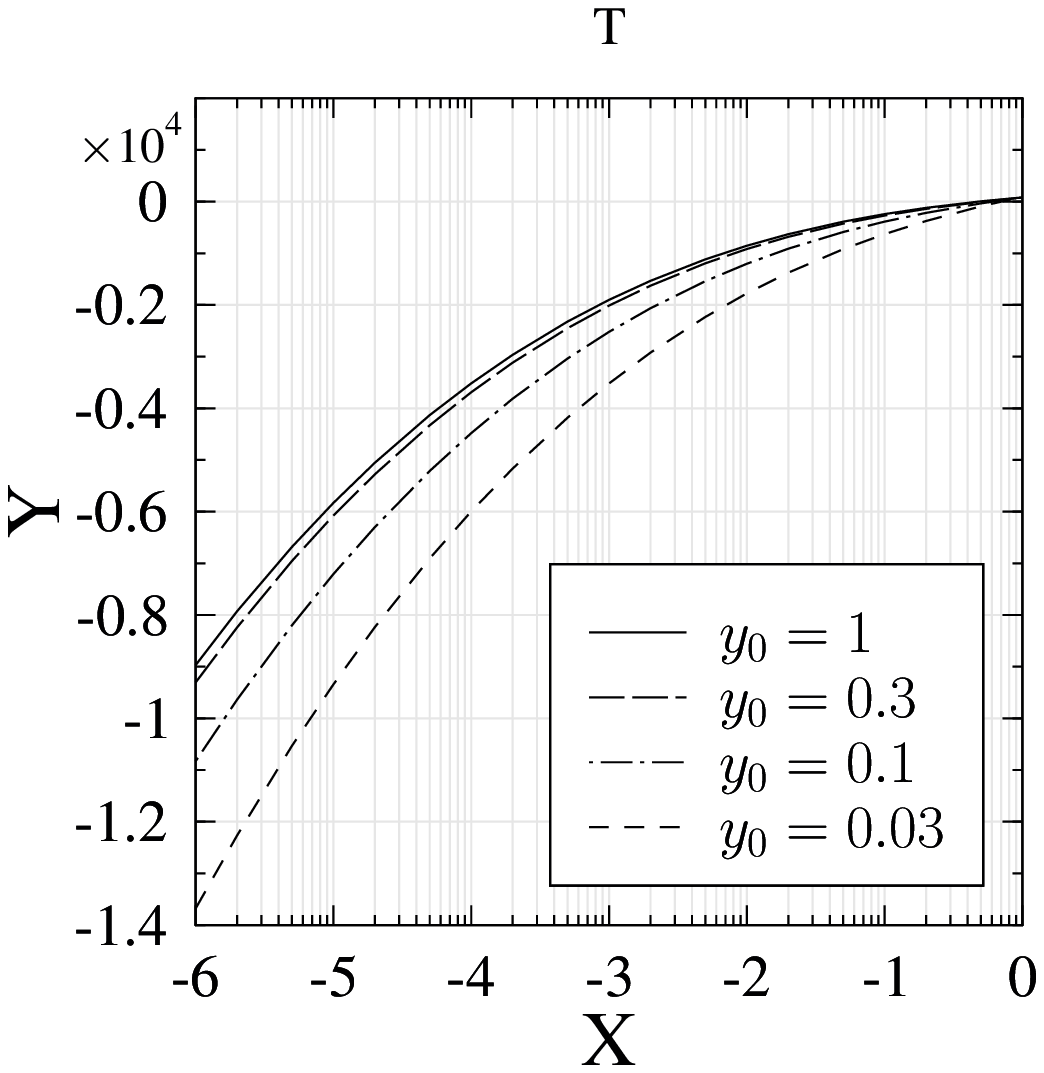}
}
\hspace{-1.5em}
\makebox{
\psfrag{T}[b]{\raisebox{0.5em}{Order: $\eps^0$}}
\includegraphics[scale=0.42]{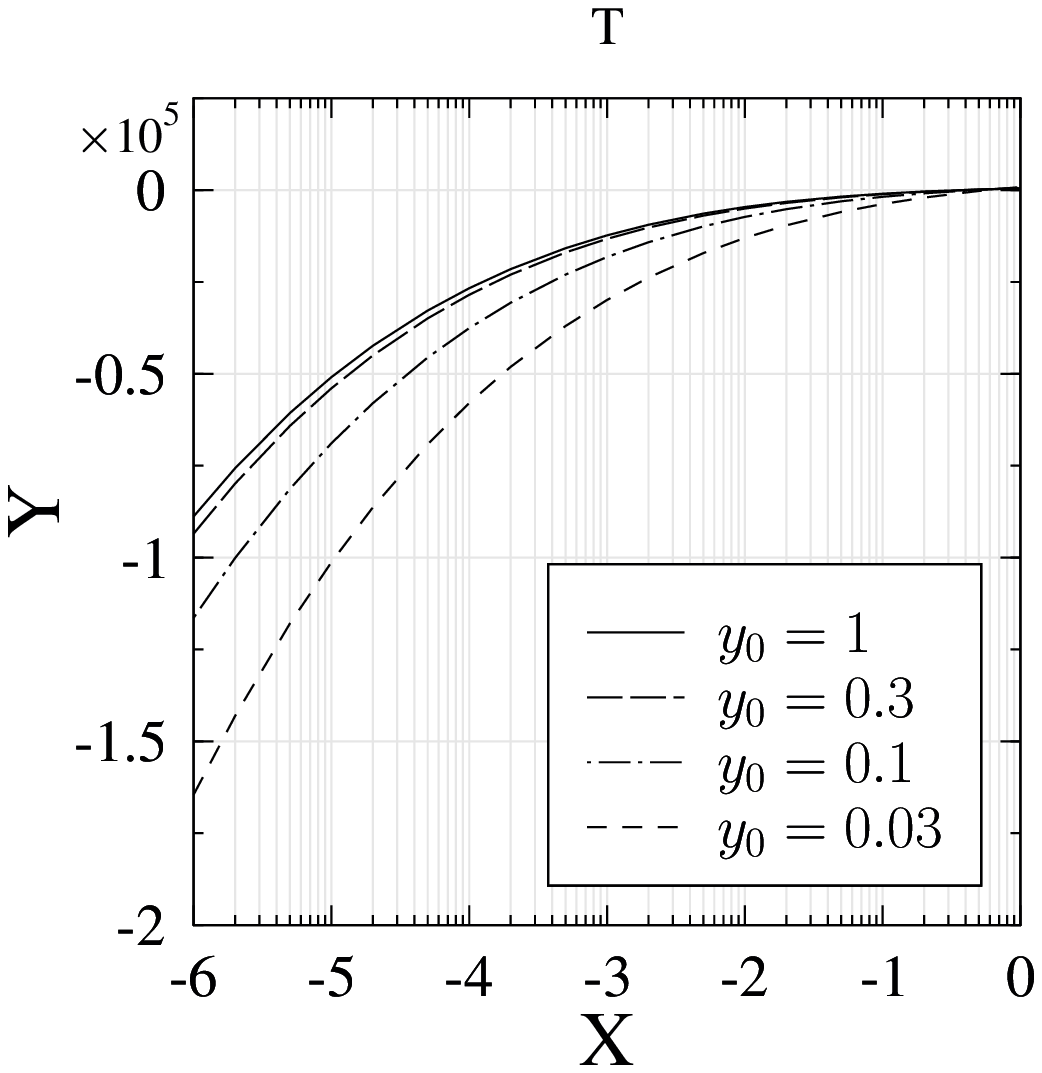}}
}
\caption{Expansion coeffiecients  of the function 
$\TcS{ik,\rm{bare}}{(1)}(Y;\eps)$ with $d'_0=d_0=3-3\eps$ and $\Nf=5$.}
}

\section{Conclusions}
\label{sec:conclude}

In this paper we computed the integrals over the phase space of the
unresolved parton of the singly-unresolved counterterms of the
subtraction scheme defined for computing QCD jet cross sections at the
NNLO accuracy in \Refs{Somogyi:2006da,Somogyi:2006db}. The results are
given in the form of Laurent-expansions in the regularization parameter
$\ep$ keeping the first five terms in the expansion. We computed the
coefficients of the two leading poles analytically, and the remaining
three coefficients numerically. The final forms of the integrals are
combined into insertion operators times various cross sections with the
same number of external legs as in the doubly-virtual cross section so
their combination into a single numerical integration is straightforward. 

The necessary integrations were carried out using iterated sector
decomposition and residuum subtraction. In order to check the rather
cumbersome computations the same integrals were computed by analytical
techniques as well. The details of those works are presented in separate
articles \cite{Aglietti:2008,Bolzoni:2008}.

We presented the expansion coefficients of the integrated subtraction
terms in the form of plots. These plots only serve to demonstrate that
in spite of the large complexity of the integrands, the resulting
functions are smooth, therefore, high-precision approximations can be
easily obtained. In an actual computation of a physical observable at
NNLO accuracy one needs only the $\Oe{0}$ piece in its expansion, for
which one can use any approximation that describes the function with
better than the expected relative accuracy of the observable in the
kinematical region of interest.  The coefficients of the lower orders
are needed only to check the cancellation of the epsilon poles, which
has to be done only once independently of the observable. Since the
numerical integrations that compute the integrated counterterms
converge quickly, any resonable precision of the cancellation can be
achieved.

\section*{Acknowledgments}
This research was supported by the Hungarian Scientific Research Fund
grant OTKA K-60432 and by the Swiss National Science Foundation (SNF)
under contract 200020-117602.  We are grateful to our collaborators in
computing the integrals: U.~Aglietti, P.~Bolzoni, V.~Del Duca, C.~Duhr
and S.~Moch.  Z.T.~is grateful to the members of the Institute of
Theoretical Physics at the University of Z\"urich for their hospitality.
We thank the Galileo Galilei Institute for Theoretical Physics for the
hospitality and the INFN for partial support during the completion of
this work.

%%%
%%% Appendix
%%%

\appendix

%%%
%%% Generalized subtraction terms
%%%

\section{Modified subtraction terms}
\label{app:modification}

In this Appendix, we describe a simple modification of the NNLO 
subtraction scheme presented in \Refs{Somogyi:2006da,Somogyi:2006db}.
The purpose of the modification is twofold: first, we wish to make
the integrated subtraction terms independent of the number of hard
partons, $m$. Secondly, to save CPU time and have a better control on the 
numerical calculation, we restrict the phase space on which the various 
counterterms are subtracted. This has been found useful in the 
context of NLO calculations
\cite{Nagy:1998bb,Nagy:2003tz,Campbell:2004ch}.

We recall the details of our NNLO subtraction scheme only to the extent 
we need to define the modification of the singly-unresolved counterterms.
For further details, we refer the reader to the original papers.

%
% Generalized doubly-real singly-unresolved subtraction terms
%

\subsection{Modification of singly-unresolved subtraction terms}
\label{app:genSU}

We first consider the singly-unresolved approximate cross section
$\dsiga{RR}{1}_{m+2}$ appearing in \eqn{eq:sigmaNNLOm+2}. We write
this term in the following symbolic form:
\beq
\dsiga{RR}{1}_{m+2} = \PS{m+1}[\rd p_1] \bcA{1}\M{m+2}{(0)}\,,
\eeq
where the singly-unresolved approximation $\bcA{1}\M{m+2}{(0)}$
is a sum of collinear, soft and soft-collinear terms (see \eqn{eq:dsigRRA1}).
The precise definition of these three terms involves the specification
of two momentum mappings (see \eqns{eq:cmap}{eq:smap})
\beq
\mom{} \cmap{ir} \momti{(ir)}\,,
\qquad
\mom{} \smap{r} \momti{(r)}\,,
\label{eq:single_mappings}
\eeq
both of which lead to an exact factorization of phase space in the
form
\beq
\PS{m+2}(\mom{}) = \PS{m+1}(\momt{})[\rd p_{1,m+1}]\,.
\label{eq:single_PSfact}
\eeq
The exact form of the one-particle factorized phase space 
$[\rd p_{1,m+1}]$ is irrelevant for the present argument, its
only feature which is important for us now is that it depends
on the number of hard partons $m$ through factors of 
$(1-\alpha_{ir})^{2m(1-\ep)-1}$ and $(1-y_{rQ})^{m(1-\ep)-1}$
for the collinear and soft mappings respectively, i.e. 
\beeq
~[\rd p_{1,m+1}^{(ir)}] &\propto& (1-\alpha_{ir})^{2m(1-\ep)-1}\,,
\label{eq:CirPS(1-a)}
\\
~[\rd p_{1,m+1}^{(r)}] &\propto& (1-y_{rQ})^{m(1-\ep)-1}\,
\label{eq:SrPSl}
\eeeq
(see \eqns{eq:dp1Cir}{eq:dp1Sr}).
The subtraction terms, as originally defined in \Ref{Somogyi:2006da}, 
are $m$-independent, therefore, the $m$-dependence of $[\rd p_{1,m+1}]$
is carried over to an $m$-dependence of the integrated subtraction terms.
However, this dependence on $m$ enters in a rather cumbersome way
in the integrated counterterms (see e.g.\ Eqs.\ (A.9) and (A.10) 
of \Ref{Somogyi:2006cz}).

It is then worthwhile to reshuffle the $m$-dependence of the integrated
counterterms into the subtraction terms themselves, where it appears in
a very straightforward and harmless way through factors of
$(1-\alpha_{ir})$ and $(1-y_{rQ})$ raised to $m$-dependent powers
multiplying the original subtraction terms as in Eqs.\ (\ref{eq:Cir00}),
(\ref{eq:Sr00}) and (\ref{eq:CirSr00}). These factors do not influence
the behavior of the subtraction terms in the infrared limits because
\beq
\bC{ir}(1-\alpha_{ir})=1\,,
\qquad\mbox{and}\qquad
\bS{r}(1-y_{rQ})=1\,,
\label{eq:limits_1}
\eeq
where $\bC{ir}$ and $\bS{r}$ are the symbolic operators to take the
collinear limit of momenta $p_i^\mu$ with $p_r^\mu$, or the soft limit of
momentum $p_r^\mu$, as defined precisely in \Ref{Somogyi:2005xz}.
Furthermore, the modified soft-collinear term still cancels the overlap
of the soft and collinear terms correctly due to 
\beq
\bS{r}(1-\alpha_{ir})=1\,.
\label{eq:limits_2}
\eeq
Thus, we are free to include the factors of $(1-\alpha_{ir})$ and 
$(1-y_{rQ})$ as in Eqs.\ (\ref{eq:Cir00}), (\ref{eq:Sr00}) 
and (\ref{eq:CirSr00}). 

The $\Theta$-functions that also appear in Eqs.\ (\ref{eq:Cir00}), 
(\ref{eq:Sr00}) and (\ref{eq:CirSr00}) control the region of the
$(m+2)$-particle phase space over which the subtraction is non-zero
such that $\alpha_0=1$ and/or $y_0=1$ corresponds to subtracting 
over the full phase space. In a NNLO computation the (very) large
number of subtraction terms and their complicated analytic structure
makes the evaluation of the approximate cross sections rather time
consuming. Constraining the phase space over which the subtraction
is non-zero can result in large gains in CPU time. The introduction
of the cutoffs also provides a strong consistency check on the whole
of the calculation: the final results should be independent of
$\alpha_0$ and $y_0$. Finally, suitably chosen values of the cut
parameters can actually improve the numerical behavior of the code.

%
% Modified real-virtual subtraction terms
%

\subsection{Modified real-virtual subtraction terms}
\label{app:genRV}

The real-virtual subtraction terms $\dsiga{RV}{1}_{m+1}$ and
$\Big(\int_1\dsiga{RR}{1}_{m+2}\Big)
\strut^{{\rm A}_{\scriptscriptstyle 1}}$ which appear in 
\eqn{eq:sigmaNNLOm+1} read symbolically
\beq
\dsiga{RV}{1}_{m+1} = 
\PS{m}{}[\rd p_1]\,\bcA{1}
\,2\Real \bra{m+1}{(0)}{}\ket{m+1}{(1)}{}\,,
\label{eq:dsigRVA1appx}
\eeq
and
\beq
\Big(\int_1\dsiga{RR}{1}_{m+2}\Big)\strut^{{\rm A}_{\scriptscriptstyle 1}} =
\PS{m}{}[\rd p_1] \bcA{1}
\Big(\M{m+1}{(0)}\otimes\bI^{(0)}(\mom{}_{m+1};\ep) \Big)
\,.
\label{eq:dsigRRA11}
\eeq
The singly-unresolved approximations
\beq
\bcA{1}\,2\Real \bra{m+1}{(0)}{}\ket{m+1}{(1)}{}
\qquad\mbox{and}\qquad
\bcA{1}\Big(\M{m+1}{(0)}\otimes\bI^{(0)}_{m+1}(\mom{};\ep) \Big)
\eeq
are sums of collinear, soft and soft-collinear terms (see
\eqns{eq:dsigRVA1}{eq:dsigRRA1A1}).
The subtraction terms in \eqns{eq:dsigRVA1}{eq:dsigRRA1A1} 
are all defined using the singly-unresolved momentum mappings 
of \eqn{eq:single_mappings}, which now map the original set of 
$m+1$ momenta appearing in the one-loop squared matrix element 
into a set of $m$ momenta. The appropriate phase space factorization 
reads
\beq
\PS{m+1}(\mom{}) = \PS{m}(\momt{})[\rd p_{1,m}]\,,
\label{eq:single_PSfact_2}
\eeq
which is of course just \eqn{eq:single_PSfact} with the substitution 
$m\to m-1$. Again the factorized one-particle phase spaces carry an
$m$-dependence through factors of $(1-\alpha_{ir})$ and 
$(1-y_{rQ})$ raised to $m$-dependent powers. For the collinear and soft 
mappings we have respectively 
\beeq
~[\rd p_{1,m}^{(ir)}] &\propto& (1-\alpha_{ir})^{2(m-1)(1-\ep)-1}\,,
\label{eq:CirPS(1-a)_2}
\\
~[\rd p_{1,m}^{(r)}] &\propto& (1-y_{rQ})^{(m-1)(1-\ep)-1}\,.
\label{eq:SrPSl_2}
\eeeq
These equations are again just \eqns{eq:CirPS(1-a)}{eq:SrPSl} with 
the replacement $m\to m-1$.

By the arguments of the previous section, we find it useful to 
modify the real-virtual counterterms similarly as done with the
doubly-real singly unresolved terms. The only difference is the shift of
$m\to m-1$. The exact definitions are presented in \sect{sec:RVA1} and 
\sect{sec:IRRA1_A1} of the present paper.

%
% Remarks on choosing $d_0$ and $d'_0$
%

\subsection{Remarks on choosing $d_0$ and $d'_0$}
\label{app:ddprimechoice}

As has been emphasized repeatedly, the integrals of the subtraction
terms as defined in the present paper are $m$-independent for any
$d_0$ and $d'_0$. In this paper we set $d_0=d'_0=3-3\eps$.
Let us make a few comments on this choice.

First of all, to avoid the introduction of spurious poles at
$\alpha_{ir}=1$ and/or $y_{rQ}=1$, we need to choose $d_0$ and $d'_0$
such that the overall powers of $(1-\alpha_{ir})$ and/or $(1-y_{rQ})$ 
that appear in any integral are non-negative. Although it is not manifest
by considering only the singly-unresolved counterterms, it actually
turns out that in the full NNLO scheme this implies
\beq
d_0\big|_{\ep=0} \, , \, d'_0\big|_{\ep=0} \ge 2\,.
\label{eq:const_1}
\eeq

Secondly, we might want to avoid the appearance of negative powers of
$(1-\alpha_{ir})$ and $(1-y_{rQ})$  in the subtraction terms themselves.
This is because away from the limits both of these factors are between
zero and one, thus if they are raised to a negative power, we are
multiplying the subtraction terms by quantities that are greater than
one, i.e.\ we are `over-subtracting' away from the limits. This leads us 
to choose
\beq
d_0\big|_{\ep=0} \, , \, d'_0\big|_{\ep=0} \ge m\,.
\label{eq:const_2}
\eeq
Since our primary interest is in 3-jet production in electron-positron
annihilation, we set $d_0\big|_{\ep=0}=d'_0\big|_{\ep=0}=3$.

Finally, to fix the $\eps$-dependent part, we note that by choosing
$d_0 = d'_0 = m(1-\eps)$
the overall powers of $(1-\alpha_{ir})$ and $(1-y_{rQ})$ in the 
doubly-real singly-unresolved subtraction terms (\eqn{eq:Cir00} and
\eqns{eq:Sr00}{eq:CirSr00}) are zero. Since we have already evaluated
the integrated subtraction terms for this case (with $\alpha_0=1$
and $y_0=1$) in \Ref{Somogyi:2006cz}, the choice of $d_0=d'_0=3-3\eps$
is a natural one.

\section{Spin-averaged splitting kernels}
\label{app:APfcns}

In this Appendix we recall the explicit expressions for the 
spin-averaged splitting kernels that enter \eqns{eq:ICir0}{eq:ICir1}.

The azimuthally averaged Altarelli--Parisi splitting kernels read
\beeq
P^{(0)}_{g_ig_r}(z_i,z_r;\ep) 
&=&
2\CA \left[\frac{1}{z_i}+\frac{1}{z_r}-2+z_i z_r\right]\,,
\label{eq:P0gg}
\\
P^{(0)}_{q_i\qb_r}(z_i,z_r;\ep) 
&=&
\TR \left[1-\frac{2}{1-\ep}z_i z_r\right]\,,
\label{eq:P0qq}
\\
P^{(0)}_{q_ig_r}(z_i,z_r;\ep) 
&=&
\CF \left[\frac{2}{z_r}-2 + (1-\ep)z_r\right]\,,
\label{eq:P0qg}
\eeeq
while their one-loop generalizations are
\beq
P^{(1)}_{f_i f_r}(z_i,z_r;\ep) 
=
r_{\S,\rm{ren}}^{f_i f_r}(z_i,z_r;\ep)
P^{(0)}_{f_i f_r}(z_i,z_r;\ep) 
+\left\{
\begin{array}{lr}
\displaystyle{
2\CA\, r_\NS^{gg}\,\frac{1-2\ep z_i z_r}{1-\ep}}\,,
& \mbox{if}\quad f_if_r = gg\,,
\\ 0\,, & \mbox{if}\quad f_if_r = q\qb\,, \\
\displaystyle{
\CF\, r_\NS^{qg}\,(1-\ep z_r)}\,, 
& \mbox{if}\quad f_if_r = qg\,.
\end{array}
\right.
\label{eq:P1av}
\eeq
The renormalized $r^{f_i f_r}_{\S,\rm{ren}}(z_i,z_r;\ep)$ functions 
that appear above are expressed in terms of the corresponding 
unrenormalized ones as
\beq
r^{f_i f_r}_{\S,\rm ren}(z_i,z_r;\ep)=
r^{f_i f_r}_\S(z_i,z_r;\ep)
- \frac{\beta_0}{2\ep}\,\frac{S_\ep}{(4\pi)^2 c_\Gamma}
\,\left[\left(\frac{\mu^2}{s_{ir}}\right)^{\ep}\cos(\pi\ep)\right]^{-1}
\,,
\label{rSUV}
\eeq
where 
the unrenormalized $r^{f_i f_r}_{\S}(z_i,z_r;\ep)$ factors may be
written in the following form
\beeq
r_{\S}^{gg}(z_i,z_r;\ep) &=& 
\frac{\CA}{\ep^2}\bigg[-\frac{\pi\ep}{\sin(\pi\ep)} 
\left(\frac{z_i}{z_r}\right)^\ep + z_i^\ep {}_2F_1(\ep,\ep,1+\ep,z_r)
\nn\\&&\qquad \quad
- z_i^{-\ep} {}_2F_1(-\ep,-\ep,1-\ep,z_r)\bigg]\,,
\label{rSgg}
\eeeq
%\\
\beeq
r_{\S}^{q\qb}(z_i,z_r;\ep) &=& 
\frac{1}{\ep^2}(\CA-2\CF)+
\frac{\CA}{\ep^2}\bigg[-\frac{\pi\ep}{\sin(\pi\ep)} 
\left(\frac{z_i}{z_r}\right)^\ep + z_i^\ep {}_2F_1(\ep,\ep,1+\ep,z_r)
\nn\\&&\qquad\qquad\qquad\qquad\quad\;\;
-\,\frac{\pi\ep}{\sin(\pi\ep)} 
\left(\frac{z_r}{z_i}\right)^\ep + z_r^\ep {}_2F_1(\ep,\ep,1+\ep,z_i)
\bigg]
\nn\\&&\qquad\quad
+\,\frac{1}{1-2\ep}\left[\frac{\beta_0-3\CF}{\ep}+\CA-2\CF
+\frac{\CA+4\TR(\Nf-\Ns)}{3(3-2\ep)}\right]\,,
\label{rSqq}
\\
r_{\S}^{qg}(z_i,z_r;\ep) &=& 
-\frac{1}{\ep^2}\bigg[2(\CA-\CF) + \CA \frac{\pi\ep}{\sin(\pi\ep)} 
\left(\frac{z_i}{z_r}\right)^\ep - \CA z_i^\ep {}_2F_1(\ep,\ep,1+\ep,z_r)
\nn\\&&\qquad \quad
-(\CA-2\CF)z_i^{-\ep} {}_2F_1(-\ep,-\ep,1-\ep,z_r)\bigg]\,.
\label{rSqg}
\eeeq
The $r^{f_i f_r}_\NS$ non-singular factors are
\beq
r_\NS^{g g} =
\frac{\CA(1-\ep)-2\TR (\Nf-\Ns)}{(1-2\ep)(2-2\ep)(3-2\ep)}\,,
\qquad
r_\NS^{q g} =
\frac{\CA-\CF}{1-2\ep}\,.
\label{rNSqg}
\eeq
For QCD, $\Ns=0$ of course.
Finally $\beta_0$ in \eqns{rSUV}{rSqq} is given in \eqn{eq:beta0}.

%%%
%%% The bibliography
%%%

\end{document}